%% file: main.tex
\documentclass{article}

\usepackage{arxiv}

\usepackage{geometry}
\usepackage[utf8]{inputenc} 
\usepackage[T1]{fontenc}    
\usepackage{hyperref}       
\usepackage{url}            
\usepackage{booktabs}       
\usepackage{amsfonts}       
\usepackage{amsmath}        
\usepackage{nicefrac}       
\usepackage{microtype}      
\usepackage{lipsum}
\usepackage{graphicx}
\usepackage{array}
\usepackage{gensymb}
\usepackage{color}
\usepackage{float}
\usepackage{bm}
\usepackage{pdflscape}
\usepackage{makecell}
\usepackage{multirow}

\usepackage{tabularx}
\newcolumntype{Y}{>{\centering\arraybackslash}X}

\usepackage[backend=biber,style=numeric-comp,sorting=none,defernumbers=true]{biblatex}

\usepackage{adjustbox}
\usepackage{threeparttable}
\usepackage{xcolor}
\usepackage{soul}
\usepackage{rotating}

\usepackage[dvipsnames]{xcolor}

\soulregister\cite7
\soulregister\parencite7
\soulregister\textcite7
\soulregister\autocite7
\soulregister\citeauthor7
\soulregister\citeyear7

\soulregister\etal7

\soulregister\ref7
\soulregister\pageref7
\soulregister\eqref7
\soulregister\autoref7
\soulregister\nameref7

\newif\ifshowhl
\showhlfalse

\newcommand{\revised}[2][yellow]{%
  \ifshowhl
    \ifmmode
      \colorbox{#1}{$\mathchoice%
        {\displaystyle #2}%
        {\textstyle #2}%
        {\scriptstyle #2}%
        {\scriptscriptstyle #2}$}%
    \else
      \begingroup
      \sethlcolor{#1}%
      \hl{#2}%
      \endgroup
    \fi
  \else
    #2%
  \fi
}

\newif\ifshowhlnew
\showhlnewfalse

\newcommand{\revisednew}[2][yellow]{%
  \ifshowhlnew
    \ifmmode
      \colorbox{#1}{$\mathchoice%
        {\displaystyle #2}%
        {\textstyle #2}%
        {\scriptstyle #2}%
        {\scriptscriptstyle #2}$}%
    \else
      \begingroup
      \sethlcolor{#1}%
      \hl{#2}%
      \endgroup
    \fi
  \else
    #2%
  \fi
}

\newif\ifshowhlfin
\showhlfinfalse

\newcommand{\revisedfin}[2][yellow]{%
  \ifshowhlfin
    \ifmmode
      \colorbox{#1}{$\mathchoice%
        {\displaystyle #2}%
        {\textstyle #2}%
        {\scriptstyle #2}%
        {\scriptscriptstyle #2}$}%
    \else
      \begingroup
      \sethlcolor{#1}%
      \hl{#2}%
      \endgroup
    \fi
  \else
    #2%
  \fi
}
\makeatletter
\def\def@donotrepeattitle{}
\makeatother

\newcommand{\convnext}{ConvNeXt}

\newcommand{\etal}{\emph{et al.}}

\newcolumntype{C}[1]{>{\centering\arraybackslash}m{#1}}

\DeclareCiteCommand{\citenum}
  {}
  {\printfield{labelnumber}}
  {}
  {}
  
\input{metadata}

\input{supplemental_refs}

\begin{document}
\savegeometry{arxivmargins}

\input{license}

\maketitle
\begin{abstract}Materials identification and structural understanding from powder X-ray diffraction (PXRD) data are long-standing challenges in materials science, fundamental to discovering and characterizing novel materials. A prerequisite for full structure solution is the accurate determination of the crystal lattice, including lattice parameters and crystallographic symmetries. Traditional methods for this are iterative and typically require expert input, and while existing deep learning approaches have shown promise, a robust, single-shot method for comprehensive lattice determination from experimental data remains a key goal. Here, we introduce AlphaDiffract, a deep learning framework \revisednew{for jointly} predicting the crystal system, space group, and lattice parameters directly from PXRD patterns. AlphaDiffract utilizes a 1D adaptation of the ConvNeXt architecture, a modern convolutional neural network that integrates key design principles from transformers, coupled with dedicated prediction heads for each crystallographic property. The model is trained and validated on the largest-to-date physics-based dataset of over 31 million simulated diffraction patterns, generated by augmenting 312,267 curated structures from the ICSD and Materials Project databases. Crucially, it demonstrates strong \revisednew{transfer from simulated training data to experimental measurements}, achieving 81.7\% crystal system accuracy and 66.2\% space group accuracy on the RRUFF dataset while additionally predicting all six lattice parameters. By providing a unified model for rapid\revisednew{, simultaneous prediction of lattice symmetry and parameters} from PXRD data, AlphaDiffract represents a significant step forward in leveraging deep learning for high-throughput materials discovery.
\end{abstract}

\section*{Introduction}

Powder X-ray diffraction (PXRD) is a fundamental characterization technique in materials science, essential for the discovery and understanding of novel materials. Determining the crystal lattice, which decodes the 1D diffraction pattern into the 3D unit cell geometry, is often a prerequisite for full structure solution from PXRD data \cite{RN27}. While this is routine for single-crystal measurements, powder diffraction presents significant challenges. Traditional indexing algorithms can accomplish this for high-quality data \cite{RN111,RN188,RN190}, and experimental advances have further improved reliability; for instance, synchrotron sources eliminate sample offset uncertainties, while high-resolution measurements enable accurate determination of peak positions even for closely spaced reflections \cite{RN7}. \revised{Once lattice parameters have been determined, probabilistic methods can assist in identifying the compatible space group from systematic absences and reflection conditions \cite{markvardsen2001probabilistic}.} However, these methods still struggle in many realistic scenarios, particularly when materials contain impurities or when experimental conditions introduce peak broadening and overlap.

Recent deep learning approaches have demonstrated promise for automated crystallographic analysis from PXRD data. Convolutional neural networks (CNNs) are the most widely adopted architecture \cite{Park_iucrj_2017, Vecsei_prb_2019, Zaloga_mtc_2020, Chitturi_jac_2021, Lee_ais_2022, Lee_ais_2023, Corriero_jac_2023, Gomez_jpca_2023, Salgado_npj_2023, Riesel_jacs_2024} as they capture both local peak features and global pattern structure through multiscale receptive fields. Alternative architectures such as multi-layer perceptrons \cite{Vecsei_prb_2019, Liang_prm_2020, Lee_ais_2022} and transformers \cite{Lee_ais_2022} have also been explored. These methods typically train on large synthetic datasets generated from crystallographic databases such as ICSD \cite{zagorac_appliedcryst_2019}, Materials Project \cite{jain_aplmat_2013}, \revisedfin{and Crystallography Open Database (COD) \cite{gravzulis2009crystallography}}, often incorporating data augmentation strategies to improve robustness \cite{Lee_ais_2023, Salgado_npj_2023}. The RRUFF database of predominantly experimental patterns \cite{lafuente_hmc_2015} has emerged as a key benchmark for evaluating generalization to real-world data \cite{Vecsei_prb_2019, Lee_ais_2023, Salgado_npj_2023,Riesel_jacs_2024}. More recently, generative approaches, including large language models \cite{Choudhary_jpcl_2025,johansen2025decifer} and diffusion models \cite{Riesel_jacs_2024}, have been applied to obtain full structure predictions from PXRD patterns, though the former depend sensitively on knowledge of the chemical formula while the latter require separate predictions of the composition and lattice parameters. \revised{A recent approach by Guo \textit{et al.}~\cite{guo2025pxrdnet} extends this further to nanocrystalline structure determination, using a diffusion model conditioned on chemical formula and diffraction patterns with significant finite-size broadening to generate structural candidates across all seven crystal systems, though restricted to materials with at most 20 atoms per unit cell.}

Despite this progress, significant challenges remain for lattice determination from PXRD data. Supplementary Table 1 provides a detailed comparison of recent PXRD-based prediction methods, highlighting their architectures, training data sources, and predicted outputs. Most existing work focuses only on crystal system and space group classification \cite{Park_iucrj_2017, Vecsei_prb_2019, Zaloga_mtc_2020, Lee_ais_2023, Corriero_jac_2023, Salgado_npj_2023}. Approaches that predict lattice parameters either train separate models for each Bravais lattice or crystal system \cite{Liang_prm_2020, Chitturi_jac_2021}, requiring prior symmetry knowledge, depend on chemical composition as input and require subsequent refinement for experimental data \cite{Gomez_jpca_2023}, \revisednew{or transfer features learned from a broad synthetic corpus to related prediction tasks and material domains \cite{gomez2025insights}.} Moreover, models trained primarily on idealized synthetic data frequently struggle with realistic experimental conditions, including noise, peak broadening, and instrumental effects. A unified framework that simultaneously predicts crystal system, space group, and lattice parameters while maintaining robust performance on experimental data remains an open challenge.

To address these challenges, we present AlphaDiffract, a unified deep learning framework for lattice determination from PXRD patterns. Our approach achieves \revisednew{competitive benchmark} performance on the RRUFF dataset, reaching 81.7\% crystal system accuracy and 66.2\% space group accuracy while simultaneously predicting all six lattice parameters. Building on insights from prior work, our framework incorporates three key innovations:

\begin{enumerate}
    \item While CNNs remain a popular and effective choice for analyzing sequential scientific data like PXRD patterns, creating an architecture that both excels at identifying local features, such as individual diffraction peaks, and captures the complex, long-range dependencies that encode crystallographic symmetry is challenging. To address this, we employ a 1D adaptation of the \convnext{} architecture \cite{Liu_arxiv_2022}. \convnext{} is a modern CNN architecture adapted from ResNet \cite{He_arxiv_2015}, incorporating key design features of transformers that improve the parameter efficiency and modeling of long-range interactions. These improvements make \convnext{} well-suited for PXRD data analysis, as the architecture inherits key strengths of transformers for processing sequential data while maintaining the spatial inductive biases that make CNNs effective for pattern recognition.

    \item Our model is trained on simulated PXRD patterns generated using a combination of structures from the ICSD and Materials Project databases, \revisednew{while RRUFF is used as a predominantly experimental development benchmark} for evaluating generalization \revisednew{from synthetic training data to realistic measured XRD patterns}. \revisednew{Because RRUFF informs a limited number of model-development decisions, we refer to it hereafter as the RRUFF benchmark dataset.} Following the curation process detailed in Methods \revisednew{section ``\nameref{subsec:curation},''} our combined ICSD and Materials Project dataset consists of 312,267 crystal structures. For each structure, we simulated 100 augmented diffraction patterns with randomized perturbations mimicking physical effects such as microstrain and crystallite size broadening (see \revisednew{``\nameref{sec:data_gen}''}). This data augmentation strategy produced a final simulated dataset of over 31 million diffraction patterns, making it, to our knowledge, the largest and most comprehensive dataset ever compiled for this task.

    \item A key design feature of our approach is a single, end-to-end model that simultaneously predicts the crystal system, space group, and lattice parameters from the PXRD pattern alone. This unified architecture streamlines the crystallographic analysis pipeline, removing the need for separate, specialized models for each task and enabling \revisednew{simultaneous prediction of the lattice symmetry and parameters} in a single inference step.
\end{enumerate}

\revised{We note that AlphaDiffract is not designed to replace phase-identification software such as JADE or EVA, which match measured patterns against databases of known crystalline phases to identify sample composition. AlphaDiffract instead predicts crystallographic properties from the diffraction pattern alone, without reference to a known-phase database. This makes it applicable to unknown or novel materials for which no database entry exists, complementing rather than competing with existing phase-identification tools.} In the following sections, we detail our data generation pipeline, model architecture and training, and comprehensive evaluation on simulated and experimental benchmark datasets. 

\section*{Results}

\subsection*{Data preparation and physics-based simulation}
\label{sec:data_gen}

To generate powder diffraction patterns matching experimental results, we used the GSAS-II crystallographic package via the \texttt{GSASIIscriptable} API \cite{toby:gsasii,G2scripting}. Since GSAS-II is designed for Rietveld analysis, it provides quantitatively accurate simulations of real instrumental data. Patterns were simulated for structures from the NIST version of the ICSD database \cite{NIST_ICSD} and the Materials Project \cite{jain_aplmat_2013} to create our training and validation datasets. All simulations employed a monochromatic X-ray source with 20 keV energy over a $2\theta$ range of $5^\circ$ to $20^\circ$ with 8192 equally spaced points. \revised{The choice of 20 keV X-ray energy and $5-20^\circ$ $2\theta$ range is motivated by the geometry of synchrotron beamlines at the Advanced Photon Source. At 20 keV ($\lambda = 0.62$ \AA), this angular range corresponds to $d$-spacings of approximately $1.78-7.10$ \AA, which closely matches the range probed by a standard laboratory Cu K$_\alpha$ ($\lambda = 1.54$ \AA) measurement spanning approximately $12-51^\circ$ $2\theta$. Although the model is trained at 20 keV, patterns collected at other wavelengths are mapped into this equivalent $d$-spacing range via Bragg’s law; this is the procedure used for the RRUFF \revisednew{benchmark dataset}, which includes many patterns measured with Cu K$_\alpha$ radiation (see Methods \revisednew{section ``\nameref{subsec:curation}''}). While a larger $2\theta$ range could be used for training, a goal of this work was to intentionally focus on the information-dense region of $Q$-space that can be sampled in a rapid, routine measurement.} \revisednew{A direct comparison of angular ranges, sampling, and wavelengths used across benchmarked methods is provided in Supplementary Table~\tabPxrdInputComparison.} To desensitize the model to experimental parameters that vary between diffraction instruments, we generated 100 different simulations per structure with randomized instrumental and sample parameters. Values for microstrain and crystallite size (both contributing Lorentzian broadening) and Gaussian instrumental broadening parameters were randomly sampled for each simulation within the ranges listed in Table \ref{tab:sim}, producing heterogeneous peak shapes representative of diverse experimental conditions \cite{ReidProfiles}. During training \revisedfin{and evaluation of simulated datasets}, additional Poisson and Gaussian noise was applied dynamically to each pattern as detailed in Methods \revisednew{section ``\nameref{subsec:noise},''} and the intensities were subsequently standardized between 0 and 100 prior to being input to the model.

\begin{figure}[t]
    \centering
    \includegraphics[width=\linewidth]{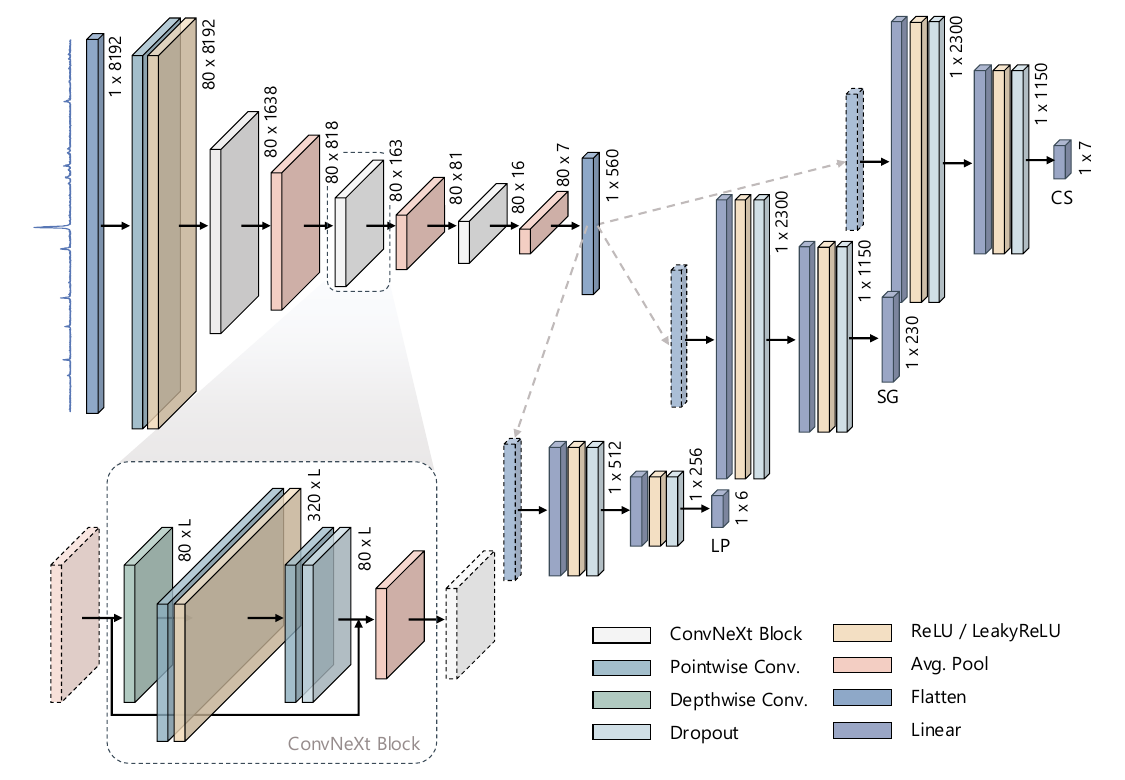}
    \caption{\textbf{AlphaDiffract model architecture.} The AlphaDiffract model consists of a 1D ConvNeXt backbone that processes input PXRD patterns through a series of ConvNeXt blocks with progressive downsampling. The composition of each ConvNeXt block is indicated in the bottom left inset. The extracted features are fed into three separate prediction heads: a crystal system (CS) classifier, a space group (SG) classifier, and a lattice parameter (LP) regressor. Each head employs a multi-layer perceptron architecture with layer dimensions as indicated.}
    \label{fig:model}
\end{figure}

\subsection*{AlphaDiffract architecture}

The AlphaDiffract model extracts crystallographic information from a PXRD pattern through a 1D ConvNeXt backbone coupled with three specialized prediction heads. As illustrated in Figure \ref{fig:model}, the backbone serves as a feature extractor that feeds into three distinct multilayer perceptron (MLP) heads for targeted predictions. The end-to-end model allows for the simultaneous prediction of the crystal system (CS), space group (SG), and all six lattice parameters (LP). \revised{The lattice parameters used for both model training and evaluation are those of the Niggli-reduced cell. The Niggli reduction} \revisednew{provides a canonical reduced-cell representation, removing the arbitrariness associated with the initial choice of unit cell setting.}

The feature extractor is a 1D adaptation of the ConvNeXt architecture \cite{Liu_arxiv_2022}, a convolutional neural network (CNN) that incorporates principles from vision transformers, such as depthwise separable convolutions, an inverted bottleneck design, and large kernel sizes. The input PXRD pattern, a 1 $\times$ 8192 vector, is processed through a series of ConvNeXt blocks. Each block, shown in the inset of Figure \ref{fig:model}, uses a depthwise convolution to capture spatial patterns (\emph{i.e.}, peak shapes and local arrangements) followed by pointwise convolutions to learn channel-wise interactions. This structure is parameter-efficient for modeling relationships across the pattern. After each block, average pooling progressively downsamples the feature map, increasing the network's receptive field and enabling it to capture long-range dependencies between distant diffraction peaks. The final output of the backbone is a 560-dimensional feature vector that encodes key features from the raw diffraction pattern to enable the downstream classification and regression tasks.

This feature vector is passed to three specialized MLP prediction heads, each tailored for a specific task: a 7-node output for crystal system classification, a 230-node output for space group classification, and a 6-node output for lattice parameter regression. \revised{The three prediction heads operate in parallel; each receives only the shared backbone feature vector as input, and no inter-head communication or cross-task conditioning is applied at inference time. This design was chosen for two main reasons:}
\begin{enumerate}
    \item \revisednew{It avoids the risk of hard-conditioning downstream lattice parameter prediction on a potentially incorrect discrete crystal system label, as would occur in a sequential pipeline.}
    \item \revisednew{The shared backbone allows the model to learn representations relevant to all three tasks jointly, without requiring separate, specialized feature extractors; we verify in ``\nameref{subsec:classification}'' that this maintains or slightly improves performance relative to the classification-only model, indicating that the shared backbone can support all three tasks without penalizing any individual one.}
\end{enumerate}

\revisednew{We note that these considerations support the parallel design choice made here but do not establish it as superior to all alternative architectures (see Discussion).} This multi-task approach allows the model to learn shared representations in the backbone while fine-tuning the predictions for each distinct crystallographic property. Details on the architectural parameters can be found in Methods \revisednew{section ``\nameref{sec:model_arch}.''}

\begin{figure}[t]
    \centering
    \includegraphics[width=\linewidth]{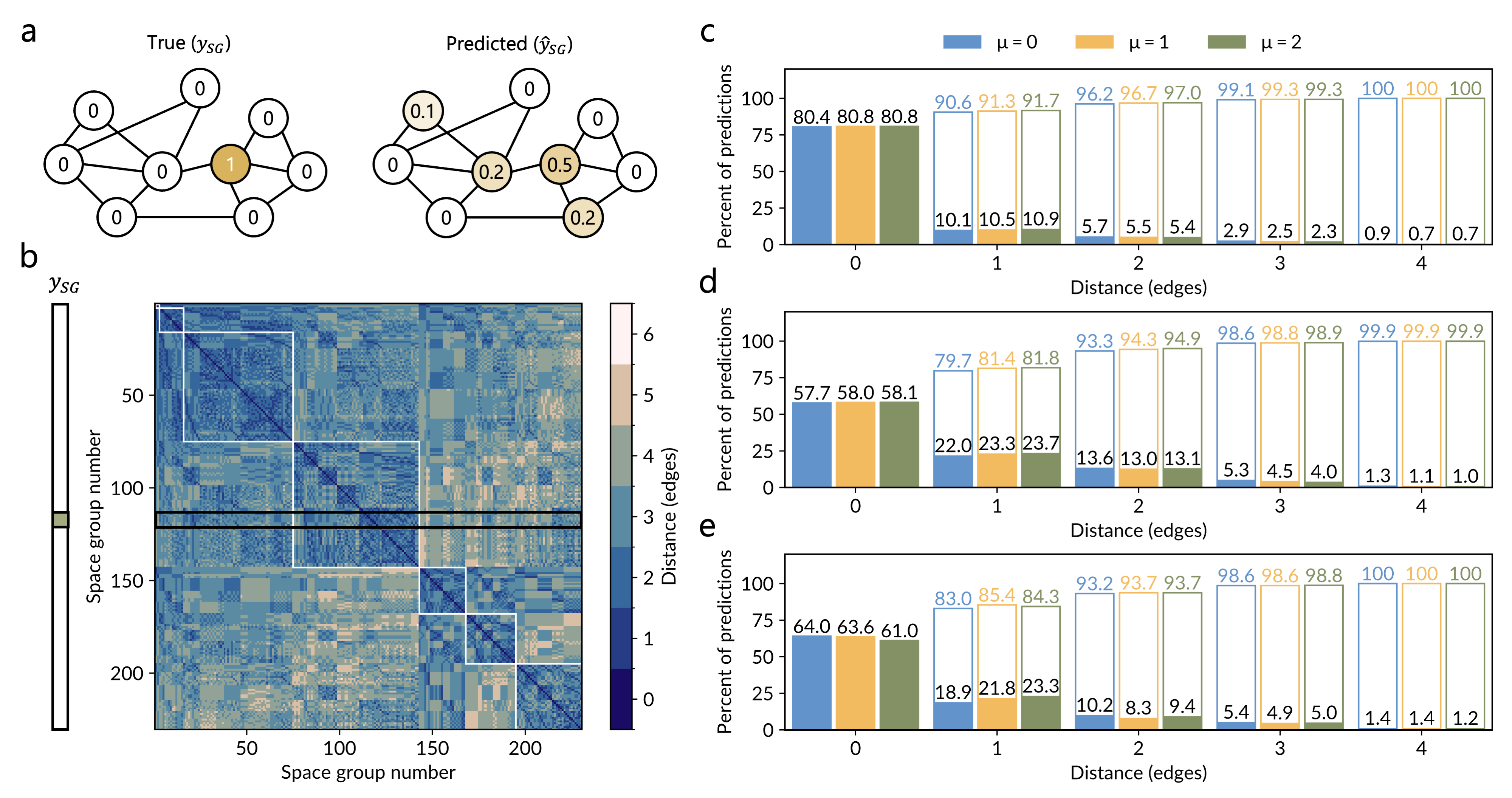}
    \caption{\textbf{Evaluation of space group predictions using Graph Earth Mover's Distance.} \textbf{a.} Illustration of true ($y_\text{SG}$) and predicted ($\hat{y}_\text{SG}$) space group probability distributions on a representative subgroup graph, where nodes represent space groups and edges indicate maximal subgroup relationships. The true label assigns probability 1 to a single node (yellow), while the predicted distribution typically spreads probability across multiple nodes. \textbf{b.} Distance matrix computed from maximal subgroup relationships between space groups, where color intensity indicates the minimum number of graph edges connecting each pair of space groups. The vertical vector ($y_\text{SG}$) represents a one-hot encoded true label that selects the corresponding row for calculating the GEMD loss against predicted distributions. \textbf{c-e.} Distribution of prediction errors as a function of graph distance (number of edges) from the true space group for three datasets: \textbf{c.} ICSD validation set, \textbf{d.} Materials Project validation set, and \textbf{e.} RRUFF \revisednew{benchmark}. \revised{Filled bars show the percentage of all predictions (including correct predictions at distance 0) that fall at each graph distance from the true space group, for three different GEMD loss weights ($\mu$ = 0, 1, 2). Distance zero indicates a correct prediction (predicted space group = true space group). Unfilled bars with labeled values indicate the corresponding cumulative percentages up to and including that distance.} With higher $\mu$ values, predictions become increasingly concentrated at shorter graph distances from the true space group.}
    \label{fig:gemd}
\end{figure}

\subsection*{Physics-aware loss function}

A key innovation in AlphaDiffract is a novel loss function for space group classification in addition to standard cross-entropy. While cross-entropy penalizes all misclassifications equally, space groups are structurally related through symmetry hierarchies where closely related sub- or supergroups are more similar, making some misclassifications less severe than others. To incorporate these relationships, we implement a Graph Earth Mover's Distance (GEMD) loss which adapts the traditional Earth Mover's Distance to the maximal subgroup graph, where nodes represent space groups and edges connect groups with direct maximal subgroup relationships, reflecting their structural similarity. This approach was inspired by the work of Vecsei \textit{et al.} that evaluated space group prediction performance using maximal subgroup distance as a metric \cite{Vecsei_prb_2019} but further develops this concept into a loss function that enforces these structural relationships during training. The GEMD loss uses a pre-computed 230 $\times$ 230 distance matrix $D$, where each element $D_{ij}$ encodes the number of hops between space groups $i$ and $j$ through this graph (Figure \ref{fig:gemd}\textbf{a} and \textbf{b}), which were determined using the MAXSUB program \cite{de2025complete} available on the Bilbao Crystallographic Server \cite{aroyo2006bilbao}. As illustrated in Figure \ref{fig:gemd}\textbf{a}, the true label assigns probability 1 to a single space group, while the predicted distribution typically spreads probability across multiple nodes. The GEMD loss calculates the cost to transport probability mass from the predicted distribution to the true distribution, where this cost is weighted by the distances in matrix $D$. This approach penalizes misclassifications based on their crystallographic dissimilarity, as predictions that are further from the true space group in the subgroup graph incur a larger penalty. The loss function includes a hyperparameter $\mu$ that controls the weight of the GEMD term relative to cross-entropy, allowing us to balance the two objectives during training. Additional details about the loss functions are provided in Methods \revisednew{section ``\nameref{subsec:loss}.''}

To determine the optimal weight for the GEMD loss term, we trained models with only the crystal system and space group prediction heads (excluding lattice parameters) using three different values of $\mu$ (0, 1, 2), where $\mu = 0$ corresponds to standard cross-entropy alone. We evaluate these models on our \revisednew{RRUFF benchmark dataset} as well as on the ICSD and Materials Project validation sets (Figures \ref{fig:gemd}\textbf{c}-\textbf{e}). \revised{In these figures, a graph distance of zero indicates an exact correct prediction; non-zero distances reflect misclassifications where the predicted space group is crystallographically separated from the true space group by the indicated number of edges in the subgroup graph.} We find that incorporating the GEMD loss shifts the error distribution toward shorter graph distances from the true space group; that is, when the model is wrong, it tends to predict space groups that are crystallographically close to the true space group in the subgroup hierarchy. This pattern is consistent across the RRUFF \revisednew{benchmark dataset} as well as the ICSD and Materials Project validation sets. \revised{The cumulative percentages demonstrate that with $\mu$ = 1, 85.4\% of predictions on RRUFF are correct or within one graph edge of the true space group, compared to 83\% with $\mu$ = 0. This improvement in crystallographic error quality comes at a marginal cost to top-1 accuracy (63.6\% vs. 64.0\% for $\mu$ = 0), reflecting the intended trade-off: The GEMD loss primarily pulls more distant errors (distance $\geq 2$) into the nearest-neighbor bin, with a small fraction of previously correct predictions shifting to higher distance bins as a side effect.} Based on these results, we select $\mu$ = 1 for training the full model with all three prediction heads.

\begin{table}[t]
    \centering
    \begin{adjustbox}{max width=\textwidth}
    \begin{tabular}{ccccc}
        \hline
        & \textbf{Model Variant} & \textbf{ICSD} & \textbf{Materials Project} & \textbf{RRUFF} \\
        \hline
        Baseline & -- & 20.13 & 20.63 & 26.98 \\
        Park \etal{} \cite{Park_iucrj_2017} & -- & \textbf{94.99} & -- & -- \\
        Vecsei \etal{} \cite{Vecsei_prb_2019} & Dense & 73 & -- & 70 \\
        Vecsei \etal{} \cite{Vecsei_prb_2019} & Convolutional & 85 & -- & 56 \\
        Lee \etal{} \cite{Lee_ais_2022} & FCN & 92.12 & \textbf{82.17} & -- \\
        Lee \etal{} \cite{Lee_ais_2023} & Large FCN & 92.10 & -- & 74.24 \\
        Salgado \etal{} \cite{Salgado_npj_2023} & Large, NPCNN & -- & -- & 74 \\
        Choudhary \cite{Choudhary_jpcl_2025} & DGPT-formula & 18.16 (19.80)$^{\dagger}$ & 30.10 (32.10)$^{\dagger}$ & 28.75 (26.57)$^{\ddagger}$ \\
        \textbf{Ours} & Cls. $(\mu = 1)$ & 90.60 \revisednew{[90.21, 90.98]}$^{\dagger}$ & 76.75 \revisednew{[76.15, 77.34]}$^{\dagger}$ & 76.70 \revisednew{[73.57, 79.70]}$^{\ddagger}$ \\
        \textbf{Ours} & Cls. + Regr. (Avg.) & 90.52 \revisednew{[90.15, 90.88]}$^{\dagger}$ & 76.47 \revisednew{[75.91, 77.03]}$^{\dagger}$ & 78.62 \revisednew{[76.02, 81.13]}$^{\ddagger}$ \\
        \textbf{Ours} & Cls. + Regr. Ensemble & 92.22 \revisednew{[91.83, 92.60]}$^{\dagger}$ & 79.00 \revisednew{[78.39, 79.63]}$^{\dagger}$ & \textbf{81.74} \revisednew{[78.88, 84.60]}$^{\ddagger}$ \\
        \hline
    \end{tabular}
    \end{adjustbox}
        \caption{\textbf{Crystal system classification accuracies (\%) of AlphaDiffract and reference models.} Some referenced works trained multiple models with different architectures and/or datasets, for which we only show the model variant giving the best result without including RRUFF data in its training set. ``Baseline'' denotes a naive classifier that always predicts the most abundant class in the training data. For our approach, ``Avg.'' denotes the average independent performance of 10 trained models, while ``Ensemble'' refers to the ensemble-averaged prediction from these models. \revisednew{Bracketed values denote 95\% confidence intervals obtained by bootstrap resampling at the level of evaluation structures.} Bold values indicate the highest classification accuracy for each dataset. \\
        \footnotesize{$^{\dagger}$Evaluated on our validation set data. For inference with the DGPT-formula model in Ref. \citenum{Choudhary_jpcl_2025}, 1000 representative examples from the validation set of each dataset were selected for evaluation. Scores in parentheses refer to results on synthetic PXRD patterns with no added Poisson or Gaussian noise. \\
        $^{\ddagger}$Evaluated on our \revisednew{RRUFF benchmark dataset}.}}
    \label{tab:cs_class_scores}
\end{table}

\begin{table}[h]
    \centering
    \begin{adjustbox}{max width=\textwidth}
    \begin{tabular}{ccccc}
        \hline
        & \textbf{Model Variant} & \textbf{ICSD} & \textbf{Materials Project} & \textbf{RRUFF} \\
        \hline
        Baseline & -- & 7.01 & 6.36 & 3.27 \\
        Park \etal{} \cite{Park_iucrj_2017} & -- & 81.14 & -- & -- \\
        Vecsei \etal{} \cite{Vecsei_prb_2019} & Dense & 57 & -- & 54 \\
        Vecsei \etal{} \cite{Vecsei_prb_2019} & Convolutional & 76 & -- & 42 \\
        Liang \etal{} \cite{Liang_prm_2020} & -- & 78.77$^*$ & -- & -- \\
        Lee \etal{} \cite{Lee_ais_2022} & FCN & 79.67 & \textbf{69.01} & -- \\
        Lee \etal{} \cite{Lee_ais_2023} & Large FCN & \textbf{84.85} & -- & 58.82 \\
        Salgado \etal{} \cite{Salgado_npj_2023} & Large, NPCNN & -- & -- & 66 \\
        Choudhary \cite{Choudhary_jpcl_2025} & DGPT-formula & 3.99 (5.43)$^{\dagger}$ & 10.05 (13.38)$^{\dagger}$ & 5.46 (5.48)$^{\ddagger}$ \\
        \textbf{Ours} & Cls. $(\mu = 1)$ & 80.79 \revisednew{[80.23, 81.33]}$^{\dagger}$ & 58.04 \revisednew{[57.32, 58.77]}$^{\dagger}$ & 63.62 \revisednew{[60.08, 67.03]}$^{\ddagger}$ \\
        \textbf{Ours} & Cls. + Regr. (Avg.) & 79.96 \revisednew{[79.41, 80.50]}$^{\dagger}$ & 57.41 \revisednew{[56.69, 58.12]}$^{\dagger}$ & 64.55 \revisednew{[61.36, 67.63]}$^{\ddagger}$ \\
        \textbf{Ours} & Cls. + Regr. Ensemble & 81.75 \revisednew{[81.19, 82.33]$^{\dagger}$} & 59.54 \revisednew{[58.77, 60.29]}$^{\dagger}$ & \textbf{66.21} \revisednew{[62.81, 69.62]}$^{\ddagger}$ \\
        \hline
    \end{tabular}
    \end{adjustbox}
        \caption{\textbf{Space group classification accuracies (\%) of AlphaDiffract and reference models.} Some referenced works trained multiple models with different architectures and/or datasets, for which we only show the model variant giving the best result without including RRUFF data in its training set. ``Baseline'' denotes a naive classifier that always predicts the most abundant class in the training data. For our approach, ``Avg.'' denotes the average independent performance of 10 trained models, while\revisednew{ }``Ensemble'' refers to the ensemble-averaged prediction from these models. \revisednew{Bracketed values denote 95\% confidence intervals obtained by bootstrap resampling at the level of evaluation structures.} Bold values indicate the highest classification accuracy for each dataset. \\
        \footnotesize{$^*$Weighted average over models specialized for each Bravais lattice.\\
        $^{\dagger}$Evaluated on our validation set data. For inference with the DGPT-formula model in Ref. \citenum{Choudhary_jpcl_2025}, 1000 representative examples from the validation set of each dataset were selected for evaluation. Scores in parentheses refer to results on synthetic PXRD patterns with no added Poisson or Gaussian noise. \\
        $^{\ddagger}$Evaluated on our \revisednew{RRUFF benchmark dataset}.}}
    \label{tab:sg_class_scores}
\end{table}

\begin{figure}[h]
    \centering
    \includegraphics[width=\linewidth]{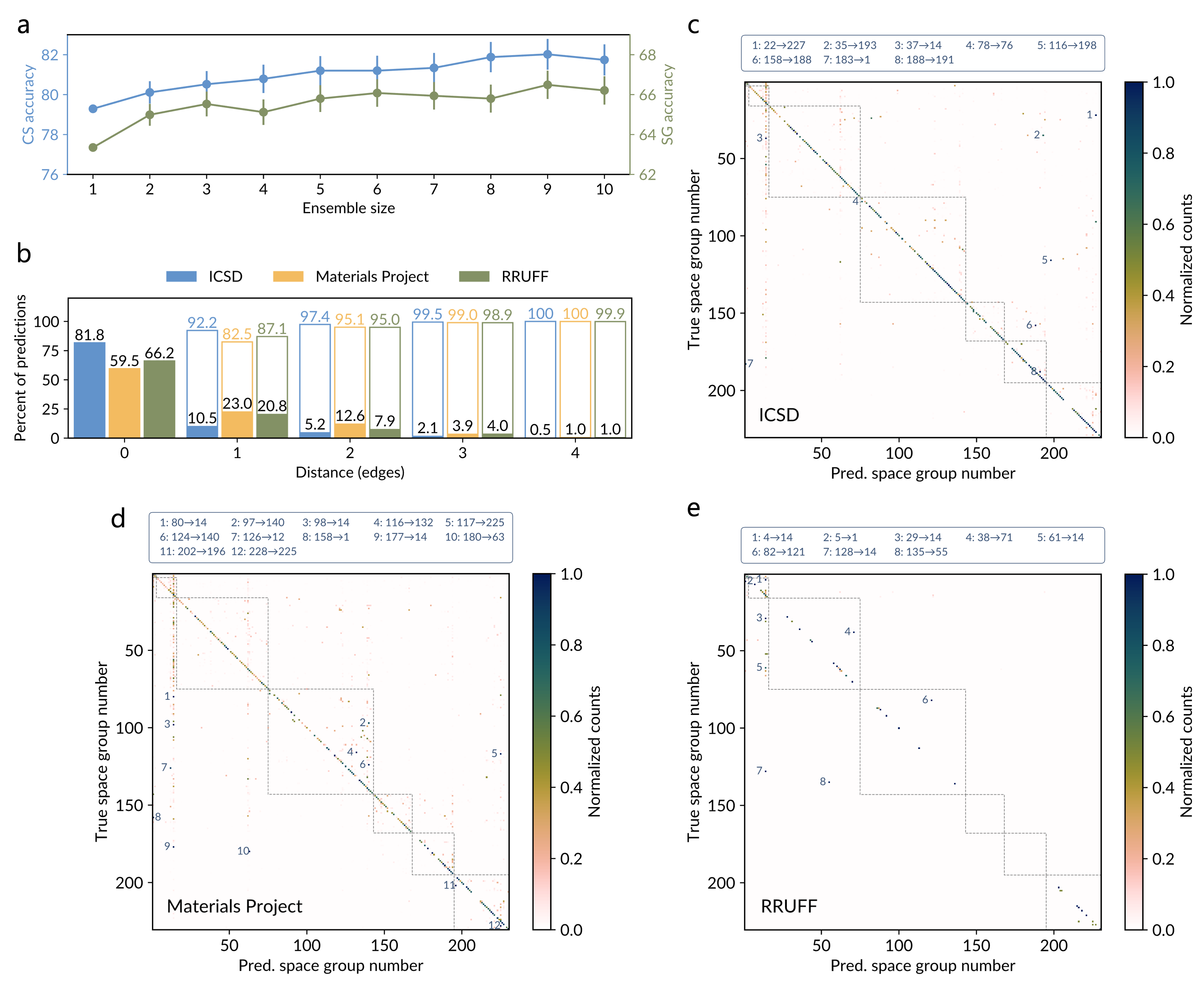}
    \caption{\textbf{AlphaDiffract ensemble model \revised{classification} performance.} \textbf{a.} Crystal system (CS) and space group (SG) prediction accuracies on the RRUFF dataset as a function of ensemble size. \revisednew{Error bars represent propagated ensemble variability derived from differences in predicted class probabilities across ensemble members and are distinct from the finite evaluation set confidence intervals reported in Tables \ref{tab:cs_class_scores} and \ref{tab:sg_class_scores}.} \textbf{b.} Distribution of prediction errors as a function of graph distance (number of edges) from the true space group for the 10-model ensemble evaluated on the three datasets. \revised{Unfilled bars with labeled values indicate the cumulative percentage of predictions falling within that graph distance of the true space group (\textit{i.e.}, the sum of all filled bars up to and including that distance).} \revised{\textbf{c--e.} Space group prediction confusion matrices for the \textbf{c.} ICSD validation set, \textbf{d.} Materials Project validation set, and \textbf{e.} RRUFF \revisednew{benchmark dataset}, normalized by the true class size. Dashed boxes delineate crystal system boundaries. Numbers annotating off-diagonal entries indicate the most common misclassification pairs relative to true class size, \textit{i.e.} with normalized counts exceeding 0.7, listed above each panel using the notation true space group $\rightarrow$ predicted space group.}}
    \label{fig:ens_class}
\end{figure}

\subsection*{AlphaDiffract performance on classification tasks}
\label{subsec:classification}

Having established $\mu$ = 1 as the optimal GEMD loss weight, we trained a 10-model ensemble with all three prediction heads (crystal system, space group, and lattice parameters). Figure \ref{fig:ens_class}\textbf{a} shows that both crystal system and space group accuracies on RRUFF improve with ensemble size with diminishing returns beyond 8-10 models, justifying our choice of a 10-model ensemble. Tables \ref{tab:cs_class_scores} and \ref{tab:sg_class_scores} summarize the classification performance across three datasets for the classification-only (Cls.) model and joint classification and regression (Cls. + Regr.) models, reporting both the independent results of the 10 models (averaged) and the ensemble results obtained by aggregating their predictions. We also include the baseline performance from a naive majority-class classifier for reference. On our RRUFF \revisednew{benchmark}, our ensemble model achieves crystal system and space group classification accuracies of \revisednew{81.74\% [95\% CI: 78.88, 84.60] and 66.21\% [95\% CI: 62.81, 69.62]}, respectively. \revisednew{The confidence intervals account for finite evaluation set variability through bootstrap resampling at the level of evaluation structures (see Methods section ``\nameref{subsec:uncertainty}''), while variability associated with the ensemble members is quantified separately in Supplementary Table~\tabClassificationUncertainty.} These results substantially exceed the baseline accuracies of 26.98\% and 3.27\%, respectively, demonstrating that the model has learned meaningful crystallographic patterns \revisednew{beyond a naive majority-class strategy.} \revisednew{We emphasize that the RRUFF evaluation uses the diffraction profiles provided directly by the RRUFF database, rather than patterns re-simulated from the associated crystal structures. Most of these profiles are directly measured PXRD patterns, while a small fraction are powder profiles calculated by the RRUFF curators from single-crystal refinement data (Methods section ``\nameref{subsec:curation}''). On our RRUFF benchmark, space group accuracy is similar for directly measured and curator-calculated profiles (65.93\% and 69.49\%, respectively; Supplementary Table~\tabRruffProfileSource). For structures with sufficient atomic information, re-simulated patterns generated under the same conditions used for training achieve substantially higher classification accuracy (15--17\% higher; see Discussion), highlighting the additional challenge posed by the predominantly experimental RRUFF benchmark.} The results also compare favorably to prior work evaluated on RRUFF data. For crystal system classification, our ensemble model outperforms Lee \textit{et al.}'s Large FCN (74.24\%) \cite{Lee_ais_2023} and Salgado \textit{et al.}'s NPCNN (74\%) \cite{Salgado_npj_2023}, while for space group classification, it surpasses Lee \textit{et al.}'s Large FCN (58.82\%) and performs comparably to Salgado \textit{et al.}'s NPCNN (66\%).

For comparison with models evaluated primarily on synthetic data, we also report results on our ICSD and Materials Project validation sets, where our ensemble model achieves crystal system accuracies of \revisednew{92.22\% [95\% CI: 91.83, 92.60] and 79.00\% [95\% CI: 78.39, 79.63], and space group accuracies of 81.75\% [95\% CI: 81.19, 82.33] and 59.54\% [95\% CI: 58.77, 60.29]}, respectively. Several prior works achieve slightly higher accuracies on ICSD synthetic data, including Park \textit{et al.} (94.99\% CS) \cite{Park_iucrj_2017} and Lee \textit{et al.} (84.85\% SG) \cite{Lee_ais_2023}, and outperform our model on Materials Project synthetic data, \textit{e.g.} Lee \textit{et al.} (82.17\% CS, 69.01\% SG) \cite{Lee_ais_2022}. However, these comparisons should be interpreted with consideration of the data generation procedures used for training. Our training data includes extensive augmentation spanning wide ranges of sample and instrument broadening parameters, as well as realistic Poisson and Gaussian noise levels derived from RRUFF patterns (Table \ref{tab:sim} and Supplementary Figure~\figNoiseEstimation). While some prior works describe parameter variation in their synthetic data generation, the specific ranges and noise models used are not always fully specified in the literature, which can affect both training robustness and generalizability to experimental data. The performance gap between synthetic and experimental datasets -- both in our work and in prior studies -- underscores the challenge of accurately modeling the full complexity of real-world PXRD measurements. Notably, incorporating lattice parameter regression alongside classification (Cls.~+ Regr.) maintains or slightly improves classification performance compared to the Cls. model, demonstrating that the additional regression task does not compromise classification accuracy. Ensembling 10 independently trained models (Cls.~+ Regr.~Ensemble) further improves performance across all datasets, achieving our best results of 81.74\% for crystal system and 66.21\% for space group classification on RRUFF. Figure \ref{fig:ens_class}\textbf{b} shows the distribution of space group prediction errors as a function of graph distance for the ensemble model across all three datasets, demonstrating that the GEMD loss successfully concentrates errors at shorter distances from the true space group, with over 87\% of predictions on RRUFF data falling within 1 edge of the true space group. \revised{Figs. \ref{fig:ens_class}\textbf{c–e} show space group prediction confusion matrices for the ICSD, Materials Project, and RRUFF datasets, normalized by true class size. The annotated off-diagonal entries represent the most common misclassification pairs relative to true class size; specifically, those with normalized counts exceeding 0.7. These pairs reveal a systematic tendency for the model to predict more abundant space groups when uncertain; for instance, space group 14 ($P2_1/c$) appears as the predicted class in a disproportionate number of annotated pairs across all three datasets, consistent with its status as the most abundant space group in the training data (Supplementary Figure~\figData).} \revisednew{To further assess classification performance across classes with different frequencies, we computed macro-averaged recall on the RRUFF benchmark, defined as the unweighted average of recall across the classes represented in the evaluation set. As shown in Supplementary Table~\tabClassImbalance, macro-averaged recall was 75.75\% across the six represented crystal systems and 60.12\% across the 72 represented space groups, compared with overall accuracies of 81.74\% and 66.21\%, respectively. The lower macro-averaged values indicate some variation in performance across classes but do not by themselves establish class imbalance as the cause, since classes may also differ intrinsically in prediction difficulty. We therefore additionally calculated the Spearman rank correlation between training set class frequency and per-class recall. Training frequency showed a positive association with recall at the crystal system level ($r_s = 0.77$), whereas essentially no monotonic association was observed across space groups ($r_s = -0.11$). Together, these results suggest that class prevalence may contribute to some error patterns, particularly at the crystal system level, but do not support a general dependence of space group recall on training set class frequency.}

\begin{table}[t]
    \centering
    \begin{adjustbox}{max width=\textwidth}
    \begin{tabular}{cccccccc}
        \hline
        \multirow{2}{*}{\textbf{Dataset}} & \multirow{2}{*}{\textbf{Model}} & \multicolumn{3}{c}{\textbf{Lattice Length}} & \multicolumn{3}{c}{\textbf{Lattice Angle}} \\
        \cline{3-8}
        & & MAE (\AA) & MAPE (\%) & $R^2$ & MAE ($^\circ$) & MAPE (\%) & $R^2$ \\
        \hline
        
        \multirow{6}{*}{ICSD}
        & Baseline & 3.1\revised{7} & 44.8\revised{5} & 0.00 & 4.56 & 4.48 & 0.00 \\
        & Chitturi \etal{} \cite{Chitturi_jac_2021} & -- & \textbf{9.2}$^*$ & -- & -- & -- & -- \\
        & Liang \etal{} \cite{Liang_prm_2020} & -- & -- & 0.45$^{**}$ & -- & -- & 0.19$^{**}$\\
        & Choudhary \cite{Choudhary_jpcl_2025}$^{\dagger}$ & 2.85 (2.79) & 34.32 (33.66) & -0.03 (-0.06) & 5.76 (6.23) & 5.82 (6.33) & -2.07 (-2.80) \\
        & \shortstack[c]{\textbf{Ours}$^{\dagger}$ \\ \strut}
        & \shortstack[c]{\textbf{1.59} \\ \revisednew{[1.56, 1.61]}}
        & \shortstack[c]{19.77 \\ \revisednew{[19.55, 19.99]}}
        & \shortstack[c]{\textbf{0.64} \\ \revisednew{[0.63, 0.65]}}
        & \shortstack[c]{\textbf{1.68} \\ \revisednew{[1.64, 1.72]}}
        & \shortstack[c]{\textbf{1.69} \\ \revisednew{[1.65, 1.73]}}
        & \shortstack[c]{\textbf{0.58} \\ \revisednew{[0.57, 0.60]}} \\
        \hline

        \multirow{4}{*}{Materials Project}
        & Baseline & 2.89 & 34.7\revised{4} & -0.01 & 5.22 & 5.55 & 0.00 \\
        & Choudhary \cite{Choudhary_jpcl_2025}$^{\dagger}$ & 2.69 (2.69) & 28.14 (27.04) & -0.01 (-0.01) & 5.97 (6.29) & 6.25 (6.61) & -1.54 (-1.99) \\
        & \shortstack[c]{\textbf{Ours}$^{\dagger}$ \\ \strut}
        & \shortstack[c]{\textbf{1.92} \\ \revisednew{[1.89, 1.95]}}
        & \shortstack[c]{\textbf{22.59} \\ \revisednew{[22.35, 22.83]}}
        & \shortstack[c]{\textbf{0.48} \\ \revisednew{[0.47, 0.50]}}
        & \shortstack[c]{\textbf{3.34} \\ \revisednew{[3.27, 3.41]}}
        & \shortstack[c]{\textbf{3.61} \\ \revisednew{[3.53, 3.69]}}
        & \shortstack[c]{\textbf{0.32} \\ \revisednew{[0.30, 0.33]}} \\
        \hline

        \multirow{4}{*}{\revisedfin{COD}}
        & \revisedfin{Baseline}
        & \revisedfin{2.78}
        & \revisedfin{39.17}
        & \revisedfin{-0.01}
        & \revisedfin{4.53}
        & \revisedfin{4.49}
        & \revisedfin{0.00} \\
        & \revisedfin{G\'{o}mez-Peralta \etal{} \cite{gomez2025insights}\textsuperscript{\P}}
        & \revisedfin{--}
        & \revisedfin{\textbf{13.51}}
        & \revisedfin{\textbf{0.77}}
        & \revisedfin{--}
        & \revisedfin{--}
        & \revisedfin{--} \\
        & \shortstack[c]{\revisedfin{\textbf{Ours}\textsuperscript{\S}} \\ \strut}
        & \shortstack[c]{\revisedfin{\textbf{1.72}} \\ \revisedfin{[1.70, 1.73]}}
        & \shortstack[c]{\revisedfin{21.58} \\ \revisedfin{[21.41, 21.75]}}
        & \shortstack[c]{\revisedfin{0.51} \\ \revisedfin{[0.50, 0.52]}}
        & \shortstack[c]{\textbf{\revisedfin{2.17}} \\ \revisedfin{[2.14, 2.20]}}
        & \shortstack[c]{\textbf{\revisedfin{2.19}} \\ \revisedfin{[2.15, 2.23]}}
        & \shortstack[c]{\textbf{\revisedfin{0.49}} \\ \revisedfin{[0.48, 0.49]}} \\
        \hline
        
        \multirow{4}{*}{RRUFF}
        & Baseline & 2.86 & 31.0\revised{2} & -0.08 & 4.19 & 4.4\revised{8} & -0.07 \\
        & Choudhary \cite{Choudhary_jpcl_2025}$^{\ddagger}$ & 2.53 (2.40) & 25.94 (25.32) & -0.11 (-0.12) & 5.32 (5.39) & 5.73 (5.72) & -2.34 (-2.31) \\
        & \shortstack[c]{\textbf{Ours}$^{\ddagger}$ \\ \strut}
        & \shortstack[c]{\textbf{2.11} \\ \revisednew{[1.87, 2.39]}}
        & \shortstack[c]{\textbf{23.50} \\ \revisednew{[21.58, 25.53]}}
        & \shortstack[c]{\textbf{0.39} \\ \revisednew{[0.32, 0.46]}}
        & \shortstack[c]{\textbf{2.72} \\ \revisednew{[2.27, 3.22]}}
        & \shortstack[c]{\textbf{2.91} \\ \revisednew{[2.38, 3.52]}}
        & \shortstack[c]{\textbf{0.25} \\ \revisednew{[0.16, 0.34]}} \\
        \hline
    \end{tabular}
    \end{adjustbox}
        \caption{\textbf{Lattice parameter prediction errors of AlphaDiffract and reference models.} Errors are quantified in terms of the Mean Absolute Error (MAE), Mean Absolute Percentage Error (MAPE), and coefficient of determination ($R^2$). For direct comparison, we limit our analysis to studies that quantify prediction accuracy using regression metrics rather than classification metrics like match rate. Due to the scarcity of works tested on RRUFF, we also list those tested on other datasets for reference only. ``Baseline'' denotes a naive model that predicts the average lattice parameters from the training data for all inputs. \revisednew{Values in brackets denote 95\% confidence intervals obtained by nonparametric bootstrap resampling at the level of evaluation structures.} Bold values indicate the lowest MAE and MAPE and the highest $R^2$ for each dataset. \\
        \footnotesize{$^*$Weighted average over models specialized for each crystal system. Note these results correspond to a combined ICSD/CSD dataset. \\
        $^{**}$Weighted average over models specialized for each Bravais lattice. \\
        $^{\dagger}$Evaluated on our validation set data. For inference with the DGPT-formula model in Ref. \citenum{Choudhary_jpcl_2025}, 1000 representative examples from the validation set of each dataset were selected for evaluation. Scores in parentheses refer to results on synthetic PXRD patterns with no added Poisson or Gaussian noise. \\
        $^{\ddagger}$Evaluated on our \revisednew{RRUFF benchmark dataset}. \\
        \revisedfin{\textsuperscript{\S}Evaluated on the 27,041 inorganic COD structures retrieved using the publicly released IDs \cite{gomez2025insights} and processed using our curation pipeline.} \\
        \revisedfin{\textsuperscript{\P}Evaluated on the authors' held-out inorganic COD dataset. MAPE is reported in Ref. }\citenum{gomez2025insights}\revisedfin{ and $R^2$ in the accompanying notebook.}}}
    \label{tab:lp_scores}
\end{table}

\subsection*{AlphaDiffract performance on regression tasks}
\label{subsec:regression}

\begin{figure}[h]
    \centering
    \includegraphics[width=\linewidth]{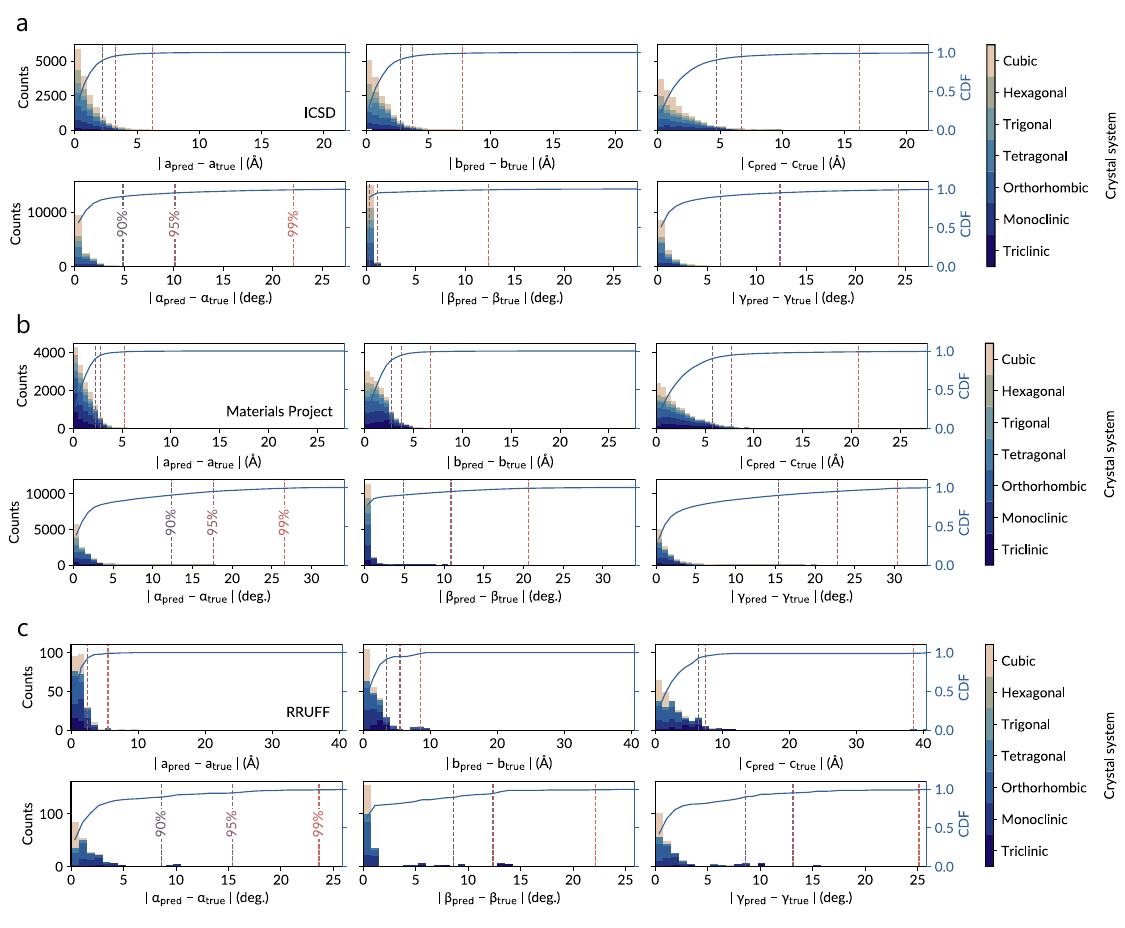}
    \caption{\revised{\textbf{AlphaDiffract ensemble model regression performance.} Histograms of absolute prediction errors for lattice lengths ($a$, $b$, $c$; top rows) and lattice angles ($\alpha$, $\beta$, $\gamma$; bottom rows) for the \textbf{a.} ICSD validation set, \textbf{b.} Materials Project validation set, and \textbf{c.} RRUFF \revisednew{benchmark dataset}. Bar colors indicate the crystal system of each structure. The blue curve shows the empirical cumulative distribution function (CDF) of errors across all crystal systems (right axis). Dashed vertical lines on all panels indicate the 90$^{\text{th}}$ (purple), 95$^{\text{th}}$ (maroon), and 99$^{\text{th}}$ (red) percentiles of the error distribution.}}
    \label{fig:ens_regr}
\end{figure}

For lattice parameter prediction, we evaluate the ensemble model's regression performance across the same three datasets. \revised{Table \ref{tab:lp_scores} reports errors and $R^2$ values averaged across all three lattice lengths and all three lattice angles separately.} As a reference, we also include the baseline performance from a naive model that predicts the average lattice parameters from the training data for all inputs. On RRUFF, the ensemble model achieves mean absolute errors (MAE) of 2.11~\AA\ \revisednew{[95\% CI: 1.87, 2.39] for lattice lengths and 2.72$^\circ$ [95\% CI: 2.27, 3.22]} for lattice angles, with corresponding mean absolute percentage errors (MAPE) of 23.50\% \revisednew{[95\% CI: 21.58, 25.53] and 2.91\% [95\% CI: 2.38, 3.52]}, respectively, representing clear improvements over the baseline \revisednew{point estimates}. The coefficients of determination ($R^2$) are 0.39 \revisednew{[95\% CI: 0.32, 0.46] for lengths and 0.25 [95\% CI: 0.16, 0.34] for angles.} \revised{Figure \ref{fig:ens_regr} shows the full distribution of absolute prediction errors per lattice parameter across all three datasets, with bars colored by crystal system and the empirical cumulative distribution function (CDF) overlaid. The majority of predictions fall within a few angstroms or degrees of the true value, with a small fraction of larger errors extending the tail of the distribution. The 90$^\text{th}$, 95$^\text{th}$, and 99$^\text{th}$ percentile markers provide practical error tolerances; for example, on RRUFF, 90\% of length predictions fall within approximately 4 \AA{} of the true value. The crystal system coloring reveals that the error tails are driven predominantly by triclinic and monoclinic structures while higher-symmetry systems contribute most of the near-zero errors, particularly for angles, where orthorhombic, tetragonal, and cubic lattice angles are correctly predicted at near 0$^\circ$ error.} Metrics evaluated per lattice parameter are also reported in Supplementary Table~\tabLpScoresSupp. \revised{To provide further context for these aggregate metrics, we also report a breakdown of these metrics per crystal system together with a crystal system-conditional baseline in Supplementary Table~\tabPerCsLpScores.} \revised{We note that while these results represent a clear improvement over the naive baseline and existing unified-model approaches, the absolute accuracy, particularly the $\sim$20\% MAPE on lattice lengths, is not yet sufficient to directly initialize whole-pattern refinement methods such as Pawley or Le Bail fits, which typically require cell parameter accuracy within a few percent. However, other approaches to initial lattice fitting with more robust minimizers than least-squares/steepest descent could be developed that embody larger convergence windows. Such software would be able to bridge between AlphaDiffract and direct pattern fitting. For now, AlphaDiffract's predictions are therefore best interpreted as rapid, coarse estimates of the lattice that can guide subsequent refinement rather than replace it.}

On synthetic data, the ensemble model also achieves strong performance, with MAE values of 1.59 \AA\ and 1.68$^\circ$ ($R^2$ = 0.64 and 0.58 for lengths and angles) on ICSD, and 1.92 \AA\ and \revisednew{3.34}$^\circ$ ($R^2$ = 0.48 and 0.32) on Materials Project. Chitturi \textit{et al.} report lower MAPE (9.2\%) on combined ICSD/CSD data; however, their approach trains separate models for each of the seven crystal systems. Since AlphaDiffract employs a single unified model for all crystal systems, it may better reflect practical application scenarios for lattice determination where the crystal system may not be known \textit{a priori}. \revisedfin{As an additional synthetic data comparison, we also evaluated AlphaDiffract on the collection of inorganic COD structures used by G\'{o}mez-Peralta \etal{} for downstream training and held-out evaluation of their lattice parameter prediction models \cite{gomez2025insights}. Here, COD refers specifically to this author-designated inorganic collection rather than the full Crystallography Open Database. On the full 27,041-structure set retained after our curation procedure, AlphaDiffract achieves MAE, MAPE, and $R^2$ values of 1.72~\AA, 21.58\%, and 0.51 for lattice lengths and 2.17$^\circ$, 2.19\%, and 0.49 for lattice angles, respectively. For comparison, the PXRD-only model of G\'{o}mez-Peralta \etal{} achieves a length MAPE of 13.51\%, while the accompanying public notebook reports an $R^2$ value of 0.77. The angle $R^2$ is calculated in a cosine-transformed target space; therefore, the reported value of 0.12 for this metric is not directly comparable to AlphaDiffract's degree-space $R^2$ metric. The authors' lower length error may be influenced by the substantially different synthetic data distributions: their evaluation uses four profiles per structure corresponding to crystallite sizes of 50, 100, and 250~nm and an infinite-size profile, with no additional noise, microstrain, or variation in instrumental broadening reported, whereas our evaluation uses 100 profiles per structure spanning randomized crystallite size, microstrain, and instrumental broadening together with dynamic Poisson and Gaussian noise. Along with differences in the evaluated subsets noted in Table~\ref{tab:lp_scores}, these differences mean that the values provide cross-study context rather than a controlled head-to-head model ranking.}

\revised{Supplementary Figure~\figRegrParity{}} shows parity plots comparing the lattice parameters predicted by the ensemble model against their ground truth values across \revisedfin{our} three \revisedfin{primary evaluation} datasets. The plots show that predictions are generally well-correlated with true values, though with greater scatter for RRUFF data compared to synthetic ICSD patterns. The $R^2$ values for lattice angles are notably lower than those for lattice lengths, which is explained by the fact that angles are constrained by symmetry to fixed values (e.g., $90\degree$) for all crystal systems except monoclinic and triclinic. The model frequently predicts these fixed values correctly, producing the dense clusters visible in \revised{Supplementary Figure~\figRegrParity}, but this simultaneously leads to low $R^2$ values as the metric measures explained variance. The performance gap between synthetic and experimental data for lattice parameter regression parallels that observed for classification, reflecting the increased difficulty of precise quantitative predictions from real-world measurements with complex experimental artifacts.

\subsection*{\revisednew{AlphaDiffract performance on structurally distinct datasets}}
\label{subsec:dedup}

\revisednew{The results reported in Tables \ref{tab:cs_class_scores} - \ref{tab:lp_scores} reflect performance on the full ICSD, Materials Project, \revisedfin{COD,} and RRUFF evaluation sets, without accounting for structural overlap between these sets and the training data. As described in Methods section ``\nameref{subsec:training}'', the training/validation split for ICSD and Materials Project was performed at the level of crystallographic structures, prior to augmentation, but without deduplication or structural equivalence-based grouping. To assess model performance on structures truly distinct from the training data, we characterized each structure using the CrystalNN site fingerprint method \cite{zimmermann2020local}, which quantifies local coordination environment, as implemented in matminer \cite{ward2018matminer}. Following the structural equivalence clustering procedure described in Methods section ``\nameref{subsec:similarity}'', we classified two structures sharing identical composition and space group as structurally equivalent if their CrystalNN fingerprints fell within a distance threshold of 0.9, the convention proposed in the Materials Project documentation \cite{jain_aplmat_2013}. Applying this criterion, we cross-matched every structure in each evaluation set against the full training set to identify whether a structural equivalent exists anywhere in the training data. For the ICSD and Materials Project evaluation sets, structural fingerprints could not be computed for a small number of structures (Supplementary Table~\tabDatasetPartitions); these are conservatively treated as matching any training set structure sharing the same composition and space group. For RRUFF, this fallback is more consequential: 473 of the 734 structures used for classification (64.4\%) lack the complete atomic coordinate information required to compute a CrystalNN fingerprint (Methods section ``\nameref{subsec:curation}'') and are therefore conservatively matched against the training set on composition and space group alone. We note that this structural equivalence criterion characterizes local coordination topology and is largely insensitive to absolute lattice scale; two structures sharing an identical structural prototype but differing in overall lattice size (\textit{e.g.}, due to measurement under different conditions) may therefore be classified as structurally equivalent under this criterion, even where their diffraction patterns differ. This is a stricter criterion for novelty than one based solely on whether two structures would be distinguishable in the diffraction pattern actually available to the model given our input angular range and sampling.}

\begin{table}[t]
    \centering
    \begin{adjustbox}{max width=\textwidth}
    \begin{tabular}{clcc}
        \hline
        \multirow{2}{*}{\revisednew{\textbf{Dataset}}} &
        \multirow{2}{*}{\shortstack[l]{\revisednew{\textbf{Evaluation Set}} \\
                                       \revisednew{\textbf{(Number of Examples)}}}} &
        \multirow{2}{*}{\revisednew{\textbf{Crystal System Accuracy (\%)}}} &
        \multirow{2}{*}{\revisednew{\textbf{Space Group Accuracy (\%)}}} \\
        & & & \\[1ex]
        \hline

        \multirow{3}{*}{\revisednew{ICSD$^\dagger$}}
        & \revisednew{Baseline}
        & \revisednew{20.13}
        & \revisednew{7.01} \\

        & \revisednew{Full (16,563)}
        & \revisednew{92.22 [91.83, 92.60]}
        & \revisednew{81.75 [81.19, 82.33]} \\

        & \revisednew{Unique (5,979)}
        & \revisednew{86.87 [86.07, 87.68]}
        & \revisednew{74.59 [73.53, 75.66]} \\
        \hline

        \multirow{3}{*}{\revisednew{Materials Project$^\dagger$}}
        & \revisednew{Baseline}
        & \revisednew{20.63}
        & \revisednew{6.36} \\

        & \revisednew{Full (14,663)}
        & \revisednew{79.00 [78.39, 79.63]}
        & \revisednew{59.54 [58.77, 60.29]} \\

        & \revisednew{Unique (9,161)}
        & \revisednew{74.34 [73.50, 75.23]}
        & \revisednew{52.44 [51.46, 53.42]} \\
        \hline

        \multirow{3}{*}{\revisednew{RRUFF$^\ddagger$}}
        & \revisednew{Baseline}
        & \revisednew{26.98}
        & \revisednew{3.27} \\

        & \revisednew{Full (734)}
        & \revisednew{81.74 [78.88, 84.60]}
        & \revisednew{66.21 [62.81, 69.62]} \\

        & \revisednew{Unique (285)}
        & \revisednew{76.14 [71.23, 81.05]}
        & \revisednew{49.82 [44.21, 55.44]} \\
        \hline
    \end{tabular}
    \end{adjustbox}

    \caption{
        \revisednew{\textbf{Crystal system and space group classification performance on structurally distinct evaluation sets.}
        Classification accuracies of the ensemble model for subsets excluding structures with potential equivalents in the training data (``Unique'') are compared against the baseline and complete evaluation sets (``Full''). ``Baseline'' denotes a naive classifier that always predicts the most abundant class in the training data. ``Unique'' denotes structures with no training set equivalent on composition, space group, and coordination pattern (see Methods section ``\nameref{subsec:similarity}''). Where structural information needed to compute coordination similarity was unavailable, structures matching on composition and space group alone were conservatively treated as potential overlap and excluded. Bracketed values denote 95\% confidence intervals obtained by bootstrap resampling at the level of evaluation structures.} \\
        \footnotesize{
        \revisednew{
        $^{\dagger}$Evaluated on data from our validation set. \\
        $^{\ddagger}$Evaluated on data from our RRUFF benchmark dataset.}
        }
    }
    \label{tab:unique_classification}
\end{table}

\begin{table}[h]
    \centering
    \begin{adjustbox}{max width=\textwidth}
    \begin{tabular}{clcccccc}
        \hline
        \multirow{2}{*}{\revisednew{\textbf{Dataset}}} &
        \multirow{2}{*}{\shortstack[l]{\revisednew{\textbf{Evaluation Set}} \\
                                \revisednew{\textbf{(Number of Examples)}}}} &
        \multicolumn{3}{c}{\revisednew{\textbf{Lattice Length}}} &
        \multicolumn{3}{c}{\revisednew{\textbf{Lattice Angle}}} \\
        \cmidrule(lr){3-5} \cmidrule(lr){6-8}
        & &
        \revisednew{MAE (\AA)} &
        \revisednew{MAPE (\%)} &
        \revisednew{$R^2$} &
        \revisednew{MAE ($^\circ$)} &
        \revisednew{MAPE (\%)} &
        \revisednew{$R^2$} \\
        \hline

        \multirow{3}{*}{\revisednew{ICSD$^{\dagger}$}}
        & \shortstack[l]{\revisednew{Baseline} \\ \strut}
        & \shortstack[c]{\revisednew{3.17} \\ \strut}
        & \shortstack[c]{\revisednew{44.85} \\ \strut}
        & \shortstack[c]{\revisednew{0.00} \\ \strut}
        & \shortstack[c]{\revisednew{4.56} \\ \strut}
        & \shortstack[c]{\revisednew{4.48} \\ \strut}
        & \shortstack[c]{\revisednew{0.00} \\ \strut} \\

        & \shortstack[l]{\revisednew{Full (16,563)} \\ \strut}
        & \shortstack[c]{\revisednew{1.59} \\ \revisednew{[1.56, 1.61]}}
        & \shortstack[c]{\revisednew{19.77} \\ \revisednew{[19.55, 19.99]}}
        & \shortstack[c]{\revisednew{0.64} \\ \revisednew{[0.63, 0.65]}}
        & \shortstack[c]{\revisednew{1.68} \\ \revisednew{[1.64, 1.72]}}
        & \shortstack[c]{\revisednew{1.69} \\ \revisednew{[1.65, 1.73]}}
        & \shortstack[c]{\revisednew{0.58} \\ \revisednew{[0.57, 0.60]}} \\

        & \shortstack[l]{\revisednew{Unique (5,979)} \\ \strut}
        & \shortstack[c]{\revisednew{1.81} \\ \revisednew{[1.77, 1.86]}}
        & \shortstack[c]{\revisednew{20.34} \\ \revisednew{[19.98, 20.70]}}
        & \shortstack[c]{\revisednew{0.58} \\ \revisednew{[0.56, 0.60]}}
        & \shortstack[c]{\revisednew{2.00} \\ \revisednew{[1.92, 2.07]}}
        & \shortstack[c]{\revisednew{2.01} \\ \revisednew{[1.93, 2.09]}}
        & \shortstack[c]{\revisednew{0.54} \\ \revisednew{[0.52, 0.56]}} \\
        \hline

        \multirow{3}{*}{\revisednew{Materials Project$^{\dagger}$}}
        & \shortstack[l]{\revisednew{Baseline} \\ \strut}
        & \shortstack[c]{\revisednew{2.89} \\ \strut}
        & \shortstack[c]{\revisednew{34.74} \\ \strut}
        & \shortstack[c]{\revisednew{-0.01} \\ \strut}
        & \shortstack[c]{\revisednew{5.22} \\ \strut}
        & \shortstack[c]{\revisednew{5.55} \\ \strut}
        & \shortstack[c]{\revisednew{0.00} \\ \strut} \\

        & \shortstack[l]{\revisednew{Full (14,663)} \\ \strut}
        & \shortstack[c]{\revisednew{1.92} \\ \revisednew{[1.89, 1.95]}}
        & \shortstack[c]{\revisednew{22.59} \\ \revisednew{[22.35, 22.83]}}
        & \shortstack[c]{\revisednew{0.48} \\ \revisednew{[0.47, 0.50]}}
        & \shortstack[c]{\revisednew{3.34} \\ \revisednew{[3.27, 3.41]}}
        & \shortstack[c]{\revisednew{3.61} \\ \revisednew{[3.53, 3.69]}}
        & \shortstack[c]{\revisednew{0.32} \\ \revisednew{[0.30, 0.33]}} \\

        & \shortstack[l]{\revisednew{Unique (9,161)} \\ \strut}
        & \shortstack[c]{\revisednew{1.93} \\ \revisednew{[1.90, 1.97]}}
        & \shortstack[c]{\revisednew{22.72} \\ \revisednew{[22.40, 23.05]}}
        & \shortstack[c]{\revisednew{0.43} \\ \revisednew{[0.41, 0.45]}}
        & \shortstack[c]{\revisednew{3.36} \\ \revisednew{[3.27, 3.44]}}
        & \shortstack[c]{\revisednew{3.63} \\ \revisednew{[3.52, 3.73]}}
        & \shortstack[c]{\revisednew{0.29} \\ \revisednew{[0.27, 0.31]}} \\
        \hline

        \multirow{3}{*}{\revisedfin{COD\textsuperscript{\S}}}
        & \shortstack[l]{\revisedfin{Baseline} \\ \strut}
        & \shortstack[c]{\revisedfin{2.78} \\ \strut}
        & \shortstack[c]{\revisedfin{39.17} \\ \strut}
        & \shortstack[c]{\revisedfin{-0.01} \\ \strut}
        & \shortstack[c]{\revisedfin{4.53} \\ \strut}
        & \shortstack[c]{\revisedfin{4.49} \\ \strut}
        & \shortstack[c]{\revisedfin{0.00} \\ \strut} \\

        & \shortstack[l]{\revisedfin{Full (27,041)} \\ \strut}
        & \shortstack[c]{\revisedfin{1.72} \\ \revisedfin{[1.70, 1.73]}}
        & \shortstack[c]{\revisedfin{21.58} \\ \revisedfin{[21.41, 21.75]}}
        & \shortstack[c]{\revisedfin{0.51} \\ \revisedfin{[0.50, 0.52]}}
        & \shortstack[c]{\revisedfin{2.17} \\ \revisedfin{[2.14, 2.20]}}
        & \shortstack[c]{\revisedfin{2.19} \\ \revisedfin{[2.15, 2.23]}}
        & \shortstack[c]{\revisedfin{0.49} \\ \revisedfin{[0.48, 0.49]}} \\

        & \shortstack[l]{\revisedfin{Unique (6,813)} \\ \strut}
        & \shortstack[c]{\revisedfin{2.12} \\ \revisedfin{[2.09, 2.16]}}
        & \shortstack[c]{\revisedfin{23.27} \\ \revisedfin{[22.94, 23.60]}}
        & \shortstack[c]{\revisedfin{0.31} \\ \revisedfin{[0.30, 0.32]}}
        & \shortstack[c]{\revisedfin{3.15} \\ \revisedfin{[3.07, 3.23]}}
        & \shortstack[c]{\revisedfin{3.19} \\ \revisedfin{[3.10, 3.27]}}
        & \shortstack[c]{\revisedfin{0.27} \\ \revisedfin{[0.25, 0.28]}} \\
        \hline
        
        \multirow{3}{*}{\revisednew{RRUFF$^{\ddagger}$}}
        & \shortstack[l]{\revisednew{Baseline} \\ \strut}
        & \shortstack[c]{\revisednew{2.86} \\ \strut}
        & \shortstack[c]{\revisednew{31.02} \\ \strut}
        & \shortstack[c]{\revisednew{-0.08} \\ \strut}
        & \shortstack[c]{\revisednew{4.19} \\ \strut}
        & \shortstack[c]{\revisednew{4.48} \\ \strut}
        & \shortstack[c]{\revisednew{-0.07} \\ \strut} \\

        & \shortstack[l]{\revisednew{Full (240)} \\ \strut}
        & \shortstack[c]{\revisednew{2.11} \\ \revisednew{[1.87, 2.39]}}
        & \shortstack[c]{\revisednew{23.50} \\ \revisednew{[21.58, 25.53]}}
        & \shortstack[c]{\revisednew{0.39} \\ \revisednew{[0.32, 0.46]}}
        & \shortstack[c]{\revisednew{2.72} \\ \revisednew{[2.27, 3.22]}}
        & \shortstack[c]{\revisednew{2.91} \\ \revisednew{[2.38, 3.52]}}
        & \shortstack[c]{\revisednew{0.25} \\ \revisednew{[0.16, 0.34]}} \\

        & \shortstack[l]{\revisednew{Unique (91)} \\ \strut}
        & \shortstack[c]{\revisednew{2.68} \\ \revisednew{[2.16, 3.31]}}
        & \shortstack[c]{\revisednew{25.66} \\ \revisednew{[22.78, 28.79]}}
        & \shortstack[c]{\revisednew{0.31} \\ \revisednew{[0.22, 0.38]}}
        & \shortstack[c]{\revisednew{3.58} \\ \revisednew{[2.77, 4.50]}}
        & \shortstack[c]{\revisednew{3.97} \\ \revisednew{[2.93, 5.23]}}
        & \shortstack[c]{\revisednew{0.19} \\ \revisednew{[0.05, 0.30]}} \\
        \hline
    \end{tabular}
    \end{adjustbox}

    \caption{
        \revisednew{\textbf{Lattice parameter prediction performance on structurally distinct evaluation sets.}
        Lattice length and angle prediction errors of the ensemble model for subsets excluding structures with potential equivalents in the training data (``Unique'') are compared against the baseline and complete evaluation sets (``Full''). ``Baseline'' denotes a naive model that predicts the average lattice parameters from the training data for all inputs. ``Unique'' denotes structures with no training set equivalent on composition, space group, and coordination pattern (see Methods section ``\nameref{subsec:similarity}''). Where structural information needed to compute coordination similarity was unavailable, structures matching on composition and space group alone were conservatively treated as potential overlap and excluded. Values in brackets denote 95\% confidence intervals obtained by nonparametric bootstrap resampling at the level of evaluation structures.} \\
        \footnotesize{
        \revisednew{$^{\dagger}$Evaluated on data from our validation set.} \\
        \revisednew{$^{\ddagger}$Evaluated on data from our RRUFF benchmark dataset.} \\
        \revisedfin{\textsuperscript{\S}Evaluated on the inorganic COD structures retrieved using the published IDs \cite{gomez2025insights} and processed using our curation pipeline.}
        }
    }
    \label{tab:unique_regression}
\end{table}

\revisednew{Applying this procedure, we identify 36.1\% of ICSD validation structures, 62.5\% of Materials Project validation structures, \revisedfin{25.2\% of inorganic COD structures,} 38.8\% of RRUFF structures used for classification, and 37.9\% of RRUFF structures used for regression as structurally distinct from the training data. Classification and regression performance on this subset is reported in Tables \ref{tab:unique_classification} and \ref{tab:unique_regression}, respectively. Crystal system classification accuracy declines by 4.7-5.6\% across all three \revisedfin{classification} datasets. The decline is more pronounced, and more variable, for space group classification, ranging from 7.1--7.2\% for ICSD and Materials Project to 16.4\% for RRUFF. A similar \revisedfin{relative decline} is observed for lattice parameter regression (Table \ref{tab:unique_regression}),} \revisedfin{with the most pronounced changes occurring for the COD and RRUFF subsets.} \revisednew{Further stratification of the RRUFF classification benchmark by profile source shows decreased space group accuracy for both directly measured and curator-calculated profiles (Supplementary Table~\tabRruffProfileSource), suggesting that the larger RRUFF reduction is not solely driven by profile source.}

\revisednew{We emphasize that the values reported in this section should not be substituted directly into the literature comparison presented in Tables \ref{tab:cs_class_scores} - \ref{tab:lp_scores}, as the deduplication procedures used in most prior studies are substantially less rigorous than the structural equivalence criterion applied here \cite{Park_iucrj_2017, Vecsei_prb_2019, Lee_ais_2022, Lee_ais_2023, Salgado_npj_2023, Liang_prm_2020, Chitturi_jac_2021, Choudhary_jpcl_2025}. The extent of deduplication varies substantially across these works. The only study where we identified a more restrictive deduplication policy prior to splitting is Liang \textit{et al.}~\cite{Liang_prm_2020}, who selected a single representative structure per chemical composition before constructing their training and test sets; this guards against duplicate compositions appearing on both sides of the split, necessitated by their model which depends on composition as an input. We note that Liang \textit{et al.} do not evaluate on the RRUFF benchmark considered here; however, on the structurally-distinct ICSD dataset, AlphaDiffract still maintains a better $R^2$ for lattice parameter prediction (0.58 vs. 0.45 for lattice lengths and 0.54 vs. 0.19 for lattice angles) by comparison~\cite{Liang_prm_2020}.}

\revisednew{Several other studies report removing duplicate or repeated database entries before constructing their datasets \cite{Park_iucrj_2017, Vecsei_prb_2019, Chitturi_jac_2021, Salgado_npj_2023}; however, except for Chitturi \textit{et al.}, these studies do not describe explicit structural equivalence-based criteria. Chitturi \textit{et al.}~\cite{Chitturi_jac_2021} treated structures as distinct if they differed in chemical formula, space group, or lattice parameters; structurally similar entries with the same composition and space group but differing lattice parameters could therefore remain as separate samples. Lee \textit{et al.}~\cite{Lee_ais_2023} used record-level splitting so that perturbed diffraction patterns generated from the same source structure did not appear in both training and test sets, addressing augmentation-level leakage but not equivalence between distinct structural records. Other studies used conventional record-level dataset splits or did not report a structural equivalence-based deduplication procedure \revisedfin{\cite{Lee_ais_2022, Choudhary_jpcl_2025, gomez2025insights}}. For studies evaluating on RRUFF \cite{Vecsei_prb_2019, Lee_ais_2023, Salgado_npj_2023}, we found no reported cross-database filtering in which RRUFF structures were matched against the training database prior to computing the reported RRUFF performance. Consequently, the performance on our structurally-distinct RRUFF subset reflects a more restrictive evaluation protocol than that used for the published RRUFF benchmarks in Tables \ref{tab:cs_class_scores} and \ref{tab:sg_class_scores}, and is reported separately rather than substituted into that comparison.}

\begin{figure}[h]
    \centering
    \includegraphics[width=\linewidth]{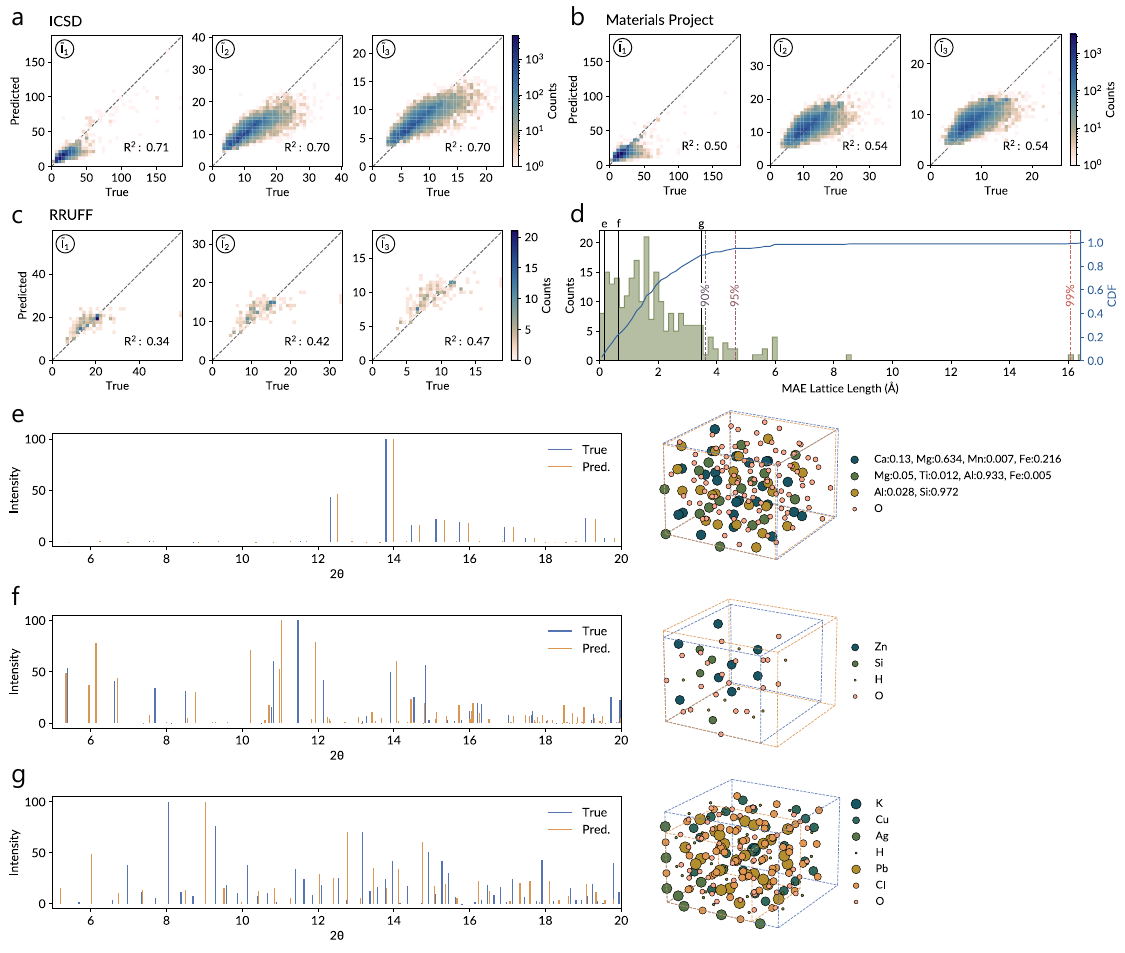}
    \caption{\revised{\textbf{Metric tensor invariant analysis and representative experimental PXRD pattern predictions.}} \revised{\textbf{a-c.} Parity plots comparing predicted versus true linearized metric tensor invariants $\tilde{I}_1$, $\tilde{I}_2$, $\tilde{I}_3$ (defined in Equations~\ref{eq:i1}--\ref{eq:i3}) for the \textbf{a.} ICSD validation set, \textbf{b.} Materials Project validation set, and \textbf{c.} RRUFF \revisednew{benchmark dataset}. Heat map coloring represents point count per bin. $R^2$ values indicate goodness of fit. \textbf{d.} Histogram of refined lattice length MAE across the RRUFF \revisednew{benchmark dataset}, with the empirical CDF overlaid (right axis). Dashed vertical lines indicate the 90$^\text{th}$ (purple), 95$^\text{th}$ (maroon), and 99$^\text{th}$ (red) percentiles. Labeled markers (e, f, g) indicate the positions of the three representative structures shown in the panels below. \textbf{e-g.} Representative comparisons of true (blue) and predicted (orange) Bragg peak positions for RRUFF benchmark structures at different levels of lattice length prediction accuracy, selected from the positions marked in panel \textbf{d}. Crystal structures shown to the right display the true (blue dashed) and predicted (orange dashed) unit cells overlaid on the fixed atomic positions.}}
    \label{fig:ens_invariants}
\end{figure}

\subsection*{\revised{Geometric analysis and symmetry-constrained refinement}}
 
\revised{To complement the per-parameter regression metrics in Table \ref{tab:lp_scores} and Figure \ref{fig:ens_regr}, we evaluate model performance using three linearized principal invariants of the metric tensor. The metric tensor $\mathbf{G}$ of a crystal unit cell is fully determined by the six lattice parameters $a$, $b$, $c$, $\alpha$, $\beta$, $\gamma$. Its three principal invariants are rotationally invariant scalars that 
characterize the geometry of the unit cell independently of axis ordering. To enable direct comparison in units of length, we define linearized versions of these invariants as follows:}

\begin{align}
    \revised{\tilde{I}_1} &\revised{= \sqrt{a^2 + b^2 + c^2}} \label{eq:i1} \\[6pt]
    \revised{\tilde{I}_2} &\revised{= \left(b^2c^2\sin^2\!\alpha + a^2c^2\sin^2\!\beta + 
                   a^2b^2\sin^2\!\gamma\right)^{1/4}} \label{eq:i2} \\[6pt]
    \revised{\tilde{I}_3} &\revised{= \left(abc\sqrt{1 - \cos^2\!\alpha - \cos^2\!\beta - 
                   \cos^2\!\gamma + 2\cos\alpha\cos\beta\cos\gamma}
                   \right)^{1/3}} \label{eq:i3}
\end{align}

\revised{$\tilde{I}_1$ is the square root of the sum of squared edge lengths; $\tilde{I}_2$ is the fourth root of the sum of squared face areas of the unit cell parallelepiped ($bc\sin\alpha$, $ac\sin\beta$, $ab\sin\gamma$); and $\tilde{I}_3$ is the cube root of the unit cell volume $V$.These invariants are insensitive to the axis-ordering convention of the Niggli reduced cell and therefore provide a complementary assessment of prediction accuracy that is free from the parameter-assignment ambiguities such as those discussed for trigonal and hexagonal systems. Figs. \ref{fig:ens_invariants}\textbf{a-c} show parity plots of the predicted versus true invariants across the ICSD, Materials Project, and RRUFF datasets. The $R^2$ values for all three invariants are generally higher than the per-parameter averages reported in Table \ref{tab:lp_scores}, particularly for ICSD ($\tilde{I}_1$: 0.71, $\tilde{I}_2$: 0.70, $\tilde{I}_3$: 0.70) and RRUFF ($\tilde{I}_1$: 0.34, $\tilde{I}_2$: 0.42, $\tilde{I}_3$: 0.47). This improvement indicates that a portion of the apparent per-parameter error arises from axis-ordering \revisednew{misassignments} such as swapping $\alpha$ and $\gamma$ in 
trigonal/hexagonal structures, rather than from incorrect predictions of the underlying unit cell geometry. The invariant-based metrics thus provide a physically meaningful measure of the model's geometric prediction capability, suggesting that the model captures the gross geometric character of the unit cell more reliably than per-parameter metrics alone imply.}

\revised{This observation motivates a closer examination of how the model's joint predictions of crystal system, space group, and lattice parameters perform when synthesized into a complete, internally consistent lattice description, the form in which they would be used in any downstream crystallographic application. As a minimal synthesis step, we apply symmetry-constrained refinement to the raw AlphaDiffract lattice parameter predictions: Based on the predicted crystal system, angles are fixed to their symmetry-required values (e.g., 90$^\circ$ for orthorhombic) and, where 
applicable, lattice lengths are averaged to enforce equality constraints (e.g., $a = b = c$ for cubic). Table \ref{tab:refined_lp} summarizes the effect of this refinement on the RRUFF \revisednew{benchmark dataset}. Angle prediction improves modestly (MAE: 2.20$^\circ$ vs. 2.72$^\circ$; MAPE: 2.31\% vs. 2.91\%), 
while length metrics are largely unchanged (MAE: 2.03~\AA{} vs. 2.11~\AA). The improvement is limited by the fact that incorrectly classified crystal systems receive wrong constraints, partially offsetting the gains from correct classifications. We also attempted to use these refined predictions to initialize Pawley and Le Bail fits on the RRUFF \revisednew{benchmark dataset}, but the fits did not converge, consistent with the $\sim$20\% MAPE on lattice lengths noted in the previous section as insufficient for whole-pattern refinement.}
 
\begin{table}[t]
    \centering
    \begin{tabular}{lcccccc}
        \hline
        \multirow{2}{*}{\textbf{Parameters}} & 
        \multicolumn{3}{c}{\textbf{Lattice Length}} & 
        \multicolumn{3}{c}{\textbf{Lattice Angle}} \\
        \cline{2-7}
        & MAE (\AA) & MAPE (\%) & $R^2$ & 
          MAE ($^\circ$) & MAPE (\%) & $R^2$ \\
        \hline
        Predicted & 2.11 & 23.50 & 0.39 & 2.72 & 2.91 & \textbf{0.25} \\
        Refined   & \textbf{2.03} & \textbf{22.70} & \textbf{0.42} & \textbf{2.20} & \textbf{2.31} & 0.23 \\
        \hline
    \end{tabular}
    \caption{\revised{\textbf{Effect of symmetry-constrained refinement on RRUFF lattice parameter prediction metrics.} Aggregate regression metrics on the RRUFF \revisednew{benchmark dataset} for raw AlphaDiffract predictions and after refinement to enforce predicted crystal system symmetry constraints. Metrics are averaged across the three lattice lengths and three lattice angles, respectively.} Bold values indicate the lowest MAE and MAPE and the highest $R^2$ across the predicted and refined results.}
    \label{tab:refined_lp}
\end{table}

\revised{To characterize the practical accuracy of the refined predictions, Figures~\ref{fig:ens_invariants}\textbf{e-g} show representative comparisons of true and predicted Bragg peak positions for three RRUFF benchmark structures drawn from different points in the lattice length MAE distribution (Figure~\ref{fig:ens_invariants}\textbf{d}). Bragg peak positions and relative intensities are computed using \texttt{pymatgen}'s XRDCalculator applied to the true crystal structure with its lattice parameters replaced by the symmetry-constrained AlphaDiffract predictions, while preserving the Cartesian 
coordinates of the atoms~\cite{ong2013python}. The crystal structures shown alongside each pattern display the true (blue dashed) and predicted (orange dashed) unit cells overlaid on the same fixed atomic positions. At low lattice length MAE (Figure~\ref{fig:ens_invariants}\textbf{e}), the predicted and true peak positions are in close agreement across the full $2\theta$ range. At moderate error (Figure~\ref{fig:ens_invariants}\textbf{f}), several peaks remain 
well-positioned, but many discrepancies emerge for reflections at higher $2\theta$ where sensitivity to lattice parameter errors is greater. At the 90$^\text{th}$ percentile of the error distribution (Figure~\ref{fig:ens_invariants}\textbf{g}), peak positions show substantial disagreement. The nature and severity of the prediction error is, at least qualitatively, readable from the pattern comparison, a property that may aid in identifying which predictions are reliable enough to guide further analysis.}

\begin{table}[b]
    \centering
    \begin{tabular}{C{3cm}C{3cm}C{3cm}C{3cm}C{3cm}}
        \hline
        \textbf{Dataset} & \textbf{Number of samples} & \textbf{Average time per sample (ms)} & \textbf{Median time per sample (ms)} & \textbf{Throughput (samples / s)} \\
        \hline
        ICSD & 10,000 & 1.33 & 1.08 & 751 \\
        Materials Project & 10,000 & 1.39 & 1.08 & 719 \\
        RRUFF & 734 & 1.15 & 1.04 & 871 \\
        \hline
    \end{tabular}
    \caption{\textbf{Inference time statistics.} Per-model inference time statistics across evaluation datasets. All measurements performed on a single NVIDIA H100 GPU with batch size 1.}
    \label{tab:throughput}
\end{table}

\subsection*{Inference time statistics}

To evaluate the computational efficiency of \revisednew{AlphaDiffract}, we measured the inference times of individual models across the three \revisedfin{primary} evaluation datasets (Table \ref{tab:throughput}). \revised{Timing measurements were performed with batch size 1 on a single NVIDIA H100 GPU and cover only the neural network forward pass; data loading, host-to-device memory transfer, and any preprocessing steps such as background correction and wavelength conversion are excluded.} A single model achieves average inference times of 1.15-1.39 ms per PXRD pattern and median times of 1.04-1.08 ms across all datasets. These results correspond to an inference rate of 700-870 samples per second on a single GPU, making AlphaDiffract highly suitable for real-time analysis and ultrahigh-throughput screening applications. The consistency of inference times across datasets of varying sizes (734 to 10,000 samples) indicates stable performance regardless of dataset scale. For the ensemble approach using 10 models, inference time scales approximately linearly to 11.5-13.9 ms per pattern (70-90 samples/second), as predictions from each model are generated sequentially. Even with this tenfold increase, the ensemble inference time remains negligible compared to typical Rietveld refinement times of several seconds to minutes per pattern. \revised{We also note that measurements at batch size 1 represent a conservative lower bound on throughput; batched inference would reduce per-sample latency further by better utilizing GPU parallelism, so the reported numbers understate the achievable throughput in a batched pipeline.} This rapid inference enables both the single-model and ensemble approaches to serve as efficient preprocessing \revisednew{tools for crystallographic analysis}, providing initial estimates of crystal system, space group, and lattice parameters without adding meaningful computational overhead to crystallographic workflows. \revisednew{While the present lattice parameter accuracy is not sufficient for direct initialization of conventional whole-pattern refinement, as discussed above, improved prediction accuracy or more robust downstream fitting methods with larger convergence windows could enable future integration of AlphaDiffract into automated refinement workflows.}

\section*{Discussion}

Despite the strong performance of AlphaDiffract, several limitations present avenues for future improvement. \revisednew{One factor that may influence classification performance is the inherent class imbalance within the training data.} Crystallographic databases like ICSD and Materials Project are heavily skewed, with a few common space groups (\textit{e.g.}, $P2_1/c$, $Pnma$) being vastly overrepresented, while dozens of others are rare. This imbalance, visible in Supplementary Figure~\figData, \revisednew{may contribute to some of the observed variation in classification performance across datasets and classes.} As shown in Supplementary Figure~\figAccuracyPerCs, performance varies significantly across crystal systems, with cubic systems achieving the highest accuracy (99.7\%, 99.3\%, and 82.4\% for crystal system prediction on ICSD, Materials Project, and RRUFF datasets, respectively, and 88.1\%, 76.8\%, and 78.2\% for space group prediction, respectively) while lower-symmetry systems show more variable performance. Triclinic and monoclinic systems, which are among the most challenging, show notably lower accuracies across all datasets. The Materials Project contains a higher fraction of these lower-symmetry systems, many of which are computationally predicted or metastable structures less common in the training data, and this compositional difference \revisednew{may contribute to the} lower overall classification accuracy on this dataset. However, the skew towards more common, higher-symmetry space groups in the training data is also commensurate with the likelihood of encountering them in real materials. \revised{This performance variability is further characterized at the lattice parameter level in Supplementary Table~\tabPerCsLpScores, which reports per-crystal system regression metrics alongside a crystal system-conditional baseline. For lattice lengths, the CS-conditional baseline has near-zero or negative $R^2$ throughout, confirming that the model's positive length $R^2$ reflects genuine predictive learning rather than distributional bias. For angles, interpretation depends critically on crystal system. For orthorhombic, tetragonal, and cubic systems, all angles are constrained to 90$^\circ$ under the Niggli reduced cell convention; thus, the CS-conditional baseline achieves trivially perfect MAE, MAPE, and $R^2$ for these systems. However, for monoclinic and triclinic structures where angles are free, the model substantially outperforms the baseline, suggesting genuine predictive capability in low-symmetry systems where it is most practically valuable. Supplementary Figure~\figPerCsBoxplots{} provides a sample-level view, showing that for most crystal systems and parameters, LP error distributions for correct and incorrect CS predictions overlap substantially, consistent with the motivation for a parallel prediction design where LP performance is not strongly contingent on CS accuracy. The most pronounced correlation between CS accuracy and LP errors is observed for trigonal and hexagonal systems, where incorrect CS predictions are associated with substantially elevated $\alpha$ and $\gamma$ errors. We hypothesize this reflects the shared difficulty of these samples: Patterns that are ambiguous for CS classification are likely also ambiguous for angle prediction, particularly given the Niggli axis-ordering convention in which the $120^\circ$ angle can appear as either $\alpha$ or $\gamma$, producing vertical clusters at both $90^\circ$ and $120^\circ$ in the parity plots for incorrectly predicted samples (Supplementary Figure~\figRegrParity). $\beta$, which lies near $90^\circ$ regardless of axis ordering, is largely unaffected. For triclinic and monoclinic structures, angle errors are large for both correct and incorrect CS groups; we attribute this primarily to the relative underrepresentation of diverse non-$90^\circ$ angle configurations in the training data, which makes precise angle prediction difficult regardless of CS accuracy. The horizontal cluster at $90^\circ$ in the parity plots likely reflects the LP head regressing toward the dominant $90^\circ$ signal in the training distribution on samples where the diffraction pattern provides insufficient information to discriminate the true angle. For lengths, the effect of CS accuracy is modest and comparatively consistent across all systems, further supporting the view that LP performance is not strongly contingent on CS accuracy for most crystal systems.}

Furthermore, the current model does not explicitly account for preferred orientation (texture), a common experimental artifact that significantly alters relative peak intensities. Since the model relies on the entire pattern, including intensities, it can be misled by experimental data with strong texture. Supplementary Figure~\figRruffExpVsSim{} compares space group classification performance on experimental versus synthetic RRUFF patterns from the same crystal structures, showing that synthetic patterns achieve 15-17\% higher accuracy than experimental patterns. This performance gap may be partly attributed to texture effects present in the experimental measurements but absent from our training data perturbations. For instance, attention analysis via Grad-CAM \cite{selvaraju2017grad} (Supplementary Figures~\figAttnCs{} and~\figAttnSg) reveals that the model sometimes focuses on very different pattern regions when processing experimental versus simulated data, particularly for structures exhibiting texture effects, suggesting the model adapts its classification strategy based on intensity variations. \revised{This divergence is most pronounced for forsterite (orthorhombic) and rutile (tetragonal), where experimental patterns show visibly different relative peak intensities consistent with preferred orientation, while minerals whose experimental and simulated patterns are more similar, such as zunyite, show more consistent attention maps between columns. For most crystal systems, attention falls primarily on peak positions; however, in some cases the model also attends to narrow regions between peaks, which may reflect the model exploiting systematic absences as diagnostic features, though these observations are qualitative and should be interpreted cautiously.} While our \revised{data and noise augmentation provide robustness across a broad range of signal-to-noise ratio (SNR) conditions, the more significant challenge for generalization is the presence of experimental artifacts not covered by the current augmentation scheme, and} future iterations could benefit from incorporating texture simulation into the training data or designing the architecture to be less sensitive to intensity ratios. \revised{Similarly, planar defects such as twinning and stacking faults, which induce anisotropic peak shifts and asymmetric broadening beyond pseudo-Voigt profiles, are not modeled in the current training data. Since neither ICSD nor Materials Project contains defect characterization information, pseudo-Voigt profiles nonetheless faithfully represent the crystallographic information available in the training databases; extending the augmentation scheme to include defect-broadening models represents a direction for future work.}

\revised{Additionally, it is worth noting that the mapping between measurements at different wavelengths via Bragg’s law preserves peak positions in $d$-spacing but does not strictly preserve other features of the diffraction pattern. In particular, peak widths and shapes depend on instrumental resolution and angular broadening effects, and therefore are not invariant under transformation between $2\theta$ domains. Relative peak intensities may also differ due to wavelength-dependent factors such as atomic form factors and absorption. To mitigate these discrepancies, the training data spans a broad range of peak broadening parameters (Table \ref{tab:sim}), exposing the model to substantial variability in line shapes. Future work could explicitly incorporate wavelength-dependent augmentation or instrument models to further improve generalization across measurement conditions.}

\revised{Regarding lattice parameter prediction, we wish to emphasize that the reported errors ($\sim$20\% MAPE on lengths, $\sim$3\% MAPE on angles for RRUFF) represent a meaningful step toward unified lattice determination but fall short of the precision required for practical Rietveld refinement initialization. The performance gap relative to crystal-system-specialized approaches such as Chitturi \textit{et al.} (9.2\% MAPE) reflects the inherent difficulty of a fully unified model that must implicitly infer crystal system before predicting lattice parameters. Improving this accuracy, for example through a sequential or hierarchically conditioned architecture, is an important direction for further study.}

Future work will focus on several directions. First, we plan to explore alternative architectures to improve multi-task learning. \revised{A hierarchically-conditioned architecture in which the lattice parameter regression head receives explicit input from the crystal system prediction is a physically intuitive extension; however, realizing this robustly presents non-trivial challenges, most notably that any misclassification of the crystal system would directly corrupt the downstream regression by applying incorrect symmetry constraints, and the benefit of conditioning is only realized when the classification is correct. Approaches that condition on the crystal system prediction in a soft or uncertainty-weighted manner may help address this and represent a promising direction for future work.} \revised{More broadly, sample composition, when available, could in principle be used not merely as an additional input feature, but to impose structure on the model's latent space, for example by encouraging the backbone to cluster representations of chemically similar structures and thereby implicitly constrain predictions toward physically plausible lattice geometries without requiring composition at inference time; this represents a further direction for future architectural exploration.} Finally, we anticipate that by introducing diffraction peaks at random locations into the training data, we can desensitize the model to the presence of additional phases, providing a unique tool that can index diffraction patterns from mixtures. Second, while our model predicts crystal system, space group, and lattice parameters, it does not yet determine atomic positions -- the final step needed for complete end-to-end crystal structure determination from powder diffraction data. However, our work addresses critical subproblems within this broader challenge. Recent approaches have explored different paradigms for this inverse design problem. For instance, the Crystalyze model by Riesel \textit{et al.} uses a diffusion model that requires accurate lattice parameters, composition, and number of atoms to initialize the unit cell before generating atomic positions \cite{Riesel_jacs_2024}. AlphaDiffract's lattice parameter predictions could \revisednew{provide useful conditioning information for} this initialization step, and the crystal system and space group predictions could be used to enforce symmetry constraints and guide the diffusion model toward atomic positions consistent with the underlying crystallographic symmetries. Alternative LLM-based approaches, such as DiffractGPT \cite{Choudhary_jpcl_2025} and deCIFer \cite{johansen2025decifer}, operate on different principles but accept optional conditioning information (\textit{e.g.}, space group information for deCIFer) that our predictions could potentially enhance. Integration of our predictions with these generative approaches represents a promising direction for future work toward fully automated structure solution. Nonetheless, \revisednew{joint} prediction of lattice parameters, crystal system, and space group from experimental PXRD data alone, as demonstrated here, already addresses a fundamental bottleneck in crystallographic analysis. This capability enables rapid \revisednew{preliminary crystallographic} characterization without requiring prior knowledge of composition or separate models for different crystal systems, advancing the goal of high-throughput materials discovery. \revised{For users who do have compositional knowledge, that information is most naturally applied downstream of AlphaDiffract; for example, by using the predicted space group and lattice parameters alongside known composition to narrow a database search}\revisednew{, inform subsequent refinement workflows,} \revised{or evaluate chemical plausibility of the prediction.}


\revisednew{In summary, AlphaDiffract provides} a unified deep learning framework for lattice determination from PXRD patterns. Leveraging a 1D ConvNeXt architecture and a physics-based dataset of over 31 million simulated patterns incorporating realistic experimental effects, AlphaDiffract simultaneously predicts crystal system, space group, and all six lattice parameters from a single diffraction pattern.

On the RRUFF benchmark, our ensemble model achieves 81.7\% crystal system and 66.2\% space group classification accuracy while simultaneously setting a new standard for lattice parameter regression from experimental PXRD data. This performance matches or exceeds prior state-of-the-art methods specialized for classification alone, demonstrating that unified prediction of symmetry and lattice parameters is feasible without compromising accuracy. Our physics-aware GEMD loss further concentrates prediction errors near the true space group in the symmetry hierarchy, providing crystallographically meaningful predictions even when exact matches are not achieved.

Determination of crystal structures from powder diffraction remains challenging, with existing \textit{ab initio} methods limited by data quality requirements and restricted applicability \cite{RN183,RN197,RN148,RN311,RN149}. AlphaDiffract addresses a critical component of this workflow by directly predicting lattice parameters and space group candidates from diffraction patterns alone, bypassing explicit peak finding and indexing steps. When predicted space groups differ from ground truth, they typically represent immediate subgroups or supergroups differing by a single symmetry generator, information that aids both conventional structure determination tools and crystallographic database searches \cite{Kaduk:an0607}. As deep learning methods continue to mature and integrate with existing crystallographic workflows, they increasingly offer practical solutions to long-standing challenges in powder diffraction structure solution.

\section*{Methods}

\subsection*{Data curation}
\label{subsec:curation}

Materials structures for training were downloaded as CIF files \cite{Brown:an0595} from two major crystallographic databases: 198,778 structures from ICSD \cite{NIST_ICSD} and 153,169 structures from the Materials Project \cite{jain_aplmat_2013}. No attempts were made to prevent structure duplication \revisednew{at this stage}. As a first refinement step, we ensured that all CIF files could be successfully parsed into \texttt{pymatgen} \cite{ong2013python} structures to enable subsequent quality checks and the writing of Niggli reduced structures. This parsing step resulted in 186,765 validated structures from ICSD and 153,169 from the Materials Project.

The second stage of refinement applied several quality checks using \texttt{pymatgen}'s SpacegroupAnalyzer \cite{ong2013python}. We confirmed that the space group of the conventional cell remained consistent across a range of site tolerances from 0.01 \AA\ to 0.1 \AA\ and angle tolerances from $0.1^\circ$ to $5^\circ$, where the upper bounds represent the default values used by the Materials Project \cite{ong2013python}. Additionally, we verified that the Niggli reduced cell was consistent with crystal system symmetry within a relative tolerance of $10^{-3}$ when comparing expected unit cell length and angle equivalences. We also applied practical constraints, requiring that structures have a unit cell volume less than 100,000 \AA$^3$\ and contain no more than 500 atoms per unit cell. After applying these screening criteria, the final dataset consisted of 165,634 valid structures from ICSD and 146,633 from the Materials Project.

Materials structures and PXRD patterns for testing were downloaded from the RRUFF database \cite{lafuente_hmc_2015}, consisting of 1,362 PXRD patterns, which are background-corrected to flat baselines but otherwise unmodified, and 2,901 DIF files, which contain structural information obtained through refinement of the experimental PXRD patterns. \revised{No RRUFF structures or patterns were included in model training.} \revisednew{However, RRUFF performance informed two discrete evaluation checkpoints during model development: assessment of the GEMD loss's effect on classification performance prior to training the full joint model, and selection of ensemble size following completion of full model training. Separately, unlabeled noise characteristics extracted from RRUFF patterns were also used to define physically realistic noise augmentations for training, as described in Methods section ``\nameref{subsec:noise}''. However, RRUFF was not used for architecture selection or other model hyperparameter optimization. We nonetheless treat RRUFF as an experimental development benchmark rather than a strictly held-out independent test set. This role differs from that of the ICSD and Materials Project validation sets, which were used for iterative model development throughout.} \revisednew{No attempts were made to deduplicate RRUFF entries against the ICSD and Materials Project training databases at this stage.} \revised{The RRUFF patterns used in this study are the background-corrected versions provided directly by the RRUFF database, where backgrounds have been subtracted to produce flat baselines by the database curators. No additional background subtraction or processing was applied by the authors beyond the wavelength conversion described below.} Initial screening removed DIF files lacking essential information (cell parameters, space group, or wavelength data), resulting in 2,572 valid files. Of these, 1,880 DIF files could be successfully parsed into \texttt{pymatgen} lattices, though not necessarily into full structures, as some lacked atom coordinates or contained non-standard atom symbols.

Further refinement verified that 1,867 structures had lattices consistent with crystal system symmetry within a relative tolerance of $10^{-3}$ (comparing expected unit cell length and angle equivalences) and a unit cell volume less than 100,000 \AA$^3$. Among these, 837 structures had corresponding PXRD patterns from the initial download of 1,362 patterns. Cross-validation between the DIF and PXRD files confirmed that 745 structures had matching crystal systems and lattice parameters. Since many of the experimental PXRD patterns were measured using Cu K$_\alpha$ radiation (8.04 keV) while our model assumes an energy of 20 keV, we converted the patterns to the desired $2\theta$ range using Bragg's law. This conversion step resulted in 734 structures with patterns that did not contain missing data in the new $2\theta$ range. Of these 734 structures, 240 had complete atom position information and yielded GSAS-II \cite{toby:gsasii} PXRD simulations with peaks consistent with those reported in the DIF files. This validation process required manually determining the origin and nonstandard settings in some cases due to incomplete data. Specifically, many RRUFF entries use nonstandard space group notation that required translation to the full Hermann-Mauguin symbols needed by GSAS-II. For structures in space groups with ambiguous symmetry center locations (Origin 1 and 2 settings), we determined the correct origin through systematic examination of special positions, interatomic distances, and visual inspection of the structure.

We utilized the full set of 734 structures for crystal system and space group classification, but only the subset of 240 structures for lattice parameter regression, as reliable parameter extraction required Niggli reduced cells derived from complete structures rather than from lattice parameters alone. We note that while the RRUFF database is widely regarded as an experimental PXRD database, some entries are calculated powder profiles derived from single crystal data \revised{(profiles generated by the RRUFF curators from single crystal refinement data, not simulated by the authors)}. Our \revisednew{benchmark} datasets include 59 such entries among the 734 patterns for classification (8.0\%) and 55 among the 240 patterns for regression (22.9\%), which we treat as real-world data consistent with their inclusion in RRUFF.

\revisedfin{To enable comparison with G\'{o}mez-Peralta \etal{} \cite{gomez2025insights}, we constructed an additional regression-only evaluation set from the 30,643 COD identifiers provided in the authors' associated GitHub repository. The COD results reported in the main text refer specifically to this author-designated inorganic collection rather than the full Crystallography Open Database \cite{gravzulis2009crystallography}. The corresponding CIFs were retrieved from COD and subjected to the same parsing and structural-quality criteria applied to ICSD and Materials Project, yielding 27,041 valid structures. For reference, the associated study reports 26,135 structures after its own preprocessing \cite{gomez2025insights}. For each retained structure, 100 augmented PXRD profiles were simulated using the same GSAS-II procedure as for ICSD and Materials Project, yielding 2,704,100 profiles. COD was used only to evaluate lattice-parameter regression with the fixed 10-model ensemble and did not affect training, model selection, or hyperparameter optimization.}

Supplementary Figure~\figData\textbf{a}, \textbf{b}, and \textbf{d} shows the distribution of crystal systems and space groups in the final ICSD, Materials Project, and RRUFF datasets, respectively, used for crystal system and space group classification. Additionally, Supplementary Figures~\figDataLengths{} and~\figDataAngles{} show the distribution of lattice lengths and angles, respectively, of each Niggli reduced cell in the final ICSD, Materials Project, and RRUFF datasets used for lattice parameter regression.

\subsection*{Structure similarity evaluation}
\label{subsec:similarity}
To evaluate structural similarity and database overlap, we employed the CrystalNN site fingerprint method \cite{zimmermann2020local} with structural order parameters as implemented in matminer \cite{ward2018matminer} to characterize each structure in \revisednew{the ICSD, Materials Project, and RRUFF} databases\revisedfin{, as well as the COD evaluation set}. We performed hierarchical clustering, as implemented in scipy \cite{virtanen2020scipy}, using complete linkage with a Euclidean distance metric on the fingerprints of structures sharing identical composition and space group. Structures were classified as ``equivalent'' if they belonged to the same cluster within a distance threshold of 0.9, following the convention proposed in the Materials Project documentation \cite{jain_aplmat_2013}. \revisednew{In addition to facilitating the evaluation of AlphaDiffract performance on structurally distinct datasets as reported in the Results, this} analysis enabled us to decompose the {combined ICSD and Materials Project} dataset into structures appearing exclusively in ICSD, exclusively in the Materials Project, or in both databases, with each dataset further subdivided into unique structures and any structural equivalents (Supplementary Figure~\figData\textbf{c}). This clustering approach revealed that the final curated dataset comprises 65,748 unique ICSD structures, 93,418 unique Materials Project structures, and 39,239 unique structures common to both databases, and that 18,683 ICSD structures, 12,616 Materials Project structures, and 82,545 structures common to both databases are structural equivalents of other structures in their respective datasets. \revisednew{Structural fingerprints could not be computed for 1 structure in the ICSD dataset and 17 structures in the Materials Project dataset and are excluded from the reported counts. For the purposes of deduplication across training and validation partitions, these structures are conservatively considered duplicates of any partition-opposite structure sharing the same composition and space group. Partition-level counts of database records, simulated patterns, structural clusters, and unfingerprinted structures for the ICSD and Materials Project training and validation splits are reported in Supplementary Table~\tabDatasetPartitions.}

\begin{table}[b]
    \centering
    \begin{tabular}{ccc}
        \hline
        \textbf{Parameter} & \textbf{Minimum} & \textbf{Maximum} \\
        \hline
        \textbf{Sample and instrumental broadening (static)} & & \\
        Microstrain & 500 & 10,000 \\
        Crystallite Size ($\mu m$) & 0.1 & 1 \\
        U (cdeg$^2$) & 0 & 3 \\
        V (cdeg$^2$) & -1 & 0 \\
        W (cdeg$^2$) & 0 & 4 \\
        \hline
        \textbf{Noise augmentation (dynamic)} & & \\
        $\lambda_{\text{max}}$ & 1 & 100 \\
        $\sigma_{\text{rel}}$ & $10^{-3}$ & $10^{-1}$ \\
    \end{tabular}
    \caption{\textbf{Parameters for PXRD pattern augmentation.} For each structure, we generated 100 augmented PXRD patterns using GSAS-II by uniformly sampling the sample and instrumental parameters between the given ranges (static augmentation). During training \revisedfin{and evaluation on simulated datasets}, noise parameters ($\lambda_{\text{max}}$, $\sigma_{\text{rel}}$) were additionally sampled uniformly in the data loader (dynamic augmentation). The unit ``cdeg'' refers to centidegrees (degrees/100).}
    \label{tab:sim}
\end{table}

\subsection*{Noise simulation}
\label{subsec:noise}

To realistically simulate noise in synthetic PXRD patterns, we characterized the noise properties of PXRD data from the RRUFF database. Background regions of each pattern were identified and used to estimate noise parameters (Supplementary Figure~\figNoiseEstimation\textbf{a}). From these background regions, we extracted two key noise characteristics: $\lambda_{\max}$, which quantifies Poisson noise, and $\sigma_{\text{rel}}$, which quantifies Gaussian noise on a relative intensity scale. Specifically, $\sigma_{\text{rel}}$ was set equal to the standard deviation of the background regions, while $\lambda_{\max}$ was set to be the inverse of the mean of the background regions. The parameter $\lambda_{\max}$ serves as the mean of a Poisson distribution from which noisy intensities are sampled from clean simulated patterns, capturing the counting statistics inherent to X-ray detection. The parameter $\sigma_{\text{rel}}$ represents the standard deviation of Gaussian noise applied to normalized intensities (i.e., after rescaling each pattern to [0,1]), making it a measure of noise relative to the pattern's maximum intensity rather than on an absolute scale. The distributions of $\lambda_{\max}$ and $\sigma_{\text{rel}}$ values across the RRUFF dataset are shown in Supplementary Figure~\figNoiseEstimation\textbf{b-c} and \textbf{f-g}, respectively. To validate our noise model, we compared RRUFF patterns at various noise levels with clean simulated patterns to which we applied corresponding amounts of Poisson and Gaussian noise (Supplementary Figure~\figNoiseEstimation\textbf{d-e} and \textbf{h-i}). While experimental patterns contain both Poisson and Gaussian noise components simultaneously (with Gaussian noise appearing dominant), the visual comparison shows that the overall noise characteristics and trends follow expected patterns, supporting the adequacy of our two-step noise model for generating realistic synthetic training data.

The noisy PXRD intensity is obtained through the following sequential steps. First, Poisson noise is applied and the intensity rescaled:

\begin{equation}
I_{\text{Pois}}(2\theta) = \frac{\max(I_{\text{clean}})}{\lambda_{\max}} \cdot \text{Poisson}\left(\lambda_{\max} \cdot \frac{I_{\text{clean}}(2\theta)}{\max(I_{\text{clean}})}\right)
\end{equation}

Then, Gaussian noise is applied to the normalized intensity:

\begin{equation}
I_{\text{Gauss}}(2\theta) = \frac{I_{\text{Pois}}(2\theta) - \min(I_{\text{Pois}})}{\max(I_{\text{Pois}}) - \min(I_{\text{Pois}})} + \mathcal{N}(0, \sigma_{\text{rel}})
\end{equation}

Finally, the pattern is renormalized to [0,1] and rescaled back to the original intensity range:

\begin{equation}
I_{\text{noisy}}(2\theta) = \left[\frac{I_{\text{Gauss}}(2\theta) - \min(I_{\text{Gauss}})}{\max(I_{\text{Gauss}}) - \min(I_{\text{Gauss}})}\right] \cdot (\max(I_{\text{Pois}}) - \min(I_{\text{Pois}})) + \min(I_{\text{Pois}})
\end{equation}

where $I_{\text{clean}}(2\theta)$ is the clean simulated intensity, $I_{\text{Pois}}(2\theta)$ is the intensity after Poisson noise application, $I_{\text{Gauss}}(2\theta)$ is the normalized intensity with Gaussian noise added, $\text{Poisson}(\lambda)$ denotes sampling from a Poisson distribution with mean $\lambda$, and $\mathcal{N}(0, \sigma_{\text{rel}})$ denotes sampling from a Gaussian distribution with mean zero and standard deviation $\sigma_{\text{rel}}$.

During training \revisedfin{and evaluation on simulated datasets}, noise augmentation was applied dynamically in the data loader by randomly sampling $\lambda_{\max}$ uniformly between 1 and 100 and $\sigma_{\text{rel}}$ uniformly between $10^{-3}$ and $10^{-1}$ for each \revisedfin{pattern}. \revised{This broad range, spanning conditions from low-noise synchrotron data to high-noise laboratory measurements, is designed to ensure robustness across the diversity of signal-to-noise ratio (SNR) conditions encountered in practice.} While these ranges are biased toward noisier samples compared to the empirical distributions shown in Supplementary Figure~\figNoiseEstimation, we found that this bias empirically improved model performance on real experimental data, likely by enhancing the model's robustness to noise.

\subsection*{Model architecture}
\label{sec:model_arch}

The AlphaDiffract model consists of a feature extractor that takes PXRD patterns and projects them to the feature space. The extracted features are passed to prediction heads, each of which is specialized to predict one of the target properties. The feature extractor is composed of 3 \convnext{} \cite{Liu_arxiv_2022} blocks adapted to 1D. An average pooling layer downsamples the feature map by a factor of 2 after each \convnext{} block. The architecture of a \convnext{} block is shown in Figure~\ref{fig:model}, where the 2D convolution layers in \cite{Liu_arxiv_2022} are replaced with the 1D counterparts, but the key block design features of \convnext{}, including the inverted bottleneck and the upstream positioned depthwise convolution layer, are preserved. The block employs a residual connection like in \convnext{} where the input is added to the output of the final pointwise convolution layer. The stem (\emph{i.e.}, non-residual) branch is subject to random drop path \cite{Larsson_arxiv_2016} with a rate of 0.3. Other parameters, including the input/output channels, kernel size and pooling stride for each block, are listed in Table \ref{tab:feat_extractor}. With an input dimension of 8192, the feature extractor outputs features with a dimension of 560.

\begin{table}[b]
    \centering
    \begin{tabular}{C{1cm}C{2.5cm}C{2.5cm}C{2.5cm}C{2.5cm}}
        \hline
        \textbf{Block} & \textbf{Input channels} & \textbf{Output channels} & \textbf{Kernel size ($k$)} & \textbf{Pooling stride ($s$)} \\
        \hline
        1 & 1 & 80 & 100 & 5 \\
        2 & 80 & 80 & 50 & 5 \\
        3 & 80 & 80 & 25 & 5 \\
        \hline
    \end{tabular}
    \caption{\textbf{Block-level architecture of the feature extractor.}}
    \label{tab:feat_extractor}
\end{table}

\begin{table}[b]
    \centering
    \begin{tabular}{C{3cm}C{3cm}C{3cm}C{3cm}C{3cm}}
        \hline
        \textbf{Predicted quantity} & \textbf{Input dimension} & \textbf{Hidden dimension} & \textbf{Output dimension} \\
        \hline
        Crystal system & 560 & 2300 $\rightarrow$ 1150 & 7 \\
        Space group & 560 & 2300 $\rightarrow$ 1150 & 230 \\
        Lattice parameters & 560 & 512 $\rightarrow$ 256 & 6 \\
        \hline
    \end{tabular}
    \caption{\textbf{Dimensions of the MLP prediction heads.}}
    \label{tab:prediction_heads}
\end{table}

Each quantity (crystal system, space \revisednew{group}, \emph{etc.}) is predicted by its own prediction head. A prediction head is a multi-layer perceptron that maps the extractor features to vectors with quantity-specific dimensions. \revised{Each prediction head receives only the shared backbone feature vector as input and operates independently of the other heads; there is no communication between heads and no conditioning of the lattice parameter head on the crystal system or space group predictions.} The input, output and hidden dimensions of the prediction heads are shown in Table \ref{tab:prediction_heads}. For classification quantities, the output vectors are passed through a softmax layer to give the predicted probability of each class. For lattice parameters, the outputs are 6-dimensional vectors corresponding to the $a$, $b$, $c$, $\alpha$, $\beta$, and $\gamma$ of the unit cell's lengths and angles; the outputs are passed through a sigmoid function and then scaled between a set lower bound and upper bound:

\begin{equation}
    \hat{y}_{i} = \sigma(p_{i}) (\mathit{UB}_i - \mathit{LB}_i) + \mathit{LB}_i
\end{equation}

where $i$ indexes the elements of the 6 lattice parameters, and $p_i$ is the element in the prediction head output corresponding to parameter $i$. $\textit{LB}_i$ and $\textit{UB}_i$ denote the lower and upper bounds and are 0 and 500 \AA{} for lengths and 0\degree{} and 180\degree{} for angles.

\subsection*{Loss function}
\label{subsec:loss}

Our overall loss function combines three components tailored to different prediction tasks. For crystal system (CS) and space group (SG) classification, we employ cross-entropy loss for both tasks. The SG loss also optionally includes the Graph Earth Mover's Distance (GEMD) loss weighted by a hyperparameter $\mu$. For our combined model with lattice parameter regression, we additionally include mean squared error (MSE) loss for lattice parameter (LP) prediction. The batch loss is computed as the average over $N$ samples:

\begin{equation}
\mathcal{L} = \mathcal{L}_{CS} + \mathcal{L}_{SG} + \mathcal{L}_{LP} = \frac{1}{N} \sum_{i=1}^{N} \ell^{(i)}
\end{equation}

where the loss for a single sample is given by:

\begin{equation}
    \ell^{(i)} = -\sum_{j=1}^{7} y_{CS,j}^{(i)} \log(\hat{y}_{CS,j}^{(i)}) - \sum_{j=1}^{230} y_{SG,j}^{(i)} \log(\hat{y}_{SG,j}^{(i)}) + \mu \sum_{k=1}^{230}\sum_{j=1}^{230} y_{SG,j}^{(i)} D_{jk} \hat{y}_{SG,k}^{(i)} + \frac{1}{6}\sum_{j=1}^{6} (y_{LP,j}^{(i)} - \hat{y}_{LP,j}^{(i)})^2
\end{equation}

Here, $y^{(i)}$ denotes the ground truth labels or values while $\hat{y}^{(i)}$ represents the model predictions of the $i^{\text{th}}$ data sample. The third term represents the GEMD loss, which leverages the hierarchical structure of space groups by penalizing misclassifications based on their distance in the maximal subgroup graph represented by the distance matrix $D$, where $D_{jk}$ encodes the number of hops between space groups $j$ and $k$ through the maximal subgroup graph. \revised{With the exception of the hyperparameter $\mu$, the unweighted sum reflects an empirical finding that the three loss components are naturally on comparable scales during training: cross-entropy losses for crystal system and space group classification, and MSE for lattice parameters in their normalized form via the sigmoid scaling.}

\subsection*{Model training}
\label{subsec:training}

The model is trained using the AdamW optimizer \cite{Loshchilov_arxiv_2017} with a learning rate of $2 \times 10^{-4}$ and weight decay factor of 0.01. We used a cyclic learning rate schedule \cite{smith2017cyclical} that linearly increases the learning rate from 10\% of its nominal value to the full value over six epochs (a half cycle), then decreases it back, with each subsequent cycle's amplitude reduced by half. Training for each model was performed on a single NVIDIA H100 GPU using a batch size of 64. \revisednew{For each of ICSD and Materials Project, 10\% of the curated structures were randomly assigned to the validation set, with the remaining 90\% assigned to training; this split was performed at the level of crystallographic structures, prior to the generation of the 100 augmented diffraction patterns per structure, such that all augmented patterns derived from a given structure were assigned to the same partition. No deduplication or structural equivalence-based clustering was applied prior to this split (see Methods section ``\nameref{subsec:curation}''); the extent of resulting structural overlap between training and validation partitions, and its effect on reported performance, is characterized in ``\nameref{subsec:dedup}.''} \revised{Each model was trained for 15 epochs, corresponding to approximately 6.6 million gradient updates per model (computed from the training set size of $\sim$28 million patterns and batch size of 64).}

\subsection*{Uncertainty estimation}
\label{subsec:uncertainty}

\revisednew{We distinguish uncertainty associated with finite evaluation set sampling from variability in predictions associated with model initialization and input perturbation. Confidence intervals on classification accuracy and regression metrics are estimated by nonparametric bootstrap resampling at the level of evaluation structures. Variability across independently trained ensemble members and, where applicable, augmented diffraction patterns is quantified separately as described below. Neural network inference itself is deterministic for a fixed model and input; these latter variability estimates therefore characterize differences associated with model initialization and input perturbation rather than stochasticity at inference time.}

For \revisednew{crystal system and space group classification}, we first compute the standard deviation of class probabilities across ensemble members, $\sigma_{model}^{(i)}$, \revisednew{for each prediction}. \revisednew{We summarize this per-sample variability on the scale of the aggregate accuracy metric using the propagated quantity,}
\begin{equation}
    \sigma_{model} = \frac{1}{N} \sqrt{\sum_{i=1}^{N} \left(\sigma_{model}^{(i)} \right)^2},
\end{equation}
where $N$ is the total number of samples. \revisednew{For the ensemble-variability analysis, we report $1.96\sigma_{\mathrm{model}}$ as a propagated variability margin. This quantity reflects variation in predicted class probabilities across independently trained ensemble members and is not interpreted as a confidence interval on classification accuracy.} \revisednew{For ICSD and Materials Project (augmented) datasets, we first define a structure-level classification score as the fraction of the 100 augmented diffraction patterns for structure $j$ that are classified correctly,}
\begin{equation}
    a^{(j)} = \frac{1}{100}\sum_{k=1}^{100} \mathbb{I}\left(y_k^{(j)}=\hat{y}_k^{(j)}\right).
\end{equation}
\revisednew{For RRUFF, where each structure is represented by a single diffraction profile, the corresponding structure-level score is simply binary correctness.}
\revisednew{The overall classification accuracy is then the mean of these structure-level scores,}
\begin{equation}
A = \frac{1}{n}\sum_{j=1}^{n} a^{(j)},
\end{equation}
\revisednew{where $n$ is the number of evaluation structures ($n=N/100$ for augmented ICSD and Materials Project validation sets and $n=N$ for the RRUFF benchmark).}
\revisednew{To estimate finite evaluation set uncertainty, we perform nonparametric bootstrap resampling of the $n$ structures with replacement. For each of 10,000 bootstrap replicates, the classification accuracy is recalculated from the resampled structure-level scores. The 2.5$^\mathrm{th}$ and 97.5$^\mathrm{th}$ percentiles of the resulting bootstrap distribution define the reported 95\% confidence interval.} \revisednew{For the single classification model (``Cls.''), bootstrap resampling is applied directly to its structure-level scores. For the ``Avg.'' result, the same resampled structure indices are applied to all 10 independently trained models, the accuracy of each model is recalculated for that replicate, and the 10 accuracies are then averaged. For the ensemble result, bootstrap resampling is applied to the structure-level scores obtained from the ensemble prediction.} \revisednew{The structure-level bootstrap confidence intervals are reported with the classification accuracies in Tables \ref{tab:cs_class_scores} and \ref{tab:sg_class_scores}, while the separately propagated ensemble-variability measure is used in Figure \ref{fig:ens_class}\textbf{a} and is compared directly with the bootstrap intervals for the final ensemble model in Supplementary Table~\tabClassificationUncertainty.}

\revisednew{For lattice parameter regression, finite evaluation set confidence intervals and prediction variability are estimated separately. To quantify prediction variability arising from ensemble members and input augmentation, we propagate these sources through the continuous} regression metrics using standard error propagation. Let $N$ denote the total number of \revisednew{evaluation} samples and $n$ the number of \revisednew{evaluation} structures, with each structure $j$ \revisednew{represented by 100 augmented diffraction patterns for ICSD and Materials Project validation sets \revisedfin{and the COD evaluation set}, or by a single diffraction profile for the RRUFF benchmark.} For each augmented sample $i$ belonging to structure $j$, the \revisednew{ensemble variability} $\sigma_{model}^{(i)}$ \revisednew{is defined as} the standard deviation of predictions across ensemble members. The augmentation \revisednew{variability} $\sigma_{aug}^{(j)}$ captures variation in the ensemble-averaged predictions across the 100 augmented patterns for structure $j$. \revisednew{To obtain a combined measure of prediction variability for each structure, we combine these two contributions in quadrature,}
\begin{equation}
    \sigma_{total}^{(j)} = \sqrt{\frac{1}{100}\sum_{k=1}^{100}\left( \sigma_{model,k}^{(j)}\right)^2 + \left(\sigma_{aug}^{(j)}\right)^2}.
\end{equation}
\revisednew{We then propagate this prediction variability to each regression metric using the delta method.} For mean absolute error (MAE), the \revisednew{propagated variability} is
\begin{equation}
    \sigma_\text{MAE} = \frac{1}{n}\sqrt{\sum_{j=1}^n\left( \sigma_{total}^{(j)}\right)^2}.
\end{equation}
For mean absolute percentage error (MAPE), the \revisednew{propagated variability} scales by the inverse of the true values:
\begin{equation}
    \sigma_\text{MAPE} = \frac{100}{n}\sqrt{\sum_{j=1}^n\left( \frac{\sigma_{total}^{(j)}}{|y^{(j)}|}\right)^2}.
\end{equation}
For the coefficient of determination ($R^2$), we use the derivative $\partial R^2/\partial \hat{y}^{(j)} = -2(\hat{y}^{(j)} - y^{(j)}) / SS_{tot}$, where $SS_{tot} = \sum_{i=1}^{n}(y^{(j)} - \bar{y})^2$, giving
\begin{equation}
    \sigma_{R^2} = \sqrt{\sum_{j=1}^n\left( \frac{2(\hat{y}^{(j)} - y^{(j)})}{SS_{tot}} \sigma_{total}^{(j)}\right)^2}
\end{equation}
Here $y^{(j)}$ and $\hat{y}^{(j)}$ denote the true and predicted values for structure $j$ (averaged over its augmentations \revisednew{where applicable}), respectively.

\revisednew{To estimate confidence intervals on regression performance, we additionally perform nonparametric bootstrap resampling at the level of evaluation structures. For ICSD, \revisedfin{Materials Project, and COD}, the true and ensemble-predicted lattice parameters are first averaged across the 100 augmented diffraction patterns associated with each structure; for RRUFF, each structure contributes a single profile. Evaluation structures are then resampled with replacement, and the MAE, MAPE, and $R^2$ metrics are recalculated for each bootstrap replicate. For the aggregate lattice length and lattice angle metrics reported in Tables~\ref{tab:lp_scores} and \ref{tab:unique_regression}, the three corresponding parameter-specific metrics are averaged within each bootstrap replicate. We use 10,000 bootstrap replicates and report the 2.5$^\mathrm{th}$ and 97.5$^\mathrm{th}$ percentiles of the resulting metric distributions as the 95\% confidence interval.}

\section*{Data availability}
Crystal structures are available from the Materials Project \cite{jain_aplmat_2013}, \revisedfin{COD \cite{gravzulis2009crystallography},} and RRUFF \cite{lafuente_hmc_2015} databases (open access) and ICSD \cite{NIST_ICSD} (requires license). \revised{Model weights and predictions derived from publicly available datasets} \revisednew{are made available at \texttt{https://github.com/AdvancedPhotonSource/OpenAlphaDiffract}.}

\section*{Code availability}
The code used in this study is made available at \revised{\texttt{https://github.com/AdvancedPhotonSource/OpenAlphaDiffract}}.

\section*{Acknowledgements}
The authors thank Dr. Laurent Chapon for insightful discussions on the application of machine learning methods to crystallographic analysis of PXRD data. This research used resources of the Advanced Photon Source, a U.S. Department of Energy (DOE) Office of Science User Facility operated for the DOE Office of Science by Argonne National Laboratory under Contract No. DE-AC02-06CH11357.

\section*{Funding}
This research is based on work supported by the U.S. DOE Office of Science-Basic Energy Sciences, under Contract No. DE-AC02-06CH11357.

\section*{Author contributions}
M.J.C. conceived the study, supervised the research, and contributed to data analysis and manuscript preparation. N.A. developed the symmetry-constrained training approach, led data processing, and contributed to model training and evaluation, analysis, and manuscript preparation. M.D. led model architecture development and implementation and contributed to data processing, model training and evaluation, analysis, and manuscript preparation. H.S. developed the PXRD simulation pipeline and generated diffraction patterns for training. J.P.H. performed crystal structure data curation and contributed to PXRD pattern generation. A.L. contributed to data processing, analysis, and manuscript preparation. X.Y. conducted benchmarking against prior methods and contributed to analysis. M.P. developed user-facing model deployment tools, containerized the codebase, and performed validation testing. B.H.T. provided crystallographic expertise, performed data curation, contributed to PXRD pattern generation, and assisted with manuscript preparation. All authors reviewed and approved the final manuscript. N.A. and M.D. contributed equally to this work.

\section*{Competing interests}
\revisednew{The authors declare no competing financial or non-financial interests.}

\section*{Additional information}
Correspondence and requests for materials should be addressed to Nina Andrejevic (\texttt{nandrejevic@anl.gov}), Ming Du (\texttt{mingdu@anl.gov}), or Mathew J. Cherukara (\texttt{mcherukara@anl.gov}).

\printbibliography[heading=bibintoc,title={References}]

\input{supplementary}

\end{document}

%% file: metadata.tex
\usepackage{authblk}

\title{AlphaDiffract: Automated Crystallographic Analysis of Powder X-ray Diffraction Data}
\date{}

\author[1*$\dagger$]{Nina Andrejevic}
\author[1*$\dagger$]{Ming Du}
\author[1]{Hemant Sharma}
\author[1]{James P. Horwath}
\author[1]{Aileen Luo}
\author[1]{Xiangyu Yin}
\author[1]{Michael Prince}
\author[1]{Brian H. Toby}
\author[1*]{Mathew J. Cherukara}

\affil[1]{\textit{Advanced Photon Source, Argonne National Laboratory, Lemont, IL 60439, USA}}

\affil[$\dagger$]{Co-first authors with equal contribution.}

\affil[*]{E-mail: \texttt{nandrejevic@anl.gov, mingdu@anl.gov, mcherukara@anl.gov}}

%% file: supplemental_refs.tex
\newcommand{\tabPxrdInputComparison}{2}

\newcommand{\tabClassificationUncertainty}{3}

\newcommand{\tabClassImbalance}{5}

\newcommand{\tabRruffProfileSource}{6}

\newcommand{\tabLpScoresSupp}{7}

\newcommand{\tabPerCsLpScores}{8}

\newcommand{\tabDatasetPartitions}{9}

\newcommand{\figData}{1}

\newcommand{\figDataLengths}{2}

\newcommand{\figDataAngles}{3}

\newcommand{\figNoiseEstimation}{4}

\newcommand{\figAccuracyPerCs}{5}

\newcommand{\figRegrParity}{6}

\newcommand{\figPerCsBoxplots}{7}

\newcommand{\figRruffExpVsSim}{8}

\newcommand{\figAttnCs}{9}

\newcommand{\figAttnSg}{10}

%% file: license.tex
{\large\bfseries GOVERNMENT LICENSE}
 
The submitted manuscript has been created by UChicago Argonne, LLC, Operator of Argonne National Laboratory (“Argonne”). Argonne, a U.S. Department of Energy Office of Science laboratory, is operated under Contract No. DE-AC02-06CH11357. The U.S. Government retains for itself, and others acting on its behalf, a paid-up nonexclusive, irrevocable worldwide license in said article to reproduce, prepare derivative works, distribute copies to the public, and perform publicly and display publicly, by or on behalf of the Government. The Department of Energy will provide public access to these results of federally sponsored research in accordance with the DOE Public Access Plan. http://energy.gov/downloads/doe-public-access-plan.

\clearpage

%% file: supplementary.tex
\clearpage
\clearpage

\setcounter{section}{0}
\setcounter{equation}{0}
\setcounter{figure}{0}
\setcounter{table}{0}
\setcounter{page}{1}
\makeatletter
\renewcommand{\figurename}{Supplementary Figure}
\renewcommand{\tablename}{Supplementary Table}

\title{\vspace{-1.5cm}\Large Supplementary Information for ``AlphaDiffract: Automated Crystallographic Analysis of Powder X-ray Diffraction Data''}
\date{}
\maketitle
\vspace{-20pt}

\newrefsection

\section*{Supplementary Note: Literature survey on deep learning methods for PXRD analysis}

\subsection*{Architecture design}
1D convolutional neural networks (CNNs) have been the predominant architectural choice for PXRD analysis \cite{Park_iucrj_2017, Vecsei_prb_2019, Zaloga_mtc_2020, Chitturi_jac_2021, Lee_ais_2022, Lee_ais_2023, Corriero_jac_2023, Gomez_jpca_2023, Salgado_npj_2023, Riesel_jacs_2024}. CNNs are well-suited to PXRD data because they capture both local features (individual peaks) through inductive bias and long-range relationships (peak patterns) through hierarchical downsampling—both critical for structure identification. Typical implementations use CNNs as feature extractors that project PXRD patterns to latent representations, followed by one or more multilayer perceptron (MLP) heads that map features to classification or regression outputs. These extractors commonly employ multiscale architectures with pooling or strided convolutions between blocks to increase receptive field size. However, Salgado \textit{et al.} \cite{Salgado_npj_2023} found that removing pooling layers while maintaining strided convolutions improved performance on some datasets, attributed to reduced information compression, though at the cost of decreased receptive fields and increased parameters in downstream heads.
A key consideration for CNNs in PXRD analysis is shift-equivariance: spatial shifts in input produce corresponding shifts in output. Combined with pooling operations \cite{Lin_arxiv_2013}, this can lead to shift-invariance—desirable for image classification but problematic for PXRD, where shifted patterns in 2$\theta$ space represent different structures. This issue is mitigated when MLP prediction heads follow CNN extractors, as MLPs are inherently shift-sensitive. Some works \cite{Vecsei_prb_2019, Liang_prm_2020, Lee_ais_2022} have explored pure MLP architectures to exploit this sensitivity, though results are mixed: Vecsei \textit{et al.} \cite{Vecsei_prb_2019} found MLPs performed better on experimental data but worse on synthetic data compared to CNNs. In practice, MLPs face computational challenges due to large parameter counts and expensive matrix operations.
Transformers, highly successful in natural language processing \cite{Vaswani_arxiv_2017} and computer vision \cite{Dosovitskiy_arxiv_2020}, have seen limited application in PXRD. Lee \textit{et al.} \cite{Lee_ais_2022} found transformer-based models underperformed CNNs even with PXRD-specific pretraining, attributing this to transformers' data hunger: without CNN-like inductive biases, they require substantially more training samples. However, recent work on DiffractGPT \cite{Choudhary_jpcl_2025} demonstrates that large pretrained language models (\textit{e.g.}, Mistral 7B \cite{Jiang_arxiv_2023}) can be effectively fine-tuned for PXRD-based lattice and atomic position prediction, suggesting a promising direction for leveraging models pretrained on extensive language data.

\subsection*{Training data sources and scale}
ICSD \cite{zagorac_appliedcryst_2019} and Materials Project \cite{jain_aplmat_2013} are the primary crystallographic databases used for training. After curation to remove problematic structures, both databases provide large-scale training sets and have been widely adopted \cite{Park_iucrj_2017, Vecsei_prb_2019, Zaloga_mtc_2020, Liang_prm_2020, Chitturi_jac_2021, Lee_ais_2022, Lee_ais_2023, Salgado_npj_2023, Riesel_jacs_2024}. While ICSD and Materials Project provide simulated data, the RRUFF database \cite{lafuente_hmc_2015} offers predominantly experimental PXRD patterns, serving as a benchmark for model performance under realistic conditions \cite{Vecsei_prb_2019, Lee_ais_2023, Salgado_npj_2023, Riesel_jacs_2024}. The largest reported training set contains 263,000 structures from ICSD and Materials Project \cite{Lee_ais_2022}. To model experimental imperfections such as lattice strain and peak broadening, multiple augmented patterns are often generated per structure with varied perturbations. For example, Lee \textit{et al.} \cite{Lee_ais_2023} generated 20 augmentations per structure, yielding 3.7 million training samples.

\subsection*{Prediction targets and model scope}
Most works focus on crystal symmetry classification, predicting crystal systems and space groups. Some extend to continuous quantities including lattice parameters \cite{Liang_prm_2020, Chitturi_jac_2021, Gomez_jpca_2023, Riesel_jacs_2024, Choudhary_jpcl_2025} and material properties such as band gap and formation energy \cite{Lee_ais_2022}. However, many approaches require separate specialized models for different tasks or structure types. For instance, Chitturi \textit{et al.} \cite{Chitturi_jac_2021} trained seven separate models -- one per crystal system -- for lattice parameter prediction, while Lee \textit{et al.} \cite{Lee_ais_2022} used distinct models for classification and property regression. While dedicated models simplify training, unified models that predict multiple quantities simultaneously streamline inference workflows.

\clearpage
\newgeometry{bottom=1.5cm}

\begin{sidewaystable}
\begin{threeparttable}[h]
    \footnotesize
    \centering
    \begin{tabular}{C{2cm}C{2cm}C{2cm}C{2cm}C{2cm}C{2cm}C{2cm}C{2cm}C{3cm}C{4cm}}
        \Xhline{2\arrayrulewidth}
        \textbf{Reference} & \textbf{Architecture} & \textbf{Training database} & \textbf{Test database} & \textbf{Number of training inorganic structures} & \textbf{Number of training organic structures} & \textbf{Total number of training structures} & \textbf{Total number of training diffraction patterns} & \textbf{Predicted quantities} & \textbf{Remarks} \\
        \Xhline{2\arrayrulewidth}
        Park \etal{} \cite{Park_iucrj_2017} & 1D CNN & ICSD & ICSD & 120,000 & 0 & 120,000 & 120,000 & CS, SG, EG & \\
        \hline
        Vecsei \etal{} \cite{Vecsei_prb_2019} & 1D CNN, MLP & ICSD & ICSD, RRUFF & 114,404 & 0 & 114,404 & 114,404 & CS, SG & \\
        \hline
        Zaloga \etal{} \cite{Zaloga_mtc_2020} & 1D CNN & ICSD & ICSD & 153,603 & 0 & 153,603 & \revisednew{153,603} & CS, SG\revisednew{, EG} & \\
        \hline
        Liang \etal{} \cite{Liang_prm_2020} & MLP & ICSD & ICSD & 110,813 & 0 & \revisednew{110,813} & \revisednew{N/A} & BL, SG, LP & \revisednew{Used chemical composition (Magpie descriptors), not PXRD patterns, as input.} After predicting BL, trained separate models to predict SG and LP for each BL. \\
        \hline
        Chitturi \etal{} \cite{Chitturi_jac_2021} & 1D CNN & ICSD, CSD & ICSD, CSD & $\sim$137,000 & $\sim$825,000 & 960,000 & 960,000 & LP & Trained a separate model for each CS. Inorganic/organic numbers are calculated from Fig.~1 of cited paper. \\
        \hline
        Lee \etal{} \cite{Lee_ais_2022} & 1D CNN, transformer (structures), MLP (properties)  & ICSD, MP & ICSD, MP & 263,000 & 0 & 263,000 & 263,000 & CS, \revisednew{S}G, \revisednew{E}G, $E_g$, $E_f$, $E_h$ & \\
        \hline
        Lee \etal{} \cite{Lee_ais_2023} & 1D CNN & ICSD & RRUFF & 187,131 & 0 & 187,131 & 3,742,620 & CS, SG, EG & Listed numbers are based on 20 texture-involving perturbations generated for each structure. A separate dataset of 10 texture-free patterns per structure was also generated. \\
        \hline
        Corriero \etal{} \cite{Corriero_jac_2023} & 1D CNN & POW\_COD & Private database & 21,783 & 261,223 & 283,006 & 283,006 & CS & Listed numbers are total data size; train/test ratio unavailable. \\
        \hline
        G\'{o}mez-Peralta \etal{} \cite{Gomez_jpca_2023} & 1D CNN & COD & COD & 0 & 83,000 & 83,000 & 332,000 & LP & \\
        \hline
        \revisedfin{G\'{o}mez-Peralta \etal{} \cite{gomez2025insights}} & \revisedfin{1D CNN (pretraining), MLP (transfer learning)} & \revisedfin{COD} & \revisedfin{COD} & \revisedfin{20,908} & \revisedfin{205,331} & \revisedfin{226,239} & \revisedfin{904,956} & \revisedfin{LP} & \revisedfin{Frozen CNN pretrained on four crystallite-size profiles per organic structure; downstream MLPs trained on four profiles per inorganic structure.} \\
        \hline
        Salgado \etal{} \cite{Salgado_npj_2023} & 1D CNN & ICSD & RRUFF & 171,006 & 0 & 171,006 & 1,200,000 & CS, SG & \\
        \hline
        Riesel \etal{} \cite{Riesel_jacs_2024} & 1D CNN & MP-20 & MP-20, AMCSD, RRUFF & 36,185 & 0 & 36,185 & 144,740 & CP, LP, NA & Train\revisednew{ed} on MP-20 subset of MP dataset containing only structures with 1-20 atoms in the primitive unit cell. \\
        \hline
        Choudhary \cite{Choudhary_jpcl_2025} & Mistral 7B & JARVIS-DFT & JARVIS-DFT & 72,990 & 0 & 72,990 & 72,990 & LP, AP \\
        \hline
        \textbf{Ours} & \textbf{1D ConvNeXt} & \textbf{ICSD, MP} & \textbf{RRUFF} & \textbf{312,267} & \textbf{$\sim$0$^{\dagger}$} & \textbf{312,267} & \textbf{31,226,700} & \textbf{CS, SG, LP} & \\
        \Xhline{2\arrayrulewidth}
    \end{tabular}
    \caption{\textbf{Survey of deep learning methods for PXRD analysis.} A survey on the data sources, sizes of training data, and predicted quantities for models that predict crystal structures and material properties from PXRD patterns. We only include works where the models were trained on a comprehensive collection of structures rather than ones only trained for certain materials systems. ``Number of structures'' refers to the distinct structures in the training data, while ``number of diffraction patterns'' means the actual number of samples in the dataset that include all augmented variants. Since multiple augmented variants can be generated from each structure, numbers in this column are greater than or equal to the numbers of structures. Abbreviations and symbols: CS -- crystal system, SG -- space group, EG -- extinction group, LP -- lattice parameters, BL -- Bravais lattice, AP -- atomic positions, $E_g$ -- bandgap, $E_f$ -- formation energy, $E_h$ -- energy above the convex hull, CP -- chemical composition, NA -- number of atoms. \\
    \footnotesize{
    \revisednew{$^{\dagger}$The ICSD and Materials Project databases are predominantly inorganic and do not provide explicit organic/inorganic/hybrid labels. A rough composition-based heuristic identified only a small fraction of potentially organic or hybrid structures (1.29\% for ICSD and 0.99\% for Materials Project); these estimates are not validated classifications.}
    }}
\end{threeparttable}
\end{sidewaystable}

\loadgeometry{arxivmargins}
\clearpage

\begin{table}[h]
    \centering
    \begin{adjustbox}{max width=\textwidth}
    \begin{tabular}{cccccc}
        \hline
        \revisednew{\textbf{Reference}} &
        \revisednew{\textbf{Reported $2\theta$ range ($^\circ$)}} &
        \revisednew{\textbf{Input points}} &
        \revisednew{\textbf{Sampling ($^\circ$)}} &
        \revisednew{\textbf{Wavelength (\AA)}} &
        \revisednew{\textbf{Equivalent $2\theta$ range at 20 keV ($^\circ$)}} \\
        \hline

        \revisednew{Park \etal{} \cite{Park_iucrj_2017}} &
        \revisednew{10--110} &
        \revisednew{10,001} &
        \revisednew{0.01$^*$} &
        \revisednew{Not specified$^{**}$} &
        \revisednew{4.02--38.49} \\

        \revisednew{Vecsei \etal{} \cite{Vecsei_prb_2019}} &
        \revisednew{5--90} &
        \revisednew{8,501$^{\dagger}$} &
        \revisednew{0.01} &
        \revisednew{1.54} &
        \revisednew{2.01--33.08} \\

        \revisednew{Chitturi \etal{} \cite{Chitturi_jac_2021}} &
        \revisednew{0--90} &
        \revisednew{9,000} &
        \revisednew{$\sim$0.01$^*$} &
        \revisednew{1.54} &
        \revisednew{0--33.06} \\

        \revisednew{Lee \etal{} \cite{Lee_ais_2022}} &
        \revisednew{5--86.91} &
        \revisednew{8,192} &
        \revisednew{0.01$^*$} &
        \revisednew{Not specified$^{**}$} &
        \revisednew{2.01--32.13} \\

        \revisednew{Lee \etal{} \cite{Lee_ais_2023}} &
        \revisednew{5--86.91} &
        \revisednew{8,192} &
        \revisednew{0.01} &
        \revisednew{Not specified$^{**}$} &
        \revisednew{2.01--32.13} \\

        \revisedfin{G\'{o}mez-Peralta \etal{} \cite{gomez2025insights}} &
        \revisedfin{3--60} &
        \revisedfin{2,780} &
        \revisedfin{0.02$^*$} &
        \revisedfin{1.5406} &
        \revisedfin{1.21--23.21} \\
        
        \revisednew{Salgado \etal{} \cite{Salgado_npj_2023}} &
        \revisednew{5--90} &
        \revisednew{8,501$^{\dagger}$} &
        \revisednew{0.01} &
        \revisednew{1.54} &
        \revisednew{2.01--33.08} \\

        \revisednew{Choudhary \cite{Choudhary_jpcl_2025}} &
        \revisednew{Not specified$^{\ddagger}$} &
        \revisednew{180} &
        \revisednew{0.5} &
        \revisednew{1.54} &
        \revisednew{--} \\

        \revisednew{\textbf{Ours}} &
        \revisednew{\textbf{5--20}} &
        \revisednew{\textbf{8,192}} &
        \revisednew{\textbf{0.00183}} &
        \revisednew{\textbf{0.6199}} &
        \revisednew{\textbf{5--20}} \\
        \hline
    \end{tabular}
    \end{adjustbox}

    \caption{
    \revisednew{\textbf{PXRD input ranges and sampling used in AlphaDiffract and reference studies.}
    Reported angular ranges, input dimensionalities, sampling intervals, and X-ray wavelengths are listed as specified in the original publications. The equivalent $2\theta$ ranges at 20 keV were calculated using Bragg's law to preserve the corresponding $d$-spacing range. For studies that did not report the X-ray wavelength, Cu K$\alpha$ radiation was assumed only for calculating the equivalent 20 keV angular range.} \\
    \footnotesize{
    \revisednew{$^*$Sampling inferred from the reported angular range and number of input points rather than explicitly stated by the authors.} \\
    \revisednew{$^{**}$The original publication does not specify the X-ray wavelength; Cu K$\alpha$ radiation ($\lambda = 1.5406$~\AA) was assumed solely for calculating the equivalent 20 keV angular range.} \\
    \revisednew{$^{\dagger}$Number of input points inferred from the reported angular range and sampling assuming both endpoints are included.} \\
    \revisednew{$^{\ddagger}$Choudhary reports a 180-point representation sampled at 0.5$^\circ$ $2\theta$, but the angular endpoints of the input grid are not explicitly stated; an equivalent 20 keV range is therefore not reported.}
    }}
    \label{tab:pxrd_input_comparison}
\end{table}

\begin{table}[h]
    \centering
    \begin{adjustbox}{max width=\textwidth}
    \begin{tabular}{clcccccc}
        \hline
        \multirow{2}{*}{\revisednew{\textbf{Dataset}}} &
        \multirow{2}{*}{\shortstack[l]{\revisednew{\textbf{Evaluation Set}} \\
                                       \revisednew{\textbf{(Number of Examples)}}}} &
        \multicolumn{3}{c}{\revisednew{\textbf{Crystal System}}} &
        \multicolumn{3}{c}{\revisednew{\textbf{Space Group}}} \\
        \cmidrule(lr){3-5} \cmidrule(lr){6-8}
        & &
        \revisednew{Acc. (\%)} &
        \revisednew{95\% CI} &
        \revisednew{Ens. Var. (\%)} &
        \revisednew{Acc. (\%)} &
        \revisednew{95\% CI} &
        \revisednew{Ens. Var. (\%)} \\
        \hline

        \multirow{2}{*}{\revisednew{ICSD$^\dagger$}} &
        \revisednew{Full (16,563)} &
        \revisednew{92.22} &
        \revisednew{[91.83, 92.60]} &
        \revisednew{0.02} &
        \revisednew{81.75} &
        \revisednew{[81.19, 82.33]} &
        \revisednew{0.02} \\

        &
        \revisednew{Unique (5,979)} &
        \revisednew{86.87} &
        \revisednew{[86.07, 87.68]} &
        \revisednew{0.03} &
        \revisednew{74.59} &
        \revisednew{[73.53, 75.66]} &
        \revisednew{0.03} \\
        \hline

        \multirow{2}{*}{\revisednew{Materials Project$^\dagger$}} &
        \revisednew{Full (14,663)} &
        \revisednew{79.00} &
        \revisednew{[78.39, 79.63]} &
        \revisednew{0.02} &
        \revisednew{59.54} &
        \revisednew{[58.77, 60.29]} &
        \revisednew{0.02} \\

        &
        \revisednew{Unique (9,161)} &
        \revisednew{74.34} &
        \revisednew{[73.50, 75.23]} &
        \revisednew{0.03} &
        \revisednew{52.44} &
        \revisednew{[51.46, 53.42]} &
        \revisednew{0.03} \\
        \hline

        \multirow{2}{*}{\revisednew{RRUFF$^\ddagger$}} &
        \revisednew{Full (734)} &
        \revisednew{81.74} &
        \revisednew{[78.88, 84.60]} &
        \revisednew{0.78} &
        \revisednew{66.21} &
        \revisednew{[62.81, 69.62]} &
        \revisednew{0.71} \\

        &
        \revisednew{Unique (285)} &
        \revisednew{76.14} &
        \revisednew{[71.23, 81.05]} &
        \revisednew{1.41} &
        \revisednew{49.82} &
        \revisednew{[44.21, 55.44]} &
        \revisednew{1.18} \\
        \hline
    \end{tabular}
    \end{adjustbox}

    \caption{\revisednew{\textbf{Classification performance and separate uncertainty estimates.}
    Crystal system and space group classification accuracies of the ensemble model are reported for the complete evaluation sets (``Full'') and for subsets excluding structures with potential equivalents in the training data (``Unique'').
    The reported 95\% confidence intervals were estimated by nonparametric bootstrap resampling at the level of evaluation structures. For ICSD and Materials Project, classification performance for each structure was averaged over its 100 augmented diffraction patterns before bootstrap resampling; each RRUFF structure is represented by a single diffraction profile.
    Ensemble variability (Ens. Var.) denotes the propagated $1.96\sigma_{\mathrm{model}}$ margin derived from variation in predicted class probabilities across independently trained ensemble members and characterizes model-instance variability; it should not be interpreted as a finite evaluation set confidence interval.} \\
    \footnotesize{
    \revisednew{
    $^\dagger$Evaluated on data from our validation set. \\
    $^\ddagger$Evaluated on data from our RRUFF benchmark dataset.}}}
    \label{tab:classification_uncertainty}
\end{table}

\begin{table}[h]
    \centering

    \begin{adjustbox}{max width=\textwidth}
    \begin{tabular}{clcccccc}
        \hline
        \multicolumn{8}{c}{\revisednew{\textbf{Lattice Length}}} \\
        \hline
        \multirow{2}{*}{\revisednew{\textbf{Dataset}}} &
        \multirow{2}{*}{\shortstack[l]{\revisednew{\textbf{Evaluation Set}} \\
                                       \revisednew{\textbf{(Number of Examples)}}}} &
        \multicolumn{2}{c}{\revisednew{MAE (\AA)}} &
        \multicolumn{2}{c}{\revisednew{MAPE (\%)}} &
        \multicolumn{2}{c}{\revisednew{$R^2$}} \\
        \cmidrule(lr){3-4} \cmidrule(lr){5-6} \cmidrule(lr){7-8}
        & &
        \revisednew{Value (95\% CI)} &
        \revisednew{Pred. Var.} &
        \revisednew{Value (95\% CI)} &
        \revisednew{Pred. Var.} &
        \revisednew{Value (95\% CI)} &
        \revisednew{Pred. Var.} \\
        \hline

        \multirow{2}{*}{\revisednew{ICSD$^{\dagger}$}}
        & \shortstack[l]{\revisednew{Full (16,563)} \\ \strut}
        & \shortstack[c]{\revisednew{1.59} \\ \revisednew{[1.56, 1.61]}}
        & \shortstack[c]{\revisednew{0.00} \\ \strut}
        & \shortstack[c]{\revisednew{19.77} \\ \revisednew{[19.55, 19.99]}}
        & \shortstack[c]{\revisednew{0.04} \\ \strut}
        & \shortstack[c]{\revisednew{0.64} \\ \revisednew{[0.63, 0.65]}}
        & \shortstack[c]{\revisednew{0.00} \\ \strut} \\

        & \shortstack[l]{\revisednew{Unique (5,979)} \\ \strut}
        & \shortstack[c]{\revisednew{1.81} \\ \revisednew{[1.77, 1.86]}}
        & \shortstack[c]{\revisednew{0.01} \\ \strut}
        & \shortstack[c]{\revisednew{20.34} \\ \revisednew{[19.98, 20.70]}}
        & \shortstack[c]{\revisednew{0.06} \\ \strut}
        & \shortstack[c]{\revisednew{0.58} \\ \revisednew{[0.56, 0.60]}}
        & \shortstack[c]{\revisednew{0.00} \\ \strut} \\
        \hline

        \multirow{2}{*}{\revisednew{Materials Project$^{\dagger}$}}
        & \shortstack[l]{\revisednew{Full (14,663)} \\ \strut}
        & \shortstack[c]{\revisednew{1.92} \\ \revisednew{[1.89, 1.95]}}
        & \shortstack[c]{\revisednew{0.00} \\ \strut}
        & \shortstack[c]{\revisednew{22.59} \\ \revisednew{[22.35, 22.83]}}
        & \shortstack[c]{\revisednew{0.04} \\ \strut}
        & \shortstack[c]{\revisednew{0.48} \\ \revisednew{[0.47, 0.50]}}
        & \shortstack[c]{\revisednew{0.00} \\ \strut} \\

        & \shortstack[l]{\revisednew{Unique (9,161)} \\ \strut}
        & \shortstack[c]{\revisednew{1.93} \\ \revisednew{[1.90, 1.97]}}
        & \shortstack[c]{\revisednew{0.00} \\ \strut}
        & \shortstack[c]{\revisednew{22.72} \\ \revisednew{[22.40, 23.05]}}
        & \shortstack[c]{\revisednew{0.05} \\ \strut}
        & \shortstack[c]{\revisednew{0.43} \\ \revisednew{[0.41, 0.45]}}
        & \shortstack[c]{\revisednew{0.00} \\ \strut} \\
        \hline

        \multirow{2}{*}{\revisedfin{COD\textsuperscript{\S}}}
        & \shortstack[l]{\revisedfin{Full (27,041)} \\ \strut}
        & \shortstack[c]{\revisedfin{1.72} \\ \revisedfin{[1.70, 1.73]}}
        & \shortstack[c]{\revisedfin{0.00} \\ \strut}
        & \shortstack[c]{\revisedfin{21.58} \\ \revisedfin{[21.41, 21.75]}}
        & \shortstack[c]{\revisedfin{0.03} \\ \strut}
        & \shortstack[c]{\revisedfin{0.51} \\ \revisedfin{[0.50, 0.52]}}
        & \shortstack[c]{\revisedfin{0.00} \\ \strut} \\

        & \shortstack[l]{\revisedfin{Unique (6,813)} \\ \strut}
        & \shortstack[c]{\revisedfin{2.12} \\ \revisedfin{[2.09, 2.16]}}
        & \shortstack[c]{\revisedfin{0.00} \\ \strut}
        & \shortstack[c]{\revisedfin{23.27} \\ \revisedfin{[22.94, 23.60]}}
        & \shortstack[c]{\revisedfin{0.05} \\ \strut}
        & \shortstack[c]{\revisedfin{0.31} \\ \revisedfin{[0.30, 0.32]}}
        & \shortstack[c]{\revisedfin{0.00} \\ \strut} \\
        \hline
        
        \multirow{2}{*}{\revisednew{RRUFF$^{\ddagger}$}}
        & \shortstack[l]{\revisednew{Full (240)} \\ \strut}
        & \shortstack[c]{\revisednew{2.11} \\ \revisednew{[1.87, 2.39]}}
        & \shortstack[c]{\revisednew{0.02} \\ \strut}
        & \shortstack[c]{\revisednew{23.50} \\ \revisednew{[21.58, 25.53]}}
        & \shortstack[c]{\revisednew{0.25} \\ \strut}
        & \shortstack[c]{\revisednew{0.39} \\ \revisednew{[0.32, 0.46]}}
        & \shortstack[c]{\revisednew{0.01} \\ \strut} \\

        & \shortstack[l]{\revisednew{Unique (91)} \\ \strut}
        & \shortstack[c]{\revisednew{2.68} \\ \revisednew{[2.16, 3.31]}}
        & \shortstack[c]{\revisednew{0.03} \\ \strut}
        & \shortstack[c]{\revisednew{25.66} \\ \revisednew{[22.78, 28.79]}}
        & \shortstack[c]{\revisednew{0.30} \\ \strut}
        & \shortstack[c]{\revisednew{0.31} \\ \revisednew{[0.22, 0.38]}}
        & \shortstack[c]{\revisednew{0.01} \\ \strut} \\
        \hline
    \end{tabular}
    \end{adjustbox}

    \vspace{0.8em}

    \begin{adjustbox}{max width=\textwidth}
    \begin{tabular}{clcccccc}
        \hline
        \multicolumn{8}{c}{\revisednew{\textbf{Lattice Angle}}} \\
        \hline
        \multirow{2}{*}{\revisednew{\textbf{Dataset}}} &
        \multirow{2}{*}{\shortstack[l]{\revisednew{\textbf{Evaluation Set}} \\
                                       \revisednew{\textbf{(Number of Examples)}}}} &
        \multicolumn{2}{c}{\revisednew{MAE ($^\circ$)}} &
        \multicolumn{2}{c}{\revisednew{MAPE (\%)}} &
        \multicolumn{2}{c}{\revisednew{$R^2$}} \\
        \cmidrule(lr){3-4} \cmidrule(lr){5-6} \cmidrule(lr){7-8}
        & &
        \revisednew{Value (95\% CI)} &
        \revisednew{Pred. Var.} &
        \revisednew{Value (95\% CI)} &
        \revisednew{Pred. Var.} &
        \revisednew{Value (95\% CI)} &
        \revisednew{Pred. Var.} \\
        \hline

        \multirow{2}{*}{\revisednew{ICSD$^{\dagger}$}}
        & \shortstack[l]{\revisednew{Full (16,563)} \\ \strut}
        & \shortstack[c]{\revisednew{1.68} \\ \revisednew{[1.64, 1.72]}}
        & \shortstack[c]{\revisednew{0.01} \\ \strut}
        & \shortstack[c]{\revisednew{1.69} \\ \revisednew{[1.65, 1.73]}}
        & \shortstack[c]{\revisednew{0.01} \\ \strut}
        & \shortstack[c]{\revisednew{0.58} \\ \revisednew{[0.57, 0.60]}}
        & \shortstack[c]{\revisednew{0.00} \\ \strut} \\

        & \shortstack[l]{\revisednew{Unique (5,979)} \\ \strut}
        & \shortstack[c]{\revisednew{2.00} \\ \revisednew{[1.92, 2.07]}}
        & \shortstack[c]{\revisednew{0.01} \\ \strut}
        & \shortstack[c]{\revisednew{2.01} \\ \revisednew{[1.93, 2.09]}}
        & \shortstack[c]{\revisednew{0.01} \\ \strut}
        & \shortstack[c]{\revisednew{0.54} \\ \revisednew{[0.52, 0.56]}}
        & \shortstack[c]{\revisednew{0.00} \\ \strut} \\
        \hline

        \multirow{2}{*}{\revisednew{Materials Project$^{\dagger}$}}
        & \shortstack[l]{\revisednew{Full (14,663)} \\ \strut}
        & \shortstack[c]{\revisednew{3.34} \\ \revisednew{[3.27, 3.41]}}
        & \shortstack[c]{\revisednew{0.01} \\ \strut}
        & \shortstack[c]{\revisednew{3.61} \\ \revisednew{[3.53, 3.69]}}
        & \shortstack[c]{\revisednew{0.01} \\ \strut}
        & \shortstack[c]{\revisednew{0.32} \\ \revisednew{[0.30, 0.33]}}
        & \shortstack[c]{\revisednew{0.00} \\ \strut} \\

        & \shortstack[l]{\revisednew{Unique (9,161)} \\ \strut}
        & \shortstack[c]{\revisednew{3.36} \\ \revisednew{[3.27, 3.44]}}
        & \shortstack[c]{\revisednew{0.01} \\ \strut}
        & \shortstack[c]{\revisednew{3.63} \\ \revisednew{[3.52, 3.73]}}
        & \shortstack[c]{\revisednew{0.01} \\ \strut}
        & \shortstack[c]{\revisednew{0.29} \\ \revisednew{[0.27, 0.31]}}
        & \shortstack[c]{\revisednew{0.00} \\ \strut} \\
        \hline

        \multirow{2}{*}{\revisedfin{COD\textsuperscript{\S}}}
        & \shortstack[l]{\revisedfin{Full (27,041)} \\ \strut}
        & \shortstack[c]{\revisedfin{2.17} \\ \revisedfin{[2.14, 2.20]}}
        & \shortstack[c]{\revisedfin{0.00} \\ \strut}
        & \shortstack[c]{\revisedfin{2.19} \\ \revisedfin{[2.15, 2.23]}}
        & \shortstack[c]{\revisedfin{0.00} \\ \strut}
        & \shortstack[c]{\revisedfin{0.49} \\ \revisedfin{[0.48, 0.49]}}
        & \shortstack[c]{\revisedfin{0.00} \\ \strut} \\

        & \shortstack[l]{\revisedfin{Unique (6,813)} \\ \strut}
        & \shortstack[c]{\revisedfin{3.15} \\ \revisedfin{[3.07, 3.23]}}
        & \shortstack[c]{\revisedfin{0.01} \\ \strut}
        & \shortstack[c]{\revisedfin{3.19} \\ \revisedfin{[3.10, 3.27]}}
        & \shortstack[c]{\revisedfin{0.01} \\ \strut}
        & \shortstack[c]{\revisedfin{0.27} \\ \revisedfin{[0.25, 0.28]}}
        & \shortstack[c]{\revisedfin{0.00} \\ \strut} \\
        \hline
        
        \multirow{2}{*}{\revisednew{RRUFF$^{\ddagger}$}}
        & \shortstack[l]{\revisednew{Full (240)} \\ \strut}
        & \shortstack[c]{\revisednew{2.72} \\ \revisednew{[2.27, 3.22]}}
        & \shortstack[c]{\revisednew{0.03} \\ \strut}
        & \shortstack[c]{\revisednew{2.91} \\ \revisednew{[2.38, 3.52]}}
        & \shortstack[c]{\revisednew{0.03} \\ \strut}
        & \shortstack[c]{\revisednew{0.25} \\ \revisednew{[0.16, 0.34]}}
        & \shortstack[c]{\revisednew{0.01} \\ \strut} \\

        & \shortstack[l]{\revisednew{Unique (91)} \\ \strut}
        & \shortstack[c]{\revisednew{3.58} \\ \revisednew{[2.77, 4.50]}}
        & \shortstack[c]{\revisednew{0.05} \\ \strut}
        & \shortstack[c]{\revisednew{3.97} \\ \revisednew{[2.93, 5.23]}}
        & \shortstack[c]{\revisednew{0.05} \\ \strut}
        & \shortstack[c]{\revisednew{0.19} \\ \revisednew{[0.05, 0.30]}}
        & \shortstack[c]{\revisednew{0.02} \\ \strut} \\
        \hline
    \end{tabular}
    \end{adjustbox}

    \caption{
        \revisednew{\textbf{Regression performance and separate uncertainty estimates.}
        Lattice length and angle regression metrics are reported for the complete evaluation sets (``Full'') and for subsets excluding structures with potential equivalents in the training data (``Unique''). Values in brackets denote 95\% confidence intervals estimated by nonparametric bootstrap resampling at the level of evaluation structures. Propagated prediction variability (Pred. Var.) denotes the variability in each regression metric arising from ensemble-member predictions and, where applicable, augmented input patterns. These quantities characterize prediction variability conditional on the evaluated structures and should not be interpreted as finite evaluation set confidence intervals.} \\
        \footnotesize{
        \revisednew{$^{\dagger}$Evaluated on data from our validation set.} \\
        \revisednew{$^{\ddagger}$Evaluated on data from our RRUFF benchmark dataset.} \\
        \revisedfin{\textsuperscript{\S}Evaluated on the inorganic COD structures retrieved using the published IDs \cite{gomez2025insights} and processed using our curation pipeline.}
        }
    }
    \label{tab:regression_uncertainty}
\end{table}

\begin{table}[h]
    \centering
    \begin{tabular}{lcc}
        \hline
        \revisednew{\textbf{Metric}} &
        \revisednew{\textbf{Crystal System}} &
        \revisednew{\textbf{Space Group}} \\
        \hline
        \revisednew{Classes represented in RRUFF benchmark} &
        \revisednew{6} &
        \revisednew{72} \\
        \revisednew{Macro-averaged recall (\%)} &
        \revisednew{75.75} &
        \revisednew{60.12} \\
        \revisednew{Spearman $r_s$} &
        \revisednew{0.77} &
        \revisednew{$-0.11$} \\
        \hline
    \end{tabular}
    \caption{\revisednew{\textbf{Class-balanced classification performance and association with training-set class frequency.}
    Macro-averaged recall is reported for crystal system (CS) and space group (SG) classification on the RRUFF benchmark and is defined as the unweighted mean of per-class recall across classes represented in the evaluation set. Spearman's rank correlation coefficient ($r_s$) quantifies the association between training-set class frequency and per-class recall. Six crystal systems and 72 space groups are represented in the RRUFF benchmark.}}
    \label{tab:class_imbalance}
\end{table}

\begin{table}[h]
    \centering
    \begin{tabular}{llccc}
        \hline
        \revisednew{\textbf{Evaluation Set}} &
        \revisednew{\textbf{Profile Source}} &
        \revisednew{\textbf{Number of Examples}} &
        \revisednew{\textbf{CS Accuracy (\%)}} &
        \revisednew{\textbf{SG Accuracy (\%)}} \\
        \hline
        
        \revisednew{Full}
        & \revisednew{All}
        & \revisednew{734}
        & \revisednew{81.74}
        & \revisednew{66.21} \\
        
        & \revisednew{Measured}
        & \revisednew{675}
        & \revisednew{81.19}
        & \revisednew{65.93} \\
        
        & \revisednew{Curator-calculated}
        & \revisednew{59}
        & \revisednew{88.14}
        & \revisednew{69.49} \\
        
        \hline
        
        \revisednew{Equivalent}
        & \revisednew{All}
        & \revisednew{449}
        & \revisednew{85.30}
        & \revisednew{76.61} \\
        
        & \revisednew{Measured}
        & \revisednew{412}
        & \revisednew{84.71}
        & \revisednew{76.70} \\
        
        & \revisednew{Curator-calculated}
        & \revisednew{37}
        & \revisednew{91.89}
        & \revisednew{75.68} \\
        
        \hline
        
        \revisednew{Unique}
        & \revisednew{All}
        & \revisednew{285}
        & \revisednew{76.14}
        & \revisednew{49.82} \\
        
        & \revisednew{Measured}
        & \revisednew{263}
        & \revisednew{75.67}
        & \revisednew{49.05} \\
        
        & \revisednew{Curator-calculated}
        & \revisednew{22}
        & \revisednew{81.82}
        & \revisednew{59.09} \\
        
        \hline
    \end{tabular}
    \caption{\revisednew{\textbf{RRUFF classification performance stratified by profile source.}
    Crystal system (CS) and space group (SG) classification accuracies of the ensemble model are reported for the complete RRUFF classification benchmark (``Full'') and for subsets with (``Equivalent'') or without (``Unique'') an identified structural equivalent in the training set. Within each evaluation set, performance is further stratified by directly measured PXRD profiles and powder profiles calculated by the RRUFF curators from single-crystal refinement data.}}
    \label{tab:rruff_profile_source}
\end{table}

\begin{table}[h]
    \centering
    \begin{adjustbox}{max width=\textwidth}
    \begin{tabular}{cccccccccc}
        \hline
        \multirow{2}{*}{} & \multirow{2}{*}{\textbf{Model variant}} & \multirow{2}{*}{\textbf{Test data}} & \multirow{2}{*}{\textbf{Metric}} & \multicolumn{6}{c}{\textbf{Metric value}}  \\
        \cline{5-10}
         & & & & a & b & c & $\alpha$ & $\beta$ & $\gamma$ \\
        \hline
        Choudhary \cite{Choudhary_jpcl_2025} & DGPT-formula & JARVIS-DFT & MAE & \textbf{0.17} & \textbf{0.18} & \textbf{0.27} & -- & -- & -- \\
        Choudhary \cite{Choudhary_jpcl_2025} & DGPT-formula & ICSD$^{\dagger}$ & MAE & 1.72 (1.66) & 2.40 (2.28) & 4.44 (4.44) & 5.00 (5.16) & 2.76 (3.34) & 9.51 (10.20) \\
        Choudhary \cite{Choudhary_jpcl_2025} & DGPT-formula & Materials Project$^{\dagger}$ & MAE & 1.55 (1.48) & 2.17 (2.13) & 4.36 (4.47) & 4.96 (5.45) & 2.85 (3.53) & 10.11 (9.89) \\
        Choudhary \cite{Choudhary_jpcl_2025} & DGPT-formula & RRUFF$^{\ddagger}$ & MAE & 1.72 (1.66) & 2.40 (2.28) & 4.44 (4.44) & 5.00 (5.16) & 2.76 (3.34) & 9.51 (10.20) \\
        \textbf{Ours} & Cls. + Regr. Ensemble & ICSD$^{\dagger}$ & MAE & 1.14 & 1.33 & 2.29 & \textbf{2.02} & \textbf{0.61} & \textbf{2.41} \\
        \textbf{Ours} & Cls. + Regr. Ensemble & Materials Project$^{\dagger}$ & MAE & 1.22 & 1.60 & 2.93 & 3.66 & 1.63 & 4.72 \\
        \textbf{Ours} & Cls. + Regr. Ensemble & RRUFF$^{\ddagger}$ & MAE & 1.37 & 1.76 & 3.22 & 2.95 & 2.31 & 2.90 \\
        \hline
        Chitturi \etal{} \cite{Chitturi_jac_2021} & Full range & ICSD, CSD & MAPE & \multicolumn{3}{c}{\textbf{9.20}} & -- & -- & --  \\
        Choudhary \cite{Choudhary_jpcl_2025} & DGPT-formula & ICSD$^{\dagger}$ & MAPE & 27.11 (26.97) & 34.63 (33.32) & 41.23 (40.69) & 4.88 (5.06) & 3.05 (3.66) & 9.54 (10.27) \\
        Choudhary \cite{Choudhary_jpcl_2025} & DGPT-formula & Materials Project$^{\dagger}$ & MAPE & 23.17 (22.88) & 29.59 (28.17) & 31.67 (30.06) & 5.10 (5.62) & 3.15 (3.91) & 10.51 (10.31) \\
        Choudhary \cite{Choudhary_jpcl_2025} & DGPT-formula & RRUFF$^{\ddagger}$ & MAPE & 22.72 (23.18) & 27.92 (25.40) & 27.20 (27.38) & 4.38 (4.45) & 3.81 (4.04) & 9.00 (8.68) \\
        \textbf{Ours} & Cls. + Regr. Ensemble & ICSD$^{\dagger}$ & MAPE & 19.52 & 18.54 & 21.23 & \textbf{2.03} & \textbf{0.66} & \textbf{2.39} \\
        \textbf{Ours} & Cls. + Regr. Ensemble & Materials Project$^{\dagger}$ & MAPE & 21.58 & 21.92 & 24.27 & 3.88 & 1.83 & 5.12 \\
        \textbf{Ours} & Cls. + Regr. Ensemble & RRUFF$^{\ddagger}$ & MAPE & 22.21 & 19.87 & 28.42 & 3.14 & 2.51 & 3.06 \\
        \hline
        Liang \etal{} \cite{Liang_prm_2020} & -- & ICSD & $R^2$ & 0.56$^*$ & 0.34$^*$ & 0.46$^*$ & 0.43$^*$ & 0.14$^*$ & 0.01$^*$ \\
        Choudhary \cite{Choudhary_jpcl_2025} & DGPT-formula & ICSD$^{\dagger}$ & $R^2$ & -0.04 (0.03) & 0.05 (0.12) & -0.08 (-0.31) & -0.55 (-0.66) & -4.50 (-6.41) & -1.15 (-1.32) \\
        Choudhary \cite{Choudhary_jpcl_2025} & DGPT-formula & Materials Project$^{\dagger}$ & $R^2$ & 0.11 (0.12) & -0.02 (0.03) & -0.13 (-0.18) & -0.82 (-1.00) & -2.77 (-3.93) & -1.02 (-1.03) \\
        Choudhary \cite{Choudhary_jpcl_2025} & DGPT-formula & RRUFF$^{\ddagger}$ & $R^2$ & 0.04 (-0.02) & -0.35 (-0.19) & -0.04 (-0.14) & -0.89 (-0.77) & -1.28 (-1.11) & -4.85 (-5.06) \\
        \textbf{Ours} & Cls. + Regr. Ensemble & ICSD$^{\dagger}$ & $R^2$ & \textbf{0.58} & \textbf{0.63} & \textbf{0.71} & \textbf{0.70} & \textbf{0.26} & \textbf{0.78} \\
        \textbf{Ours} & Cls. + Regr. Ensemble & Materials Project$^{\dagger}$ & $R^2$ & 0.49 & 0.48 & 0.49 & 0.38 & 0.11 & 0.45 \\
        \textbf{Ours} & Cls. + Regr. Ensemble & RRUFF$^{\ddagger}$ & $R^2$ & 0.54 & 0.39 & 0.25 & 0.33 & 0.26 & 0.18 \\
        \hline
    \end{tabular}
    \end{adjustbox}
    \caption{\textbf{Lattice parameter prediction errors of AlphaDiffract and reference models.} Errors are quantified in terms of the Mean Absolute Error (MAE), Mean Absolute Percentage Error (MAPE), and coefficient of determination ($R^2$). For direct comparison, we limit our analysis to studies that quantify prediction accuracy using regression metrics rather than classification metrics like match rate. Due to the scarcity of works tested on RRUFF, we also list those tested on other datasets for reference only. \revisednew{Point estimates are reported here to facilitate comparison across individual lattice parameters; bootstrap confidence intervals for aggregate lattice length and lattice angle performance are reported in Table~\ref{tab:lp_scores}, and propagated prediction variability is reported separately in Supplementary Table~\ref{tab:regression_uncertainty}.} \\ \footnotesize{$^*$Weighted average over models specialized for each Bravais lattice. \\
    $^{\dagger}$Evaluated on our validation set data. For inference with the DGPT-formula model, 1000 representative examples from the validation set of each dataset were selected for evaluation. Scores in parentheses refer to results on synthetic PXRD patterns with no added Poisson or Gaussian noise. \\
    $^{\ddagger}$Evaluated on our \revisednew{RRUFF benchmark dataset}.}}
    \label{tab:lp_scores_supp}
\end{table}

\clearpage

\begin{table}[h]
    \centering
    \footnotesize
    \begin{adjustbox}{max width=\textwidth}
    \begin{tabular}{cllcccccc}
        \hline
        \multirow{2}{*}{\textbf{Crystal System}} & \multirow{2}{*}{\shortstack[l]{\textbf{Dataset} \\ \textbf{(Number of Examples)}}} & \multirow{2}{*}{\textbf{Model}} & \multicolumn{3}{c}{\textbf{Lattice Length}} & \multicolumn{3}{c}{\textbf{Lattice Angle}} \\
        \cmidrule(lr){4-6} \cmidrule(lr){7-9}
        & & & MAE (\AA) & MAPE (\%) & $R^2$ & MAE ($^\circ$) & MAPE (\%) & $R^2$ \\
        \hline

        \multirow{6}{*}{Triclinic}
        & \multirow{2}{*}{\shortstack[l]{ICSD \\ (42,600)}}
            & Baseline & 2.61 & \textbf{27.25} & -0.05 & 11.18 & 12.99 & 0.00 \\
        & & \textbf{Ours} & \textbf{2.43} & 28.36 & \textbf{0.36} & \textbf{10.07} & \textbf{12.04} & \textbf{0.15} \\[6pt]
        & \multirow{2}{*}{\shortstack[l]{Materials Project \\ (237,500)}}
            & Baseline & 2.70 & 31.05 & 0.00 & 9.73 & 11.67 & 0.00 \\
        & & \textbf{Ours} & \textbf{2.19} & \textbf{26.81} & \textbf{0.32} & \textbf{9.25} & \textbf{11.16} & \textbf{0.09} \\[6pt]
        & \multirow{2}{*}{\shortstack[l]{RRUFF \\ (31)}}
            & Baseline & \textbf{1.79} & \textbf{23.82} & \textbf{-0.46} & 12.56 & 13.25 & -0.21 \\
        & & \textbf{Ours} & 2.86 & 36.85 & -2.20 & \textbf{9.99} & \textbf{11.05} & \textbf{0.16} \\

        \hline

        \multirow{6}{*}{Monoclinic}
        & \multirow{2}{*}{\shortstack[l]{ICSD \\ (215,900)}}
            & Baseline & 2.79 & 29.85 & -0.01 & 4.71 & 4.78 & 0.00 \\
        & & \textbf{Ours} & \textbf{2.07} & \textbf{21.99} & \textbf{0.40} & \textbf{3.30} & \textbf{3.27} & \textbf{0.22} \\[6pt]
        & \multirow{2}{*}{\shortstack[l]{Materials Project \\ (343,600)}}
            & Baseline & 2.47 & 28.46 & 0.00 & 4.36 & 4.46 & \textbf{0.00} \\
        & & \textbf{Ours} & \textbf{2.08} & \textbf{23.16} & \textbf{0.27} & \textbf{3.42} & \textbf{3.41} & -0.01 \\[6pt]
        & \multirow{2}{*}{\shortstack[l]{RRUFF \\ (63)}}
            & Baseline & 2.37 & \textbf{24.57} & -0.07 & 5.08 & 5.15 & -0.05 \\
        & & \textbf{Ours} & \textbf{2.28} & 24.68 & \textbf{-0.07} & \textbf{3.78} & \textbf{3.77} & \textbf{0.24} \\

        \hline

        \multirow{6}{*}{Orthorhombic}
        & \multirow{2}{*}{\shortstack[l]{ICSD \\ (333,400)}}
            & Baseline & 2.91 & 36.02 & -0.01 & \textbf{0.00} & \textbf{0.00} & \textbf{1.00}$^*$ \\
        & & \textbf{Ours} & \textbf{1.79} & \textbf{20.90} & \textbf{0.50} & 0.94 & 1.05 & 0.00$^\dagger$ \\[6pt]
        & \multirow{2}{*}{\shortstack[l]{Materials Project \\ (302,500)}}
            & Baseline & 2.64 & 30.31 & -0.01 & \textbf{0.00} & \textbf{0.00} & \textbf{1.00}$^*$ \\
        & & \textbf{Ours} & \textbf{1.87} & \textbf{21.28} & \textbf{0.41} & 1.18 & 1.31 & 0.00$^\dagger$ \\[6pt]
        & \multirow{2}{*}{\shortstack[l]{RRUFF \\ (73)}}
            & Baseline & 3.14 & 29.29 & -0.02 & \textbf{0.00} & \textbf{0.00} & \textbf{1.00}$^*$ \\
        & & \textbf{Ours} & \textbf{2.47} & \textbf{24.17} & \textbf{0.22} & 1.01 & 1.12 & 0.00$^\dagger$ \\

        \hline

        \multirow{6}{*}{Tetragonal}
        & \multirow{2}{*}{\shortstack[l]{ICSD \\ (271,600)}}
            & Baseline & 2.58 & 36.45 & 0.00 & \textbf{0.00} & \textbf{0.00} & \textbf{1.00}$^*$ \\
        & & \textbf{Ours} & \textbf{1.54} & \textbf{21.73} & \textbf{0.56} & 0.57 & 0.64 & 0.00$^\dagger$ \\[6pt]
        & \multirow{2}{*}{\shortstack[l]{Materials Project \\ (164,000)}}
            & Baseline & 2.57 & 35.18 & 0.00 & \textbf{0.00} & \textbf{0.00} & \textbf{1.00}$^*$ \\
        & & \textbf{Ours} & \textbf{1.85} & \textbf{25.80} & \textbf{0.43} & 0.88 & 0.98 & 0.00$^\dagger$ \\[6pt]
        & \multirow{2}{*}{\shortstack[l]{RRUFF \\ (19)}}
            & Baseline & 2.91 & 42.95 & -0.27 & \textbf{0.00} & \textbf{0.00} & \textbf{1.00}$^*$ \\
        & & \textbf{Ours} & \textbf{1.50} & \textbf{22.96} & \textbf{0.63} & 0.64 & 0.72 & 0.00$^\dagger$ \\

        \hline

        \multirow{4}{*}{Trigonal}
        & \multirow{2}{*}{\shortstack[l]{ICSD \\ (156,700)}}
            & Baseline & 4.92 & 49.05 & 0.00 & 4.13 & 4.01 & \textbf{0.33}$^{**}$ \\
        & & \textbf{Ours} & \textbf{2.24} & \textbf{19.14} & \textbf{0.60} & \textbf{2.99} & \textbf{2.59} & 0.18$^{**\dagger}$ \\[6pt]
        & \multirow{2}{*}{\shortstack[l]{Materials Project \\ (120,200)}}
            & Baseline & 4.88 & 48.29 & 0.00 & \textbf{4.10} & 3.99 & \textbf{0.33}$^{**}$ \\
        & & \textbf{Ours} & \textbf{2.81} & \textbf{22.66} & \textbf{0.49} & 4.37 & \textbf{3.76} & -0.20$^{**\dagger}$ \\

        \hline

        \multirow{4}{*}{Hexagonal}
        & \multirow{2}{*}{\shortstack[l]{ICSD \\ (213,700)}}
            & Baseline & 2.45 & 38.95 & 0.00 & 9.97 & 9.69 & 0.33$^{**}$ \\
        & & \textbf{Ours} & \textbf{1.10} & \textbf{17.16} & \textbf{0.67} & \textbf{2.53} & \textbf{2.23} & \textbf{0.53}$^{**\dagger}$ \\[6pt]
        & \multirow{2}{*}{\shortstack[l]{Materials Project \\ (102,400)}}
            & Baseline & 2.23 & 33.03 & -0.01 & 10.01 & 9.74 & 0.33$^{**}$ \\
        & & \textbf{Ours} & \textbf{1.26} & \textbf{19.15} & \textbf{0.59} & \textbf{4.12} & \textbf{3.69} & \textbf{0.39}$^{**\dagger}$ \\

        \hline

        \multirow{6}{*}{Cubic}
        & \multirow{2}{*}{\shortstack[l]{ICSD \\ (422,400)}}
            & Baseline & 2.48 & 41.03 & 0.00 & \textbf{0.00} & \textbf{0.00} & \textbf{1.00}$^*$ \\
        & & \textbf{Ours} & \textbf{1.12} & \textbf{17.16} & \textbf{0.70} & 0.38 & 0.42 & 0.00$^\dagger$ \\[6pt]
        & \multirow{2}{*}{\shortstack[l]{Materials Project \\ (196,100)}}
            & Baseline & 2.07 & 29.25 & -0.02 & \textbf{0.00} & \textbf{0.00} & \textbf{1.00}$^*$ \\
        & & \textbf{Ours} & \textbf{1.26} & \textbf{17.53} & \textbf{0.57} & 0.40 & 0.44 & 0.00$^\dagger$ \\[6pt]
        & \multirow{2}{*}{\shortstack[l]{RRUFF \\ (54)}}
            & Baseline & 3.61 & 38.95 & -0.52 & \textbf{0.00} & \textbf{0.00} & \textbf{1.00}$^*$ \\
        & & \textbf{Ours} & \textbf{1.24} & \textbf{13.75} & \textbf{0.62} & 0.36 & 0.40 & 0.00$^\dagger$ \\

        \bottomrule
    \end{tabular}
    \end{adjustbox}

    \caption{
    \revised{\textbf{Per-crystal system lattice parameter prediction errors of AlphaDiffract and
    crystal system-conditional baseline.}}
    \revised{The baseline predicts the mean lattice parameters of all training structures belonging to the same crystal system as the input. Errors are quantified in terms of the Mean Absolute Error (MAE), Mean Absolute Percentage Error (MAPE), and coefficient of determination ($R^2$). Trigonal and hexagonal systems are absent from the RRUFF benchmark dataset.}
    \revisednew{Point estimates are reported for this per-crystal-system diagnostic breakdown. Bootstrap confidence intervals for aggregate regression performance are reported in Table~\ref{tab:lp_scores} while propagated prediction variability is reported separately in Supplementary Table~\ref{tab:regression_uncertainty}.} \\
    \footnotesize{
    \revised{$^*$All angles are symmetry-constrained to 90$^\circ$ in these crystal systems under the Niggli reduced cell convention, giving the baseline trivially perfect angle metrics (MAE\,=\,0, MAPE\,=\,0, $R^2$\,=\,1). \\
    $^{**}$In trigonal and hexagonal systems, the obtuse angle (120$^\circ$ in the conventional cell) can appear as either $\alpha$ or $\gamma$ in the Niggli reduced cell depending on axis ordering, producing a bimodal distribution. The conditional mean sits between the two clusters, yielding $R^2 \approx 0.33$ for the baseline despite the symmetry constraint.} \\
    \revisednew{$^\dagger$For crystal systems in which at least one angle is symmetry-constrained to exactly 90$^\circ$, the corresponding $R^2$ should be interpreted with caution because zero variance in a constrained angle makes the parameter-specific $R^2$ undefined. The reported value is shown where computable; a value of 0.00 indicates that the computed $R^2$ is not meaningful due to zero variance in all angle targets.}
    }
    }
    \label{tab:per_cs_lp_scores}
\end{table}

\begin{table}[t]
    \centering
    \begin{tabularx}{\textwidth}{YYYYY}
        \hline
        \revisednew{\textbf{Partition}} &
        \revisednew{\textbf{Number of structures (database records)}} &
        \revisednew{\textbf{Number of patterns}} &
        \revisednew{\textbf{Number of structural clusters}} &
        \revisednew{\textbf{Number of unfingerprinted structures}} \\
        \hline

        \revisednew{ICSD -- Training} &
        \revisednew{149,071} &
        \revisednew{14,907,100} &
        \revisednew{96,506} &
        \revisednew{1} \\

        \revisednew{ICSD -- Validation} &
        \revisednew{16,563} &
        \revisednew{1,656,300} &
        \revisednew{14,544} &
        \revisednew{0} \\

        \revisednew{Materials Project -- Training} &
        \revisednew{131,970} &
        \revisednew{13,197,000} &
        \revisednew{119,975} &
        \revisednew{16} \\

        \revisednew{Materials Project -- Validation} &
        \revisednew{14,663} &
        \revisednew{1,466,300} &
        \revisednew{14,203} &
        \revisednew{1} \\
        \hline
    \end{tabularx}

    \caption{
        \revisednew{\textbf{Summary of training and validation dataset partitions.}
        Numbers of crystallographic database records, simulated PXRD patterns,
        structural clusters, and structures for which structural fingerprints
        could not be obtained are reported separately for the ICSD and Materials
        Project training and validation partitions.}
    }
    \label{tab:dataset_partitions}
\end{table}

\begin{figure}[b]
    \centering
    \includegraphics[width=\linewidth]{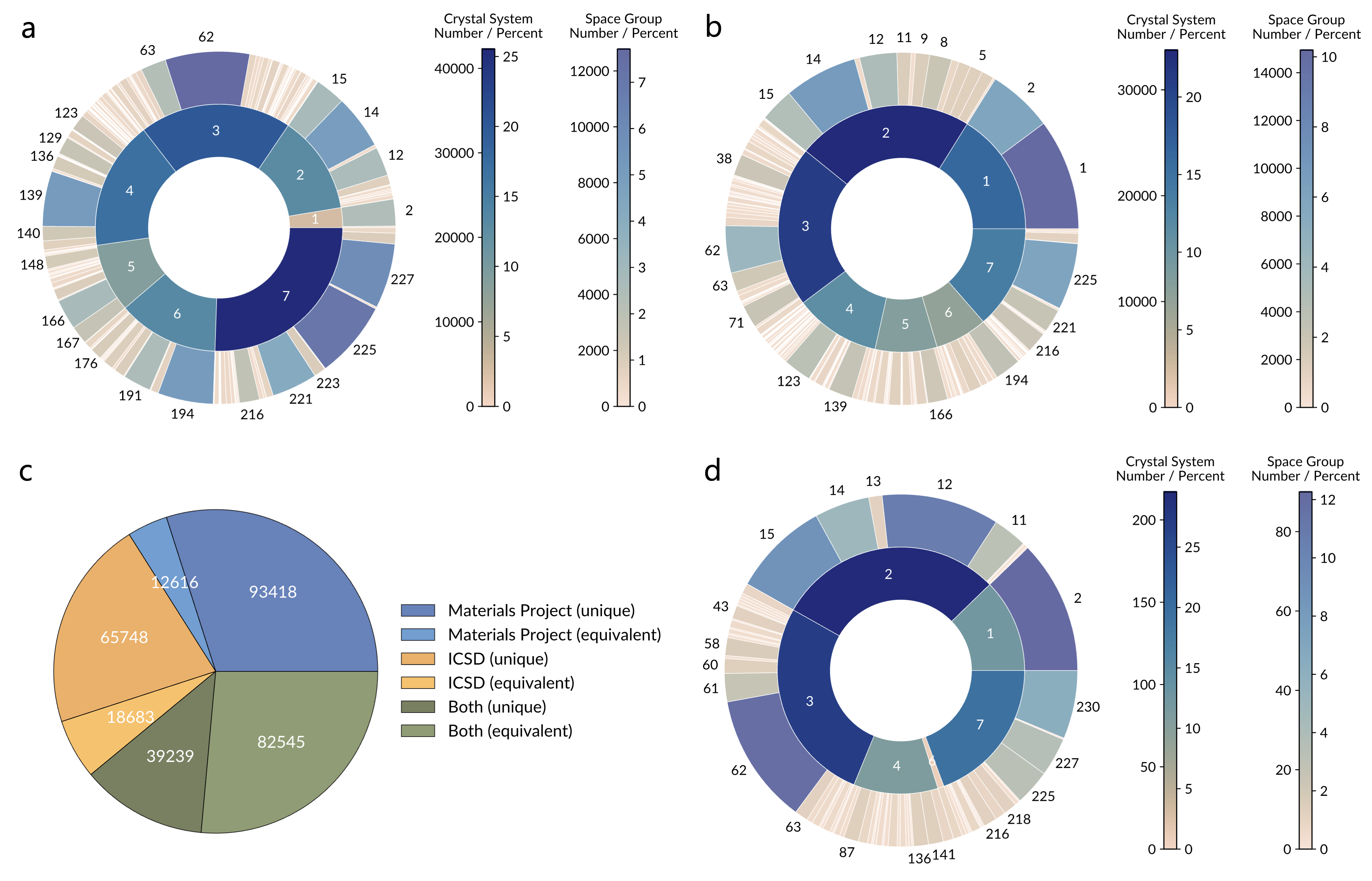}
    \caption{\textbf{Distribution of crystal systems and space groups across crystallographic databases and structural uniqueness analysis.} \textbf{a.} ICSD database showing the distribution of crystal systems (inner ring) and their associated space groups (outer ring) with color intensity indicating the number and percentage of structures. \textbf{b.} Materials Project database displaying the same hierarchical representation of crystal systems and space groups. \textbf{c.} Pie chart quantifying structural uniqueness and redundancy across ICSD and Materials Project databases, where ``unique'' structures are crystallographically distinct and ``equivalent'' structures are identified as similar to a unique structure based on the structure similarity metric detailed in \revisednew{Methods section ``\nameref{subsec:similarity}''.} \revisednew{Structural fingerprints could not be computed for 1 structure in the ICSD dataset and 17 structures in the Materials Project dataset and are excluded from the counts and percentages shown in this figure.} \textbf{d.} RRUFF database showing crystal system and space group distributions following the same visualization scheme as panels \textbf{a} and \textbf{b}. The seven crystal systems are: 1-triclinic, 2-monoclinic, 3-orthorhombic, 4-tetragonal, 5-trigonal, 6-hexagonal, and 7-cubic. Color bars indicate both absolute counts and relative percentages for crystal systems and space groups in each database.}
    \label{fig:data}
\end{figure}

\begin{figure}[h]
    \centering
    \includegraphics[width=\linewidth]{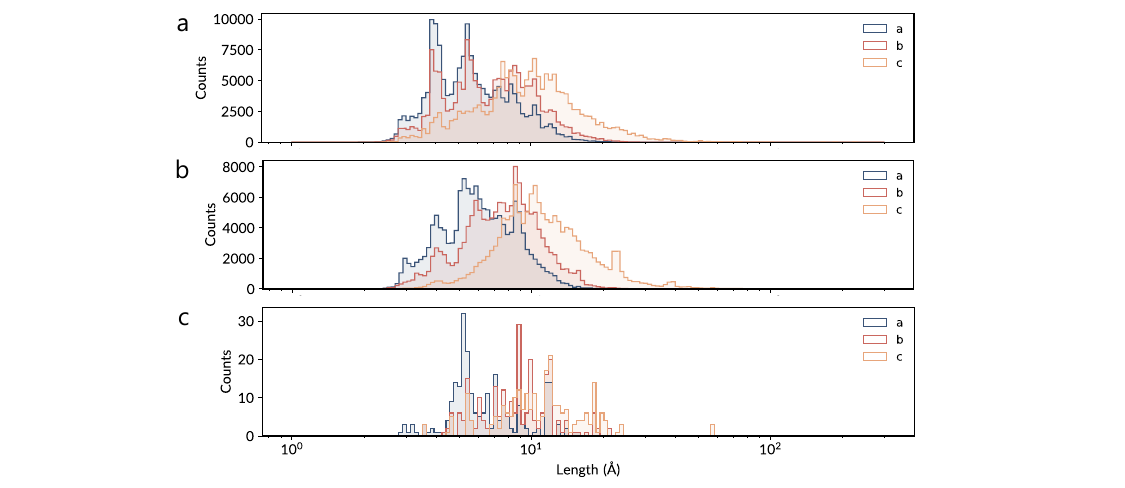}
    \caption{\textbf{Distribution of lattice lengths across crystallographic databases.} \textbf{a.} Histogram of Niggli reduced cell lattice lengths (a, b, c) for structures in the final ICSD dataset. \textbf{b.} Corresponding lattice length distributions for structures in the final Materials Project dataset. \textbf{c.} Lattice length distributions for structures in the RRUFF dataset used for regression.}
    \label{fig:data_lengths}
\end{figure}

\begin{figure}[h]
    \centering
    \includegraphics[width=\linewidth]{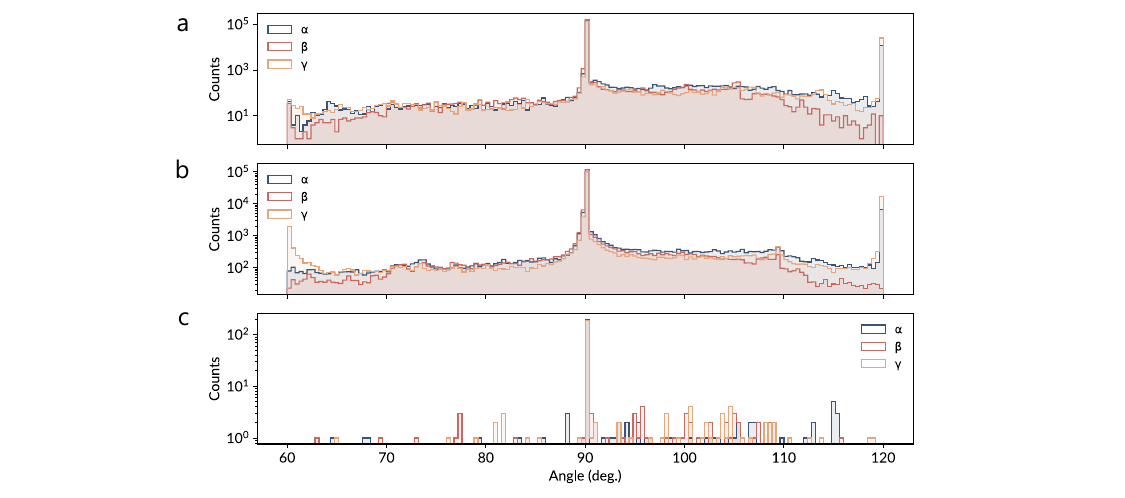}
    \caption{\textbf{Distribution of lattice angles across crystallographic databases.} \textbf{a.} Histogram of Niggli reduced cell lattice angles ($\alpha$, $\beta$, $\gamma$) for structures in the final ICSD dataset. \textbf{b.} Corresponding lattice angle distributions for structures in the final Materials Project dataset. \textbf{c.} Lattice angle distributions for structures in the RRUFF dataset used for regression.}
    \label{fig:data_angles}
\end{figure}

\begin{figure}[h]
    \centering
    \includegraphics[width=\linewidth]{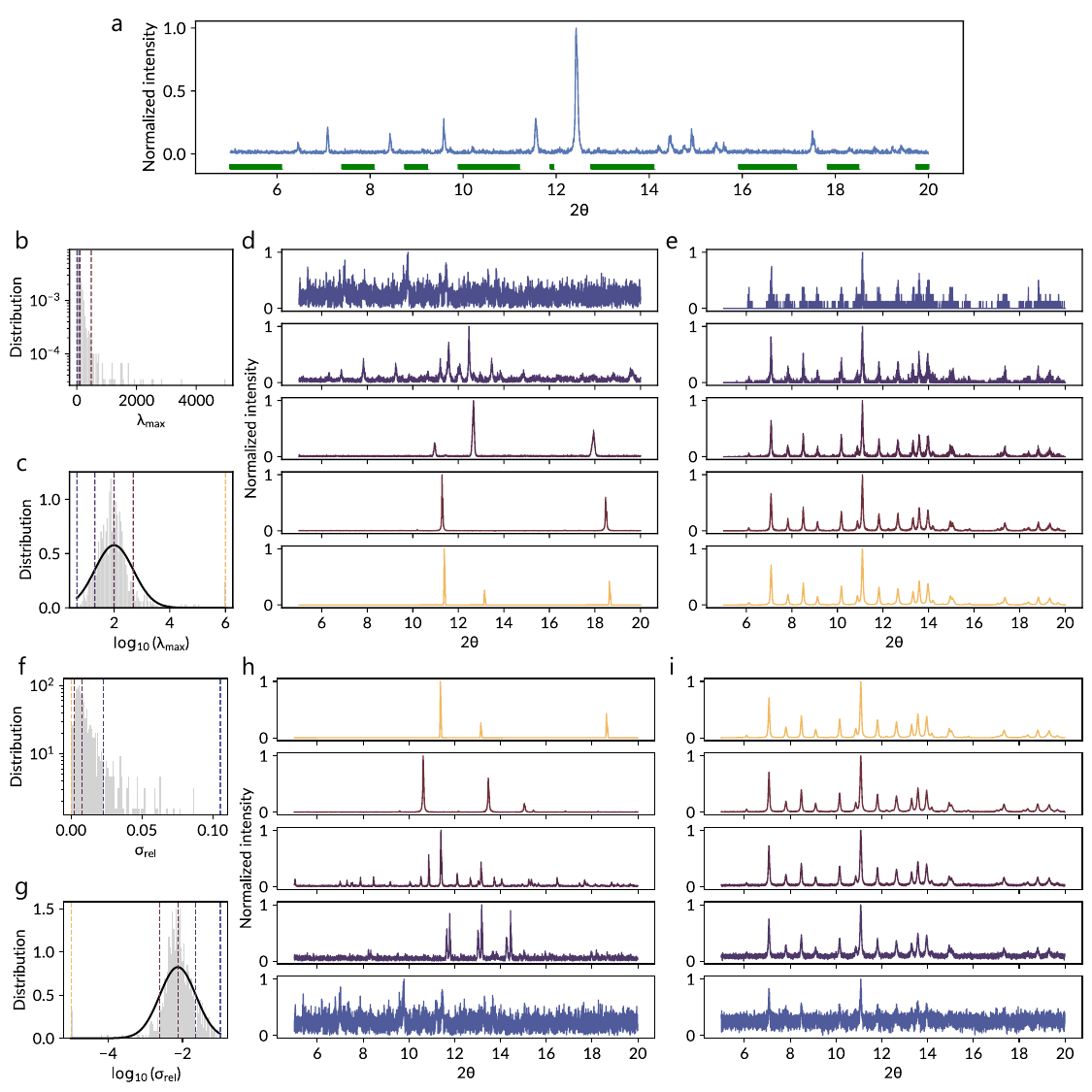}
    \caption{\textbf{Characterization and simulation of noise in X-ray diffraction patterns.} \textbf{a.} Representative PXRD pattern with background regions (underlined in green) used for noise estimation. \textbf{b-c.} Distribution of $\lambda_{\max}$ values from the RRUFF dataset in linear (\textbf{b}) and logarithmic (\textbf{c}) scales, where $\lambda_{\max}$ serves as the mean of the Poisson distribution used to sample noise values for simulation. Dashed vertical lines indicate representative $\lambda_{\max}$ values selected for visualization. \textbf{d.} Selected experimental PXRD patterns from the RRUFF database exhibiting the representative levels of Poisson noise. \textbf{e.} Simulated PXRD patterns with Poisson noise applied by sampling from distributions with means corresponding to the representative $\lambda_{\max}$ values. \textbf{f-g.} Distribution of $\sigma_{\text{rel}}$ values from the RRUFF dataset in linear (\textbf{f}) and logarithmic (\textbf{g}) scales, where $\sigma_{\text{rel}}$ represents the standard deviation of the Gaussian distribution used to sample noise values for simulation. Dashed vertical lines indicate representative $\sigma_{\text{rel}}$ values selected for visualization. \textbf{h.} Selected experimental RRUFF patterns at representative Gaussian noise levels. \textbf{i.} Simulated PXRD patterns with Gaussian noise applied at levels corresponding to the representative $\sigma_{\text{rel}}$ values. All PXRD patterns are plotted as normalized intensity versus 2$\theta$ (degrees).}
    \label{fig:noise_estimation}
\end{figure}

\begin{figure}[h]
    \centering
    \includegraphics[width=\linewidth]{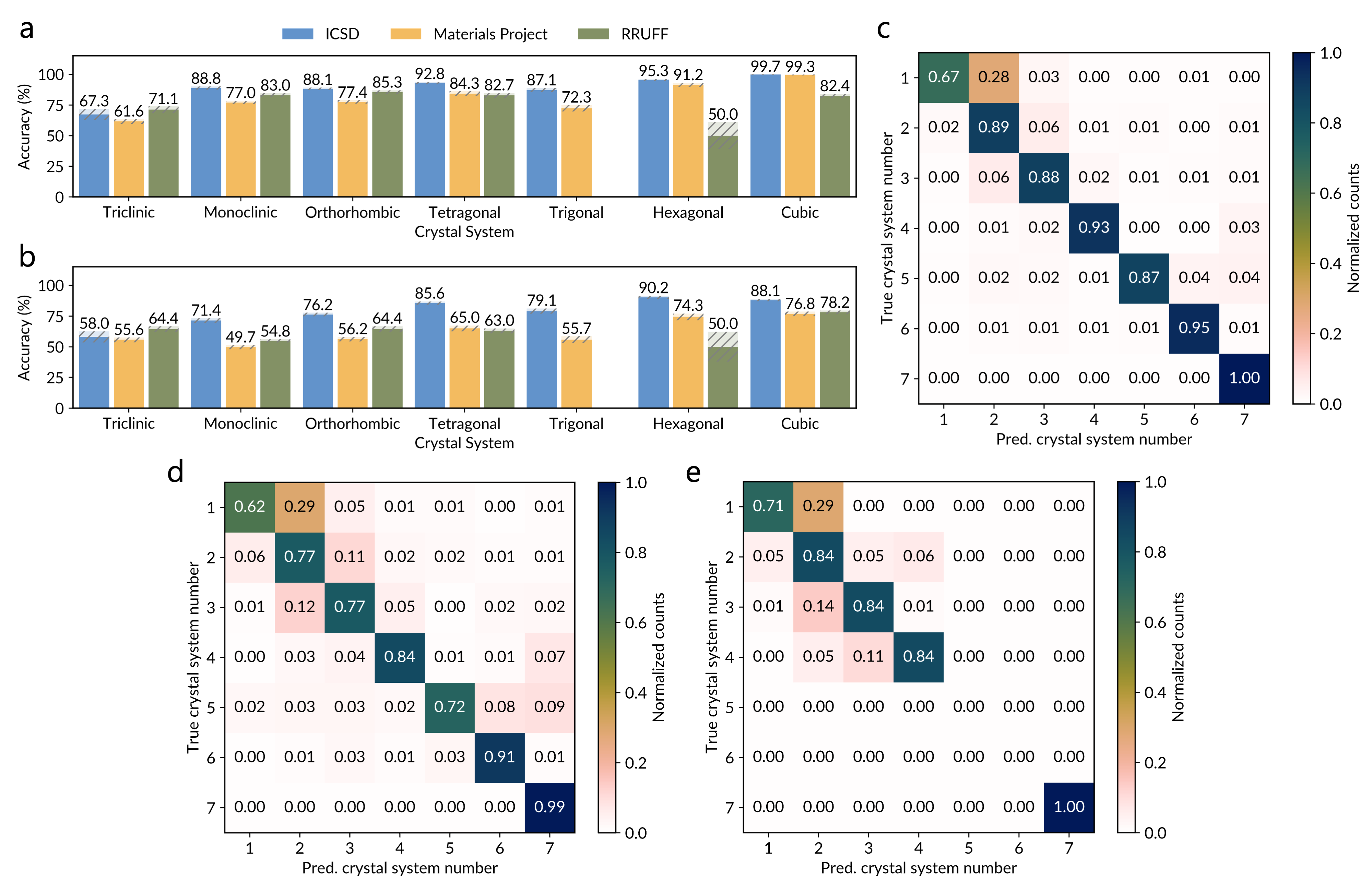}
    \caption{\textbf{Crystal system and space group classification accuracy per crystal system.} Accuracy (\%) of \textbf{a.} crystal system and \textbf{b.} space group classification, evaluated on ICSD (blue), Materials Project (orange), and RRUFF (green) datasets. Each group shows performance across the seven crystal systems. \revisednew{Hatched bars indicate propagated prediction variability associated with ensemble-member predictions and, where applicable, augmented input patterns; these quantities characterize prediction variability conditional on the evaluated structures and should not be interpreted as finite evaluation set confidence intervals.} Cubic systems achieve the highest accuracy across all datasets, while lower-symmetry systems (Triclinic, Monoclinic) show more variable performance, particularly for Materials Project and RRUFF datasets. \revised{\textbf{c--e.} Crystal system confusion matrices for the \textbf{c.} ICSD validation set, \textbf{d.} Materials Project validation set, and \textbf{e.} RRUFF \revisednew{benchmark dataset}, normalized by the true class size. Crystal system numbers correspond to: 1 -- Triclinic, 2 -- Monoclinic, 3 -- Orthorhombic, 4 -- Tetragonal, 5 -- Trigonal, 6 -- Hexagonal, 7 -- Cubic. The dominant off-diagonal pattern across all datasets is confusion between adjacent lower-symmetry systems, particularly triclinic--monoclinic and orthorhombic--monoclinic.}}
    \label{fig:accuracy_per_cs}
\end{figure}

\begin{figure}[h]
    \centering
    \includegraphics[width=\linewidth]{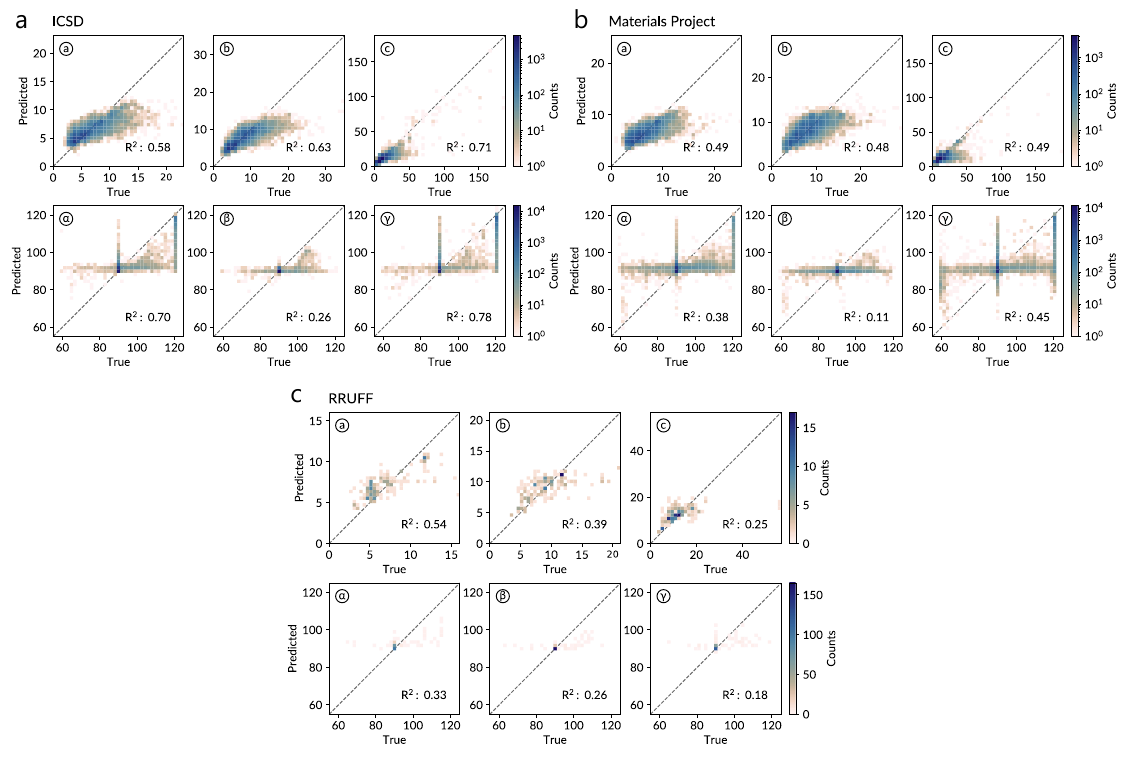}
    \caption{\revised{\textbf{Parity plots comparing predicted and true lattice parameters for ensemble model.} Parity plots for the \textbf{a.} ICSD validation set, \textbf{b.} Materials Project validation set, and \textbf{c.} RRUFF \revisednew{benchmark dataset}. Each panel group shows predictions for the three lattice lengths ($a$, $b$, $c$; top row) and three lattice angles ($\alpha$, $\beta$, $\gamma$; bottom row). Dashed lines indicate perfect agreement. $R^{2}$ values indicate goodness of fit. Heat map coloring represents point count per bin.}}
    \label{fig:regr_parity}
\end{figure}

\begin{figure}[h]
    \centering
    \includegraphics[width=\linewidth]{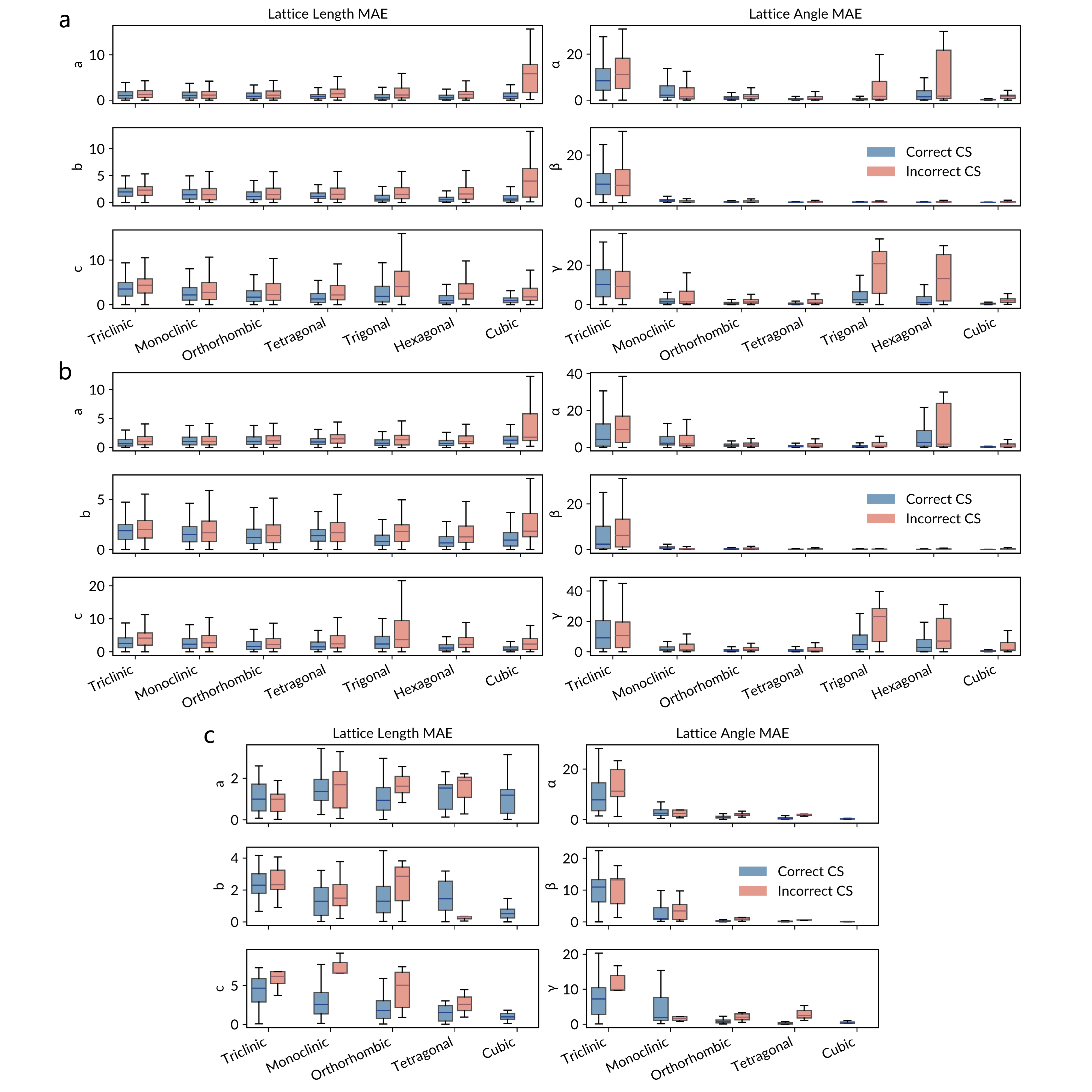}
    \caption{\revised{\textbf{Correlation between crystal system classification accuracy and lattice parameter prediction errors.} Box plots showing the distribution of per-parameter mean absolute error (MAE) for lattice lengths ($a$, $b$, $c$; left columns) and lattice angles ($\alpha$, $\beta$, $\gamma$; right columns), stratified by whether the crystal system (CS) was correctly (blue) or incorrectly (red) predicted by the ensemble model. Results are shown for \textbf{a.} ICSD validation set, \textbf{b.} Materials Project validation set, and \textbf{c.} RRUFF \revisednew{benchmark dataset}. Trigonal and hexagonal systems are absent from the RRUFF \revisednew{benchmark dataset}. Box boundaries indicate the interquartile range (IQR), horizontal lines indicate the median, and whiskers extend to 1.5$\times$IQR.}}
    \label{fig:per_cs_boxplots}
\end{figure}

\begin{figure}[h]
    \centering
    \includegraphics[width=\linewidth]{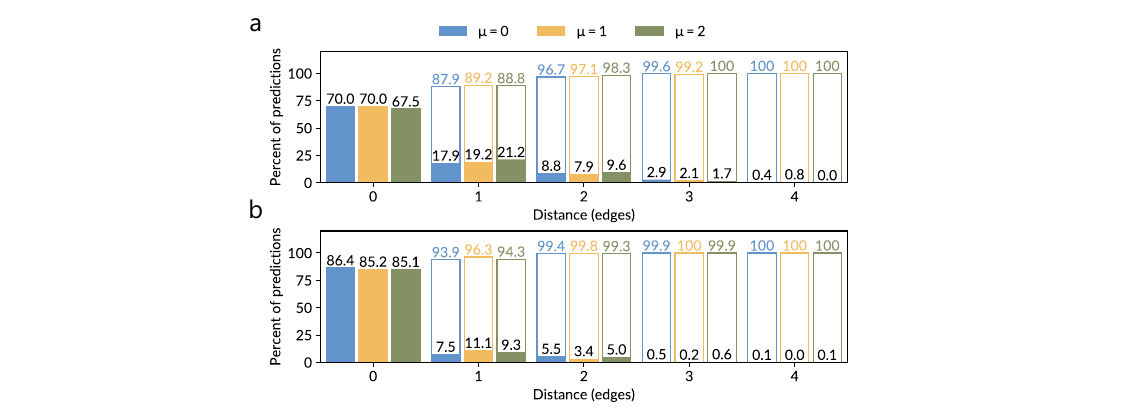}
    \caption{\textbf{Evaluation of space group predictions from experimental and synthetic RRUFF data.} Distribution of prediction errors as a function of graph distance (number of edges) from the true space group for RRUFF data subset with complete structures available (240 samples) using \textbf{a.} experimental and \textbf{b.} synthetic PXRD patterns as input. Filled bars show the percentage of predictions at each distance for three different weights applied to the GEMD loss term ($\mu$ = 0, 1, 2), while unfilled bars with labeled values indicate cumulative percentages.}
    \label{fig:rruff_exp_vs_sim}
\end{figure}

\begin{figure}[h]
    \centering
    \includegraphics[width=\linewidth]{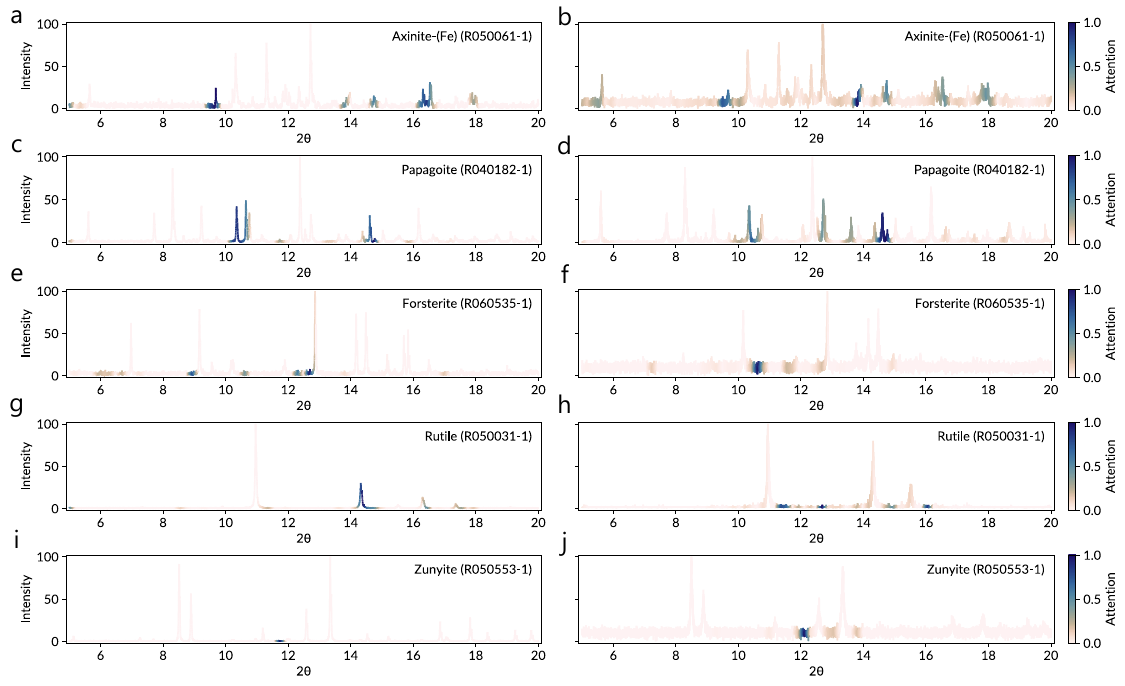}
    \caption{\textbf{GradCAM attention maps of feature importance for crystal system classification.} The left column shows experimental powder PXRD patterns from the RRUFF database, while the right shows synthetic patterns of the same mineral structures. Rows represent different crystal systems: \textbf{a-b.} triclinic, \textbf{c-d.} monoclinic, \textbf{e-f.} orthorhombic, \textbf{g-h.} tetragonal, \textbf{i-j.} cubic. The color scale indicates normalized attention weights extracted from the last ConvNeXt block.}
    \label{fig:attn_cs}
\end{figure}

\begin{figure}[h]
    \centering
    \includegraphics[width=\linewidth]{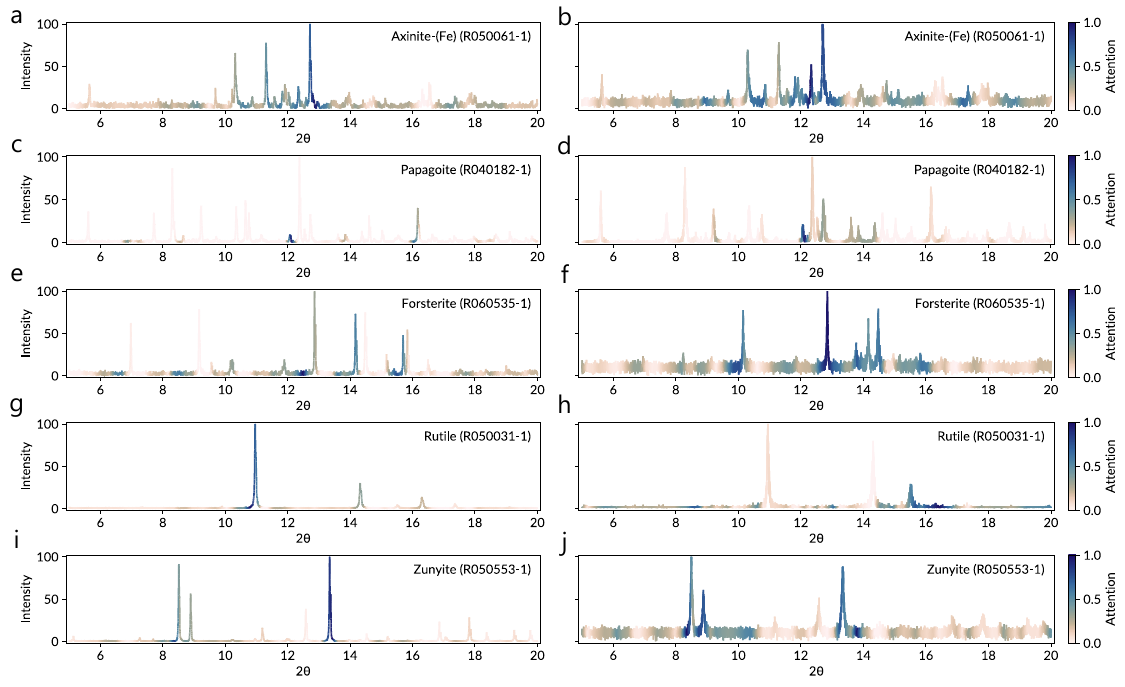}
    \caption{\textbf{GradCAM attention maps of feature importance for space group classification.} The left column shows experimental powder PXRD patterns from the RRUFF database, while the right shows synthetic patterns of the same mineral structures. Rows represent different crystal systems: \textbf{a-b.} triclinic, \textbf{c-d.} monoclinic, \textbf{e-f.} orthorhombic, \textbf{g-h.} tetragonal, \textbf{i-j.} cubic. The color scale indicates normalized attention weights extracted from the last ConvNeXt block.}
    \label{fig:attn_sg}
\end{figure}

\clearpage
\printbibliography[heading=bibintoc,title={Supplementary references}]